\documentclass[nonacm, acmsmall]{acmart}
\ifdefined\pdfpxdimen
\else\ifdefined\pxdimen
\fi\fi
\usepackage{amsmath}

\usepackage{xcolor}
\usepackage{colortbl}
\usepackage{caption}
\usepackage{subcaption}
\usepackage{contour}
\usepackage{listings}
\usepackage{algorithm}
\usepackage{algpseudocode}
\usepackage{graphicx} % Required for inserting images
\usepackage{tikz}
\usetikzlibrary{positioning, calc}
\usepackage{siunitx}
\definecolor{imgbordercolor}{HTML}{111111}
\definecolor{best}{HTML}{D4EDDA}
\definecolor{good}{HTML}{FFF3CD}
\definecolor{bad}{HTML}{F8D7DA}
\usepackage[capitalise]{cleveref}

\usepackage{microtype} % improves typesetting
\usepackage{bm}
\usepackage{minted}
\usepackage[varqu]{zi4}
\colorlet{codeboxbg}{black!3!white}
\colorlet{codeboxframe}{black!16!white}
\newsavebox{\codeboxsavebox}
\newenvironment{codebox}[2][]{%
  \VerbatimEnvironment
  \begin{center}%
  \begin{lrbox}{\codeboxsavebox}%
  \begin{minipage}{\dimexpr0.97\linewidth-2mm-0.6mm\relax}%
  \begin{minted}[
    fontsize=\footnotesize,
    breaklines,
    autogobble,
    fontfamily=tt,
    escapeinside=@@,
    style=tango, %trac
    linenos,
    mathescape=true,
    baselinestretch=1.08,
    #1]{#2}%
}{%
  \end{minted}%
  \end{minipage}%
  \end{lrbox}%
  \begin{tikzpicture}%
    \node[draw=codeboxframe, fill=codeboxbg, line width=0.3mm,
          rounded corners=2pt, inner xsep=1.8mm, inner ysep=1mm]
      {\usebox{\codeboxsavebox}};%
  \end{tikzpicture}%
  \end{center}%
}
\floatstyle{plain} % figure-like: caption below
\newfloat{code}{htbp}{loc}
\floatname{code}{Code} % caption label says "Code"
\crefname{code}{code}{codes}
\Crefname{code}{Code}{Codes}

\AtBeginDocument{%
  }

\usepackage{xcolor}

\newif\ifmarkups
\markupsfalse

\ifmarkups
  \newcommand{\rev}[2]{{\color{red}{#1}}}
  \newcommand{\revlegal}[2]{{\color{orange}{#1}}}
\else
  \newcommand{\rev}[2]{{#1}}
  \newcommand{\revlegal}[2]{{#1}}
\fi

\setcopyright{cc}
\acmDOI{10.1145/3820015}

\title{Deferred Software Rasterization for Efficient Real-time Hair Rendering}
\author{Lukas Lipp}
\email{{lukas.lipp@cg.tuwien.ac.at}}
\affiliation{%
  \institution{Meta}
  \city{Z\"urich}
  \country{Switzerland}
}

\affiliation{%
  \institution{TU Wien}
  \department{Rendering and Modeling Group}  
  \city{Vienna}
  \country{Austria}
}

\author{Adrian Jarabo}
\affiliation{%
  \institution{Meta}
  \city{Zaragoza}
  \country{Spain}
}

\author{Michael Wimmer}
\affiliation{%
  \department{Rendering and Modeling Group}
  \institution{TU Wien}
  \state{Vienna}
  \country{Austria}
}

\author{Lukas Bode}
\affiliation{%
  \institution{Meta}
  \city{Z\"urich}
  \country{Switzerland}
}
\date{February 2026}

\begin{document}

\begin{abstract}
In this work we propose an efficient deferred software rasterization pipeline for real-time rendering of strand-based hair using hair meshes. 
Hair plays a crucial role in creating expressive 3D characters, yet strand-based approaches are often restricted to high-end hardware and typically applied to only a small number of hero characters. Hair meshes have proven to be an efficient representation capable of handling a wide variety of groom styles, but existing mesh shader-based implementations still suffer from significant bottlenecks.
In this work, we address these limitations with a software rasterization approach that improves performance and compatibility. Our method enables efficient far-field strand-based hair rendering—even at a single sample per pixel—by combining deferred shading with a strand filtering and reconstruction step, while requiring only minimal hardware support. To further enhance scalability, we introduce a level-of-detail (LOD) scheme that adapts hair representation and shading complexity based on viewing distance and screen-space coverage, reducing computational cost further while preserving visual fidelity. To the best of our knowledge, this is the first approach to achieve this combination of efficiency, flexibility, scalability, and broad hardware compatibility.
\end{abstract}


\begin{teaserfigure}%
\centering%
\begin{tikzpicture}%
    \node[anchor=south west, inner sep=0, draw=black] (A) at (0,0)
        {\includegraphics[width=\textwidth]{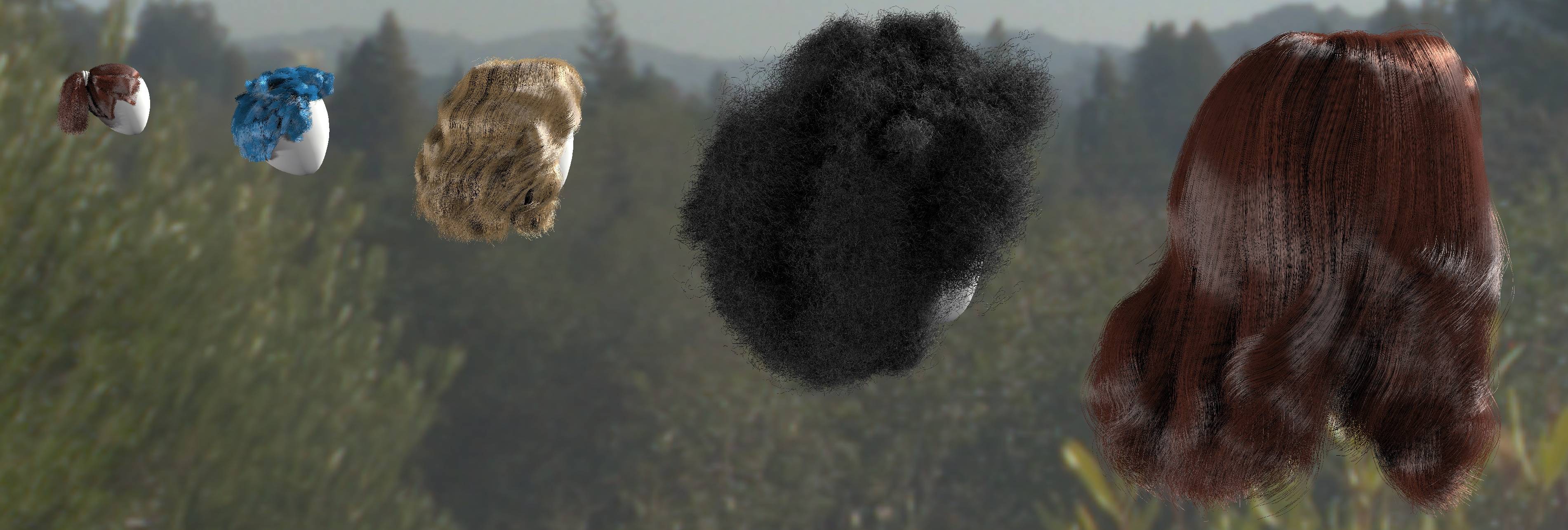}};
    \begin{scope}
        \clip ([yshift=-18pt]A.north west) -- ([yshift=60pt]A.south east) -- (A.south east) -- (A.south west) -- cycle;
        \node[anchor=south west, inner sep=0, draw=black] at (0,0)
            {\includegraphics[width=\textwidth]{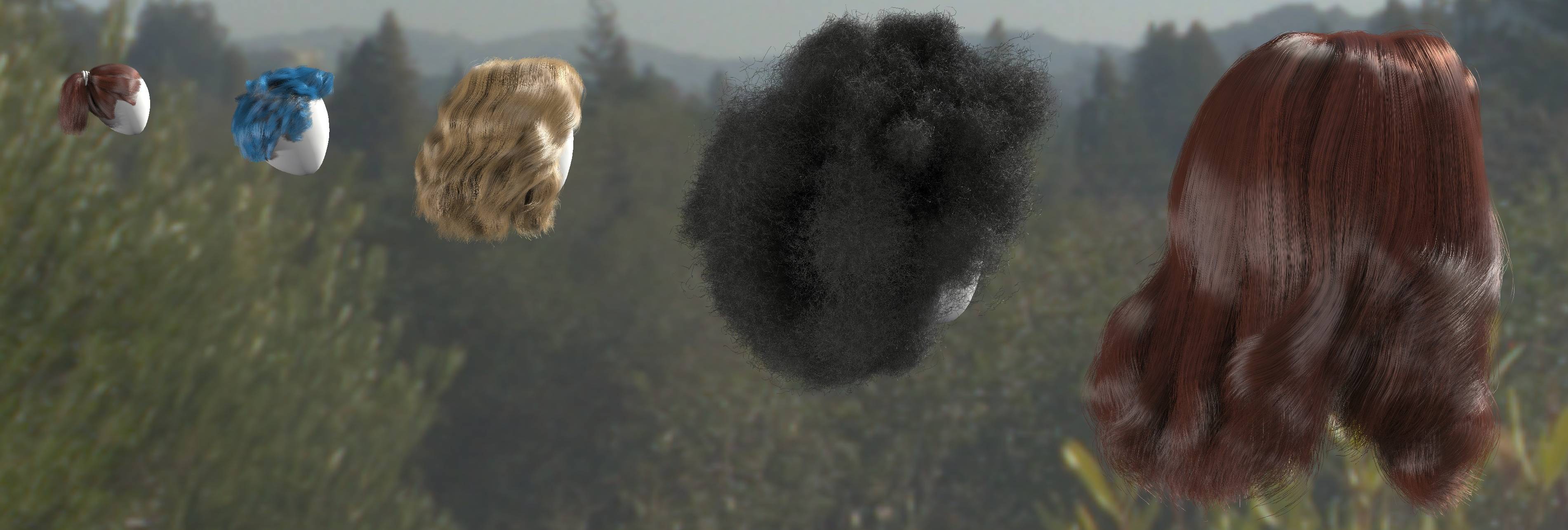}};
    \end{scope}
    \draw[white, line width=0.7pt, cap=round] ([yshift=-18pt]A.north west) -- ([yshift=60pt]A.south east);
    % 3 square zoom insets along bottom-left — same crop from each image
    % Zoom parameters: center (cx,cy) in bp, half-size hs in bp
    \def\zcx{1060}% center x in bp (image coords)
    \def\zcy{772}% center y from bottom in bp
    \def\zhs{100}% half-size of square crop in bp (200x200 total)
    \pgfmathsetlengthmacro{\ztl}{\zcx-\zhs bp}%       left trim
    \pgfmathsetlengthmacro{\ztr}{3802bp-\zcx bp-\zhs bp}%  right trim
    \pgfmathsetlengthmacro{\ztb}{\zcy-\zhs bp}%        bottom trim
    \pgfmathsetlengthmacro{\ztt}{1284bp-\zcy bp-\zhs bp}%  top trim
    \pgfmathsetlengthmacro{\zs}{0.13\textwidth}%
    \pgfmathsetlengthmacro{\zg}{3pt}%
    \node[anchor=south west, inner sep=0, draw=white, line width=1pt] (Z1) at ([shift={(\zg,\zg)}]A.south west)
        {\includegraphics[width=\zs, trim={\ztl{} \ztb{} \ztr{} \ztt{}}, clip]{figure/teaser/mesh_lod_msaa0_noao2.jpg}};
    \node[anchor=south west, inner sep=0, draw=white, line width=1pt] (Z2) at ([shift={(\zg,0)}]Z1.south east)
        {\includegraphics[width=\zs, trim={\ztl{} \ztb{} \ztr{} \ztt{}}, clip]{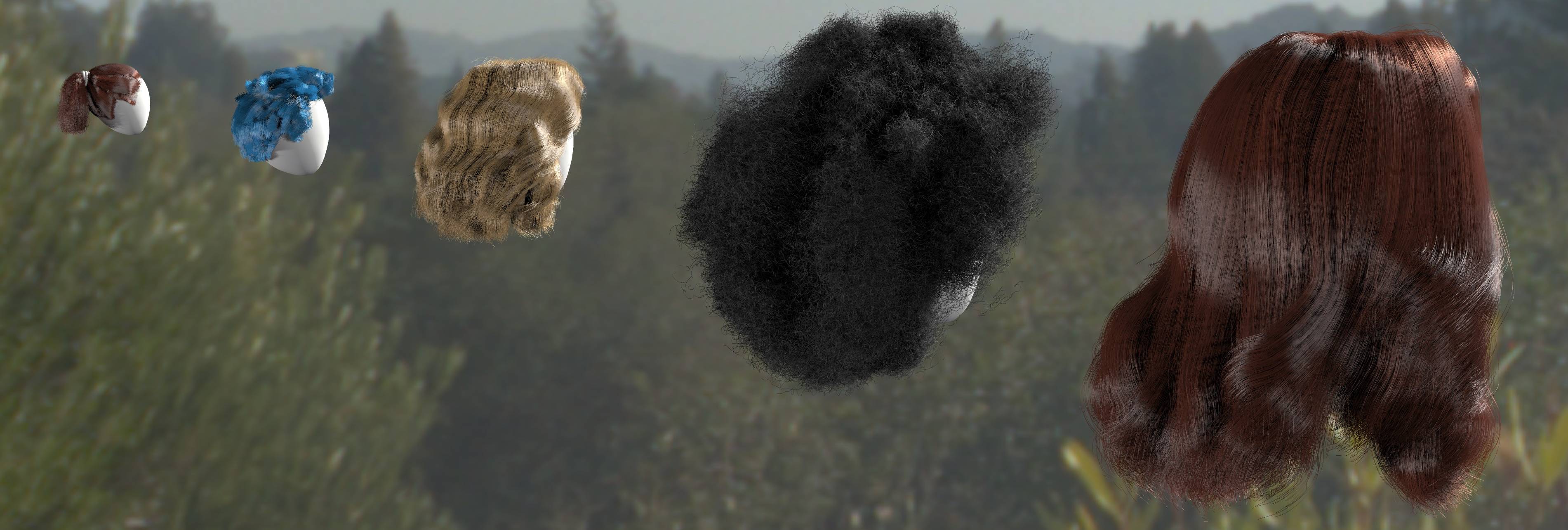}};
    \node[anchor=south west, inner sep=0, draw=white, line width=1pt] (Z3) at ([shift={(\zg,0)}]Z2.south east)
        {\includegraphics[width=\zs, trim={\ztl{} \ztb{} \ztr{} \ztt{}}, clip]{figure/teaser/swr_lod_msaa0_filter_ao2.jpg}};
    % Zoom labels — white text with black outline (no background box)
    \node[anchor=south, inner sep=1pt, font=\tiny\bfseries] at (Z1.north) {\contour{black}{\textcolor{white}{Mesh 1$\times$ - 37.6ms}}};
    \node[anchor=south, inner sep=1pt, font=\tiny\bfseries] at (Z2.north) {\contour{black}{\textcolor{white}{Mesh 4$\times$ - 50.5ms}}};
    \node[anchor=south, inner sep=1pt, font=\tiny\bfseries] at (Z3.north) {\contour{black}{\textcolor{white}{Ours 1$\times$ - \textcolor{green!60!white}{9.4ms}}}};
    % Main labels — white text with black outline
    \node[anchor=south east, inner sep=2pt, font=\footnotesize\bfseries] at ([shift={(-2pt,2pt)}]A.south east) {\contour{black}{\textcolor{white}{Ours, 1$\times$ MSAA}}};
    \node[anchor=north west, inner sep=2pt, font=\footnotesize\bfseries] at ([shift={(2pt,-2pt)}]A.north west) {\contour{black}{\textcolor{white}{Mesh Shader, 1$\times$ MSAA}}};
    % Dotted rectangle marking the zoom crop area (drawn last, on top)
    % Uses same zcx, zcy, zhs variables as the zoom insets
    \pgfmathparse{\zcx-\zhs}\edef\zrxl{\pgfmathresult}%
    \pgfmathparse{\zcx+\zhs}\edef\zrxr{\pgfmathresult}%
    \pgfmathparse{\zcy-\zhs}\edef\zryb{\pgfmathresult}%
    \pgfmathparse{\zcy+\zhs}\edef\zryt{\pgfmathresult}%
    \path let \p1=(A.south west), \p2=(A.north east),
              \n{w}={\x2-\x1}, \n{h}={\y2-\y1} in
        coordinate (ZBL) at (\x1+\zrxl/3802*\n{w}, \y1+\zryb/1284*\n{h})
        coordinate (ZTR) at (\x1+\zrxr/3802*\n{w}, \y1+\zryt/1284*\n{h});
    % Draw 4 sides separately — each starts with a dot at the corner
    \coordinate (ZTL) at (ZBL |- ZTR);
    \coordinate (ZBR) at (ZBL -| ZTR);
    \draw[white, line width=0.8pt, line cap=round, dash pattern=on 2pt off 2pt, dash phase=0.5pt] (ZBL) -- (ZBR);
    \draw[white, line width=0.8pt, line cap=round, dash pattern=on 2pt off 2pt, dash phase=0.5pt] (ZBR) -- (ZTR);
    \draw[white, line width=0.8pt, line cap=round, dash pattern=on 2pt off 2pt, dash phase=0.5pt] (ZTR) -- (ZTL);
    \draw[white, line width=0.8pt, line cap=round, dash pattern=on 2pt off 2pt, dash phase=0.5pt] (ZTL) -- (ZBL);
\end{tikzpicture}%
\caption{Multiple grooms (total of 483k strands, 4K resolution) rendered at different distances to the camera using our deferred software rasterization pipeline with level-of-detail and baked \rev{ambient}{} occlusion, compared against the reference mesh shader-based mplementation~\cite{bhokareRealTimeHairRendering2024a}. Our method is 4x faster than the mesh shader, with quality comparable to 4x MSAA. Per-style LOD strength (left to right) $\lambda{=}\{4.0, 5.5, 2.0, 2.8, 3.0\}$.}
\label{fg:teaser}
\end{teaserfigure}
% Total/Remaining Strands = 483,416/244,154
\maketitle

\section{Introduction}
Rendering hair and fur is an important long-standing problem in computer graphics. Its intricate volumetric look, formed by hundreds of thousands of individual thin strands, makes it one of the most challenging appearances to render: It requires a large amount of small sub-pixel-scale semi-transparent primitives with a high-frequency scattering function. 
This results into a significant rendering \revlegal{computational}{} cost due to memory bandwidth and overdraws, but also is prone to aliasing due to rendering thin primitives with glinty appearance. As a result, a common approach in real-time rendering is to severely simplify hair rendering, either by means of textured surfaces~\cite{scheuermann2004practical} or cards~\cite{jiang2016volumetric_uncharted4}, and only recently explicit strand rendering has become feasible in real-time~\cite{tafuri2019strand,taillandier2020every_strand_counts}. However, these current strand-based approaches are usually restricted to very high-end hardware and for a limited number of hero characters, while on the other hand there is a growing demand for rendering large numbers of characters, each with unique and detailed hairstyles, efficiently across a wide range of hardware.

To partially alleviate some of these problems, \citet{bhokareRealTimeHairRendering2024a} proposed to represent hair strands implicitly using a coarse cage-like structure of the groom (the \emph{hair mesh}) and styling operators, and only generate them on-the-fly at rendering time. This significantly reduced bandwidth requirements, becoming significantly more efficient than the commonly used linear hair skinning (LHS)~\cite{somasundaram2015dynamically}. 
However, this only addressed part of the problems: A significant number of overdraws were still required, and aliasing was still an issue, which degraded the rendering quality. Moreover, Bhokare's work~\citep{bhokareRealTimeHairRendering2024a} uses a mesh shader-based implementation, which is not supported by all existing hardware. 

In this work, we introduce a new solution for efficient real-time hair rendering, which builds upon the idea of~\citet{bhokareRealTimeHairRendering2024a}, but implements it as a highly efficient deferred software rasterizer. This has several benefits, including better support across platforms, improved efficiency on the strand generation by carefully designing the GPU warp synchronization, and a significant reduction of the shading \revlegal{computational}{} cost by shading only the visible hair strands via a deferred scheme, while at the same time reducing aliasing by post-filtering from strand-pixel intersection data. 
We accelerate this deferred pipeline by using a level-of-detail scheme that reduces both the number of strands and their complexity without visual penalty.  
Finally, we leverage the hair mesh structure as a proxy for occlusion, allowing us to use complex image-based lighting with occlusions, significantly increasing the type of lighting that can be used to shade explicit strands in real time. 
All of these contributions combined are a step forward for efficient hair rendering in real-time, allowing efficient rendering and shading of hair grooms with complex lighting and reduced visual artifacts (\cref{fg:teaser}). In particular, our contributions are: 
\begin{itemize}
    \item A fast indirect dispatch based LOD-enabled hair software rasterization pipeline, with parallel construction of styled strands via SIMD-wide register communication and deferred shading using an atomic-enabled compressed G-buffer;
    \item \rev{a statistically driven depth-map visibility correction method for on-the-fly level-of-detail adaptation of groom complexity}{a level-of-detail technique that adapts the complexity of the rendered groom on the fly, and statistically corrects visibility via depth map};
    \item a \rev{single-pass}{} strand connectivity reconstruction and filtering \rev{method, enabled by}{involving} conservative rasterization, that produces smooth, anti-aliased transparent strands even in our single-sample pipeline;
    \item and an ambient occlusion term directly encoded on the hair mesh that allows volumetric complex probe-based lighting in grooms.
\end{itemize}

%\adrian{I need a better picture of the whole method to continue this part; right now, it seems all Method has the same level of importance/detail}\LL{I would say so. We don't really have a core contribution and I think the most unique ones are probably the depth correction, the strand connect filter using the ignore sample layer instead, the compressed gbuffer/deferred rendering and the ao/sh lighting. Everything else is either a remix of something existing or it is not clear if we improved the method because there is no open source reference and we don't know how they did it. So the contribution is the whole package. At the same time I am not sure if going super deep with any of these things would help, because most of it should be know. E.g., writing a swr for hair is not something that needs to be explained line for line, and the higher level insights we share are more helpful overall. Also the paper is already quite long. }. 

%The text width is: \printinunitsof{pt}\prntlen{\textwidth}

\section{Related Work \& Background}
%\adrian{Maybe instead of "related work" we can do something different as the usual approach in graphics (though it is not uncommon), and make it a "Background" section. This would allow us explaining hair meshes and software rasterization. In the hair meshes part, we can mention for context LHS (and the animation papers for hair meshes), as well as hair shading models. Thoughts?}\LL{I am open to this, I think having related work and background could bloat this kind of paper too much. Also some background parts are also in the method section (e.g lod and filter)}

\paragraph*{Hair rendering}
Early methods for rendering of hair in real-time leveraged textured hair surfaces~\cite{scheuermann2004practical} or cards~\cite{koh2001simple}. With the improvement of GPUs, rasterizing explicit strands have become feasible, with two main problems: On one hand, the complexity of grooms — with over 100K splines per groom — makes it unpractical to store and transfer the explicit strands. On the other hand, the amount of overdraw of semi-transparent strands results in a very \revlegal{computationally}{} expensive, and aliasing-prone, rendering and shading phase. 
To avoid the former problem, the common approach is to only store, edit, simulate, and transfer a set of representative \emph{guide hairs}, which are interpolated by each individual strand via \emph{linear hair skinning} (LHS)~\cite{somasundaram2015dynamically}, optionally applying some procedural styling operator~\cite{chang2025transforming}. This, however, requires storing the interpolation weights per strand, making it less practical in terms of memory footprint and bandwidth. Other more recent approaches have leveraged neural networks for the strand generation~\cite{rosu2022neural}, but these methods are still far from practical for real-time rendering. A more compact approach was proposed by \citet{bhokareRealTimeHairRendering2024a}, which used the \emph{hair mesh} structure \cite{yukselHairMesh} as a proxy for representing the whole groom. The hair mesh uses a proxy geometry to model the overall shape of a groom, using connected meshes to define hair bundles. Then, in render time, explicit strands are generated in a mesh shader, by interpolating within each individual bundle: In essence, this is very similar to LHS, where the edges of the bundle act as guide hairs, with the key difference that the interpolation weights are implicitly defined by the position of the strand in the bundle, and thus no weight needs to be stored, which significantly reduces the storage and bandwidth requirements. We build on Bhokare's approach, but proposing a highly-efficient method based on a deferred software rasterization method, while at the same time increasing its representative power and adding extended support for level of detail.

\paragraph*{Software rasterization}
With the advent of fully-programmable GPU pipelines, software rasterization has been demonstrated to be a viable pipeline~\rev{\cite{laine2011high_performance_software_rasterization,karis2021nanite}}{\cite{laine2011high_performance_software_rasterization}}, even superior to hardware rasterization in certain scenarios with semitransparent objects or subpixel features, such as point clouds~\cite{markus2022} or semi-transparent Gaussians~\cite{kerbl20233d}. In the context of hair, \citet{taillandier2020every_strand_counts} proposed using software rasterization for efficient hair rendering in AAA games, by leveraging tiled per-pixel depth-bucket-based order-independent transparency (OIT) ~\cite{everitt2001interactive,thibieroz2018order} and analytic line anti-aliasing. Later work by~\citet{ishihara2023resident_evil_4_hair_discussion} reduced memory consumption by implementing OIT using only \rev{three layers}{three} for hair rendering. Hair software rasterization is nowadays a standard in high-end AAA game engines~\cite{cifariellociardi2022probe_based_lighting_unity_enemies, kulikov2025indiana_jones_strand_based_hair_fur, epicgames2021unreal}. We describe an efficient single-bucket deferred software rasterization pipeline for hair, followed by a filtering pipeline to account for semi-transparency and aliasing.

\paragraph*{Hair level-of-detail}
An industry-standard for hair level-of-detail is to switch strand representations to cards, though it is usually \revlegal{a computationally}{an} expensive, manual process, with the notable exception of the work of~\citet{zheng2025}, which automatized the process using differentiable rendering. 
Stochastic simplification~\cite{stochasticSimplification} has been successfully applied to hair simplification by reducing the number of strands and thickening the remaining ones to account for the lost projected area, as well as reducing the complexity of the strands themselves by removing control points~\cite{bhokareRealTimeHairRendering2024a}. \citet{zhu2022practical, zhu2023practical} clustered groups of strands accounting for the aggregated scattering and internal scattering in the cluster. A similar approach was used by \citet{montazeri2020practical, montazeri2021practical} in the context of cloth rendering. We built upon the stochastic simplification approach for level of detail, but introducing simple yet efficient approaches to prevent for aliasing and other artifacts due to these simplifications. 

\rev{\paragraph*{Anti-Aliasing}
Due to their extremely small screen-space footprint, individual hair strands suffer from severe under-sampling and noisy rendering. Furthermore, the thin, intricate, and highly detailed sub-pixel structure of hair presents a unique spatial challenge that general anti-aliasing and modern temporal upscaling methods, such as AMD FidelityFX Super Resolution \shortcite{amd2021fsr} or NVIDIA DLSS 4 \shortcite{nvidia2025dlss4}, struggle to resolve accurately without introducing blurring or ghosting artifacts. While some methods rely on neural-networks \cite{neuralfilter2022}, recent work by \citet{huangDetailPreservingRealTimeHair2025} demonstrates that a non-neural, explicit filtering framework can offer a superior alternative. By implementing explicit strand linking and directional filtering, their algorithmic approach preserves sub-pixel continuity and fine fiber details, like flyaway strands, without the blurring typical of black-box networks.
}{}

\paragraph*{Background: Hair mesh rendering}
The key idea of \citet{bhokareRealTimeHairRendering2024a} is to utilize the implicit definition of strands using hair meshes~\cite{yukselHairMesh}. Hair meshes define the groom using a set of geometric bundles (see~\cref{fg:hm_assemble}, left) that define the coarse structure of the groom. Each bundle is formed by a set of $L$ layers, with the first and last layers placed at the hair roots and tips respectively. Each layer $i$ is defined as a polygonal shape (in our case, we use quads) with vertices $\mathbf{v}^i$ and associated per-vertex tangents $\mathbf{t}^i$ and uvw coordinates. 

During rendering, each strand is assembled as a spline with $N$ control points uniformly spaced along the length of the strand $w \in [0,1]$, with points at $w=0$ and $w=1$ placed the root and tip layers, respectively. Each strand origin is sampled using its uv coordinates which remain fixed along the strand, and its base position is computed by using spline interpolation from the bilinear interpolation of vertices and tangents in the bundle layers. Then, the final strand is obtained by applying the styling function as a function of each control point's uvw coordinates.

\begin{figure*}[t]
    \includegraphics[width=\textwidth]{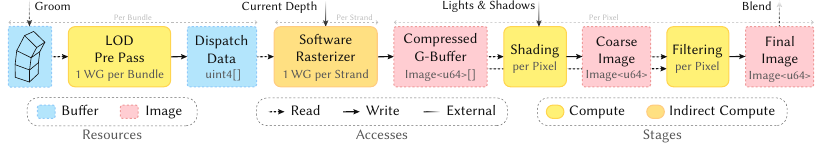}
  \caption{Our software rasterization pipeline consumes a groom represented as a simple mesh composed of multiple bundles. A level-of-detail (LOD) pre-pass first prepares an indirect dispatch buffer, determining how many strands and control points will be rendered. During the rasterization stage, the pipeline reads the existing depth buffer and \rev{updates strand data in}{writes strand data to} a \texttt{uint64} \rev{composite buffer}{image} using \texttt{atomicMin} operations. This data is subsequently used for per-pixel shading, with an optional filtering stage applied to further refine the result. The final output is then blended into the main render target. WG = workgroup. }
  \label{fg:hm_pipeline}
\end{figure*}

\section{Our Method}

Our deferred software rasterization, designed for grooms encoded as hair meshes, is described in~\cref{fg:hm_pipeline}. In its core, the software rasterizer takes the input hair mesh and a strand ID (defined, e.g., by its UV coordinate) and generates the strand on the fly, which is rasterized into a compressed G-Buffer (\cref{sec:swr}). The geometric information in the G-Buffer is later used for shading (\cref{sec:shading}). To maximize efficiency and image quality, during the strand generation phase before rasterizing, we apply a per-bundle level-of-detail that reduces both the number of strands and their complexity (\cref{sec:lod}). Finally, to avoid visual artifacts due to undersampling and lack of transparency information in the G-Buffer, we apply a filtering phase to reconstruct the final image (\cref{sec:filtering}), which is depth-composited with the rest of the scene.

\subsection{Deferred Software Rasterization}
\label{sec:swr}
To achieve high frame rates for distant hair rendering, we extend the hair mesh approach proposed by \citet{bhokareRealTimeHairRendering2024a} with a custom software rasterizer. Software rasterization has proven particularly effective in scenarios with extensive subpixel geometry as well as a huge amount of overdraw as shown by \citet{markus2022}. Hair strands are geometrically thin and often cover only a single pixel, which incurs significant \revlegal{computational}{} overhead in traditional rasterization pipelines due to the mandatory 2$\times$2 fragment shader quad dispatches — wasting up to 75\% of shader invocations on helper lanes. 
Additionally, assembling a strand requires each segment to access data from its neighboring control points; done naively, every thread would redundantly re-read and re-compute all prior vertices. Alternatively, using shared memory to store intermediate data is a better approach but still uses up unnecessary storage and bandwidth. 

To address this, our implementation leverages a workgroup-wide, register-only cooperative strand assembly to maximize efficiency and minimize memory traffic. To dynamically adapt hair complexity, we perform a prepass that computes fine-grained resolution data (details in \cref{sec:lod}) consumed by the subsequent indirect dispatch stage for spawning the correct amount of workgroups (\rev{WG}{WP}). The screen-space strand data gets compressed and stored in a per-pixel 64-bits wide G-Buffer which is then used for deferred shading. 

\subsubsection{Cooperative Strand Assembly}
\begin{figure*}[t]
    \includegraphics[width=\textwidth]{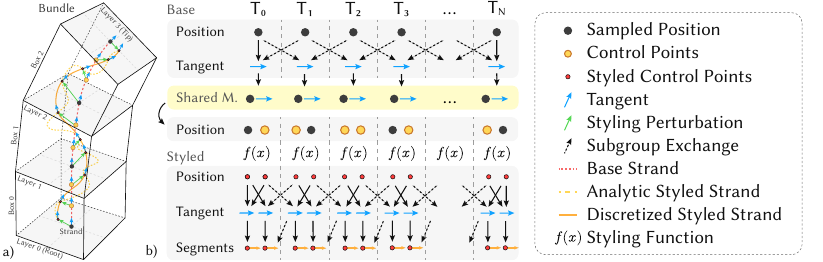}
  \caption{In order to efficiently parallelize the strand assembly each workgroup (\rev{WG}{WP}) first cooperatively loads all layer data into shared memory. Each thread shares the \rev{first and last}{} calculated positions \rev{of its strip}{} with the previous and next thread \rev{to circumvent recomputation}{}. We do this process twice: First \rev{when}{} assembling \rev{the layer data for}{} the base spline \rev{(position)}{}, and second \rev{when}{} assembling the final styled strand \rev{(position and tangent)}{}.}
  \label{fg:hm_assemble}
\end{figure*}

Every \rev{WG, consisting of subgroup (SG) size number of threads,}{WP} cooperatively assembles one strand and writes the end result to the target image via atomic min operations. While \citet{bhokareRealTimeHairRendering2024a} describe the strand assembly at a high level, they do not provide concrete implementation details on how to parallelize it efficiently. Since assembling each strand's segment requires information from previous and next segments, parallelization is not trivial. Done naively, every thread would \rev{redundantly recompute the data dependencies surrounding the current control point multiple times}{need to iterate over all vertices up to the current one twice} to get the final styled strand segment. Instead, we write the bundle's layer data into shared memory \rev{while using}{and then use} fast subgroup communication to share computed control points between neighboring threads. In this way, each control point needs to be computed only once (see \cref{fg:hm_assemble}).

The first pass assembles a base spline where each control point has a tangent calculated from its previous and next vertices \rev{via finite differences}. This tangent is then used together with the position to calculate the \rev{Hermite}{} spline between two control points. Once the base spline is built, we further subdivide it based on the desired resolution, which gives a list of final vertex positions. Each position is then used as the input for the styling function $f(x)$ resulting in the final styled strand. Note that for the styled strand we no longer use a spline, and instead render a polyline. For the styling we need to calculate new tangents for shading; otherwise we run into shading artifacts as mentioned in the original work~\cite{bhokareRealTimeHairRendering2024a}. Each tangent is again calculated from the previous and next control points. This two-pass dependency for calculating the final tangents makes it non-trivial to parallelize; once again, our approach uses sub-pass communication to share vertex and tangent data between neighboring threads (see \cref{cod:swr_code}).

\subsubsection{\rev{Compressed G-Buffer for Deferred Atomic Shading}{Deferred Atomic Shading}}

Once each styled segment is computed, we rasterize them using the digital differential analyzer (DDA) line-drawing algorithm. \rev{Because hair strands are extremely thin, we rasterize them as single-pixel-wide lines. We experimented with rendering thick strands; however, we found that this provided no visual benefit in the far field case while \revlegal{requiring much more compute}{being much more expensive}.}{} 

\rev{To support our reconstruction filter (see \cref{sec:filtering}), the rasterization stage writes to two separate G-buffer targets (or array slices): a \emph{center-sample} G-buffer that is only updated if the strand passes within a configurable diameter through the pixel center, and a \emph{conservative} G-buffer that is updated for any pixel the strand intersects.}{}

Shading each strand during this line-rasterization is prohibitively \revlegal{computationally}{} expensive due to overdraws: We evaluated a per-fragment (similar to a fragment shader) and a per-vertex shading approach, and both proved to be too slow in practice. Consequently, we adopted a deferred shading strategy.

For deferred shading, we need the following per-pixel data for shading (see details later in \cref{sec:shading}):
\begin{itemize}
    \item \textbf{Position} ($\texttt{float3}$)
    \item \textbf{Tangent} ($\texttt{float3}$)
    \item \textbf{Shading parameters}: albedo (\texttt{float3}), roughness (\texttt{float}), tilt (\texttt{float})
    \item \textbf{Ambient occlusion} ($\texttt{float}$)
\end{itemize}
Additionally, depth values are required in the high bits to allow for an atomic min operation. 

However, for deferred shading to be viable, all necessary shading data must fit within the frame-buffer, and be written simultaneously to avoid race conditions, into a 64-bit per-pixel payload. This constraint is crucial, as recomputing the necessary shading data is \revlegal{computationally}{} expensive: It would require either reconstructing the full strand geometry or storing indices for each bundle, strand, and segment, which quickly exceeds the available \revlegal{bits}{bit budget}. To address this, we compress the data needed for shading as follows:

\begin{enumerate}
    \item \textbf{Reduced depth precision:} 24 bits are used for depth.
    \item \textbf{Position reconstruction:} Instead of storing the position ($\text{float3}$), we use the common approach in deferred engines and reconstruct it from the depth value. This approach restricts us to a specific sample location within the pixel, as the position can only be accurately recovered at that point \cite{cryengine}.
    \item \textbf{Shading Parameters:} Instead of storing the shading parameters in the G-Buffer, we store the uvw styling coordinates, which we use for querying these parameters at shading time. This allows us to query these shading parameters only once per pixel, and store the whole appearance model in three floats.  
    \item \textbf{Tangent compression:} We encode the tangent vector using an octahedral unit vector encoding, reducing it from $\text{float3}$ to $\text{float2}$.
    \item \textbf{Data quantization:} All float values \rev{except for depth}{} are quantized as either $\text{uint8}$ or $\text{uint6}$ to further reduce memory usage \rev{(see inset)}{}.
\end{enumerate}

\noindent
\begin{minipage}[t]{0.77\columnwidth}
By applying this compression, we can fit all required data into a 64-bit framebuffer, enabling efficient deferred shading. Note that as with common deferred rendering, we do not handle transparency: This unfortunately is very important for hair, since strands are usually significantly smaller than the pixel footprint, and thus multiple strands might contribute to the pixel footprint, and transparency is very important for compositing with the rest of the scene. We solve this issue as a post-process filter, as we discuss later in \cref{sec:filtering}.
\end{minipage}\hfill
\begin{minipage}[t]{0.20\columnwidth}
  \vspace{-2pt}%
  \includegraphics[width=\linewidth]{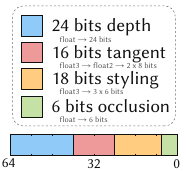}
  \vspace{0pt}
\end{minipage}

\subsection{Level-of-detail}
\label{sec:lod}

\noindent
\begin{minipage}[t]{0.77\columnwidth}
For far-field rendering we want to reduce the workload by limiting the number of line primitives being rendered, which can be achieved by both reducing the number of strands and the strand resolution. We leverage the on-the-fly nature of our strand generation to adaptively generate strands based on its importance on screen. First, we run a pre-pass per bundle, where we compute the number of strands per bundle and the accuracy of each strand, storying it to a device-only buffer. Each thread of this pass processes a single bundle and writes its corresponding data.
\end{minipage}\hfill
\begin{minipage}[t]{0.20\columnwidth}
  \vspace{-20pt}%
  \includegraphics[width=\linewidth]{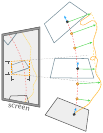}
  \vspace{0pt}
\end{minipage}

This data is then used as the source for the strand assembly pass in the software rasterizer as the source for an indirect dispatch call, dispatching a variable amount of work-groups per bundle. The level-of-detail (LOD) is calculated based on the screen-space footprint \cite{unterguggenbergerFastRenderingParametric2024, unterguggenbergerFastRenderingParametric2025} of a layer (see inlay) and the sensitivity can be controlled via a parameter $\lambda$, which can be different for different hairstyles or strand counts.

\subsubsection{Dynamic Strand Counts}
In order to reduce the complexity of the groom we use an approach inspired by stochastic simplification methods \cite{stochasticSimplification}. Essentially, we randomly remove strands in far away bundles based on their projected screen-space footprint, so that the number of strands per bundle $N_{\mathrm{LOD}}$ is
\begin{equation}\label{eq:strand-lod}
  N_{\mathrm{LOD}} = \mathrm{clamp}\bigl(
    \lceil\, L\cdot(N + \delta)\,\rceil,\;1,\;N\bigr),
\end{equation}
where $N$ is the full strand count per bundle, $\delta\sim\mathcal{U}[0,1)$ is a per-bundle random offset that smooths LOD transitions, and $L\in[0,1]$ is the LOD selection value computed as
\begin{equation}\label{eq:lod-factor}
  L = \mathrm{clamp}\left(
    \frac{\|\,\mathrm{AABB}_{\max} - \mathrm{AABB}_{\min}\,\|}{R_y}\cdot\lambda,
    \;0,\;1\right),
\end{equation}
where $\mathrm{AABB}$ is the screen-space bounding box of the bundle, which is normalized by $R_y$ the vertical resolution of the screen, and $\lambda$ is the LOD sensitivity parameter. 
The projected bounding box is computed by projecting each layer's vertices to screen space, and accumulating a single axis-aligned bounding box (AABB) as the component-wise maximum over all per-layer AABBs:
\begin{equation}
  \mathrm{AABB}_{\min} = \min_{i}\min\bigl(\mathbf{s}^i_x,\;\mathbf{s}^i_y,\;\mathbf{s}^i_z,\;\mathbf{s}^i_w\bigr),\quad
  \mathrm{AABB}_{\max} = \max_{i}\max\bigl(\mathbf{s}^i_x,\;\mathbf{s}^i_y,\;\mathbf{s}^i_z,\;\mathbf{s}^i_w\bigr),
\end{equation}
where $\mathbf{s}^i_{x,y,z,w}$ are the screen-space positions of the four quad vertices of layer~$i$. \rev{}{The screen-space diagonal of the accumulated AABB already encodes the projected size $\tfrac{h\,R}{d\,\tan\theta}$ from \cite[Eq.7]{bhokareRealTimeHairRendering2024a}, as it results from perspective division followed by viewport transformation.}
Bundles with AABBs laying entirely outside the viewport are frustum-culled ($N_{\mathrm{LOD}}=0$).
The resulting value is written to the instance count of an indirect dispatch buffer with $B$ entries, and computed via a dispatch of $B$ threads, with one thread assigned to each bundle. 

\subsubsection{Dynamic Strand Resolution}
\rev{A second form of LOD we apply is when selecting the number of control points for each strand geometry, which inherently reduces the number of rasterized segments}{A second form of LOD we apply is the downsampling of each strand geometry, which inherently reduces the number of control points (and thus, of rasterized segments)}. We use the same LOD selection $L$ value as before (\cref{eq:lod-factor}), and — following Bhokare's approach \cite[Eq.9]{bhokareRealTimeHairRendering2024a} — compute the number of control points per bundle as 
\begin{equation}\label{eq:cp-lod}
  C_{\mathrm{raw}} = \max\bigl(\lfloor\sqrt{L}\;\cdot C_{\max}\rfloor,\;
    C_{\mathrm{layers}}\bigr),
\end{equation}
where $C_{\max}=127$ is the maximum number of control points and $C_{\mathrm{layers}}$ is the number of bundle layers (a hard lower bound). We use $\sqrt{L}$ instead of $L$ to better preserve curvature at intermediate distances. To ensure smooth vertex LOD transitions, we snap $C_{\mathrm{raw}}$ to a power-of-two-plus-one value, giving a final control points count value of
\begin{equation}\label{eq:cp-snap}
  C = \min\bigl(2^{\lfloor\log_2(C_{\mathrm{raw}}-1)\rfloor}+1,\;C_{\max}\bigr),
  \qquad C_{\mathrm{raw}} \geq 3.
\end{equation}
This computation is also done with one thread per bundle, in the same pass as the strand count computation. The final control point count is stored in a GPU buffer next to the (indirect) dispatch count (fitting into a single \texttt{uint4} per bundle), which is consumed by the software rasterizer. \rev{Combined with the per-bundle random offset $\delta$ in \cref{eq:strand-lod}, which staggers strand-count transitions across bundles, this snapping rule keeps geometric LOD switches sub-pixel under typical viewing speeds; we did not observe perceptible popping in our test sequences.}{}

\begin{figure}[t]
    \begin{tikzpicture}[x=0.01\columnwidth, y=0.01\columnwidth]
        \node[anchor=south west, inner sep=0] at (0, 0) {\includegraphics[width=\columnwidth]{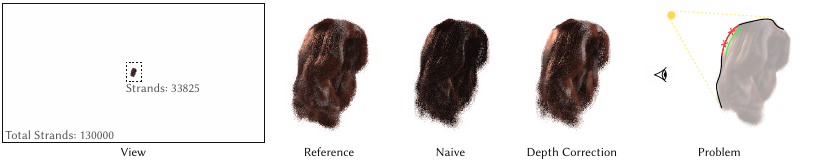}};
        \node[anchor=south west, inner sep=0] at (86, 16) {$S_1$};
        \node[anchor=south west, inner sep=0] at (89, 13) {$S_2$};
    \end{tikzpicture}
  \caption{\textbf{Shadow artifacts due to LOD-based strand prunning.} Applying stochastic strand removal results in inconsistent shadow or deep opacity maps (DOM), which might produce shadow artifacts when applying LOD. When the depth map is calculated using the full geometry set, but the front-most strand ($S_1$) is randomly culled, the shadow volume becomes ``stale''. Consequently, the remaining strand ($S_2$) incorrectly samples the high-opacity region deeper inside the groom (\emph{Problem}), leading to unnatural darkening and visual popping (\emph{Naive}). Pushing all strands closer based on the ratio of removed strands fixes this (\emph{Depth Correction}).}
  \label{fg:depth_fix}
\end{figure}

\subsubsection{Depth Comparison Artifact Fix}
An important problem of removing strands is that previously computed depth maps become invalid \rev{(recomputing them is \revlegal{computationally}{} expensive, which we want to avoid).}{} This is especially noticeable with shadow or opacity maps (\cref{fg:depth_fix}, \emph{Problem}), where stochastically removing a strand at the boundary of the groom might lead to inner strands becoming visible, but still fall behind the stored depth value. In the case of shadow maps, this results in a dark region (\cref{fg:depth_fix}, \textit{Naive}). We fix this issue by moving the inner strand closer to the camera during depth testing by a small offset $\Delta$, which depends on the distance to the camera and the proportion of remaining strands. 

To derive this offset $\Delta$ we make use of the Beer-Lambert law and statistically account for aggregated visibility as
\begin{equation}
V(d) = \exp(-\sigma\cdot d),
\end{equation}
with $\sigma$ proportional to strand density. When stochastically removing strands, only a fraction $\beta=\frac{N_{\mathrm{remaining}}}{N_{\mathrm{total}}}$ of the strands remain, so that density becomes $\sigma \cdot \beta$ and visibility becomes
\begin{equation}
V'(d) = \exp(-(\sigma \cdot \beta) \cdot d ) = \exp(-\sigma \cdot (\beta \cdot d) ).
\end{equation}
This shows that reducing the number of strands is equivalent to traversing a reduced effective depth $d_{\mathrm{eff}} = rd$. To express this compensation as an additive shift in \rev{}{log-}depth space, we invert the exponential, yielding an offset
\begin{equation}
    \Delta = -\log(\rev{\beta}{r}) = -\log\left(\frac{N_{\mathrm{remaining}}}{N_{\mathrm{total}}}\right).
\end{equation}
This shift is $\Delta=0$ when no culling occurs ($\rev{\beta}{r}=1$) and increases smoothly as more strands are removed ($\rev{\beta}{r}\downarrow 0$).

\subsection{Anti-Aliasing and Opacity Reconstruction}
\label{sec:filtering}

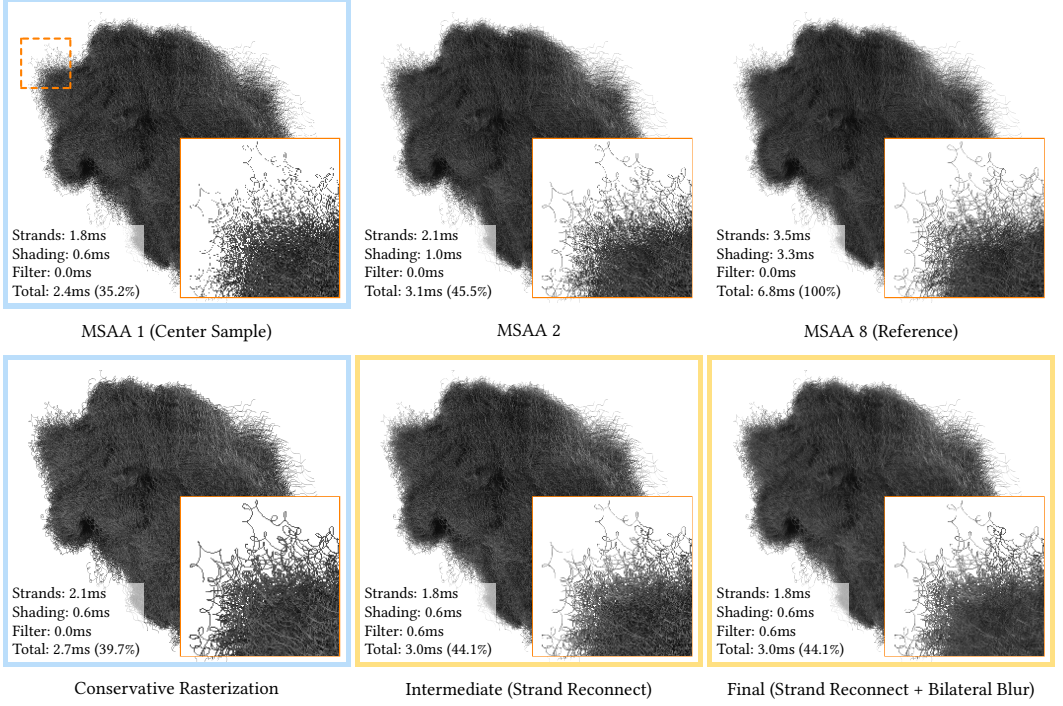
\begin{figure*}[t]
\centering
\definecolor{inputbg}{HTML}{BBDEFB}%  light blue
\definecolor{finalbg}{HTML}{FFE082}%  light gold
\setlength{\fboxsep}{2pt}%
\setlength{\fboxrule}{0.3pt}%
\newcommand{\filtimg}{0.32\textwidth}%
\newcommand{\filtzoom}{0.15\textwidth}%
% Zoom window parameters (center and half-size in pixels)
\def\fcx{530}% center x in px
\def\fcy{845}% center y from bottom in px
\def\fhs{80}% half-size of square crop in px
% Image dimensions
\def\fiw{1922}% image width in px
\def\fih{1112}% image height in px
% Main image crop
\def\fml{390}% main left trim px
\def\fmb{50}%  main bottom trim px
\def\fmr{400}% main right trim px
\def\fmt{50}%  main top trim px
% Computed zoom trims
\pgfmathsetlengthmacro{\fztl}{(\fcx-\fhs)*1px}%
\pgfmathsetlengthmacro{\fztr}{(\fiw-\fcx-\fhs)*1px}%
\pgfmathsetlengthmacro{\fztb}{(\fcy-\fhs)*1px}%
\pgfmathsetlengthmacro{\fztt}{(\fih-\fcy-\fhs)*1px}%
% Computed dotted rect fractions (relative to main visible area)
\pgfmathparse{(\fcx-\fhs-\fml)/(\fiw-\fml-\fmr)}\edef\fdrxl{\pgfmathresult}%
\pgfmathparse{(\fcx+\fhs-\fml)/(\fiw-\fml-\fmr)}\edef\fdrxr{\pgfmathresult}%
\pgfmathparse{(\fcy-\fhs-\fmb)/(\fih-\fmb-\fmt)}\edef\fdryb{\pgfmathresult}%
\pgfmathparse{(\fcy+\fhs-\fmb)/(\fih-\fmb-\fmt)}\edef\fdryt{\pgfmathresult}%
% -- helper: image cell with zoom inset and timing overlay --
% #1=bg color, #2=image path, #3=timing text, #4=caption
\newcommand{\filtcell}[4]{%
\begin{tikzpicture}[baseline=(I.south)]%
    \node[anchor=south west, inner sep=0] (I) at (0,0)
        {\colorbox{#1}{\includegraphics[width=\filtimg, trim={\fml px \fmb px \fmr px \fmt px}, clip]{#2}}};
    \node[anchor=south east, inner sep=0pt, draw=orange, line width=1.0pt] at ([shift={(-4pt,4pt)}]I.south east)
        {\includegraphics[width=\filtzoom, height=\filtzoom, trim={\fztl{} \fztb{} \fztr{} \fztt{}}, clip]{#2}};
    \node[anchor=south west, font=\tiny, align=left, fill=white, fill opacity=0.6, text opacity=1, inner sep=1.5pt] at ([shift={(2pt,2pt)}]I.south west) {#3};
    \node[anchor=north, font=\scriptsize] at ([yshift=-2pt]I.south) {#4};
\end{tikzpicture}%
}%
% Variant with dotted rectangle marking zoom area (for first image only)
\newcommand{\filtcellmark}[4]{%
\begin{tikzpicture}[baseline=(I.south)]%
    \node[anchor=south west, inner sep=0] (I) at (0,0)
        {\colorbox{#1}{\includegraphics[width=\filtimg, trim={\fml px \fmb px \fmr px \fmt px}, clip]{#2}}};
    \node[anchor=south east, inner sep=0pt, draw=orange, line width=1.0pt] at ([shift={(-4pt,4pt)}]I.south east)
        {\includegraphics[width=\filtzoom, height=\filtzoom, trim={\fztl{} \fztb{} \fztr{} \fztt{}}, clip]{#2}};
    \path let \p1=(I.south west), \p2=(I.north east),
              \n{w}={\x2-\x1}, \n{h}={\y2-\y1} in
        coordinate (ZBL) at (\x1+\fdrxl*\n{w}, \y1+\fdryb*\n{h})
        coordinate (ZTR) at (\x1+\fdrxr*\n{w}, \y1+\fdryt*\n{h});
    \coordinate (ZTL) at (ZBL |- ZTR);
    \coordinate (ZBR) at (ZBL -| ZTR);
    \draw[orange, line width=0.8pt, line cap=round, dash pattern=on 2pt off 2pt, dash phase=0.1pt] (ZBL) -- (ZBR);
    \draw[orange, line width=0.8pt, line cap=round, dash pattern=on 2pt off 2pt, dash phase=0.1pt] (ZBR) -- (ZTR);
    \draw[orange, line width=0.8pt, line cap=round, dash pattern=on 2pt off 2pt, dash phase=0.1pt] (ZTR) -- (ZTL);
    \draw[orange, line width=0.8pt, line cap=round, dash pattern=on 2pt off 2pt, dash phase=0.1pt] (ZTL) -- (ZBL);
    \node[anchor=south west, font=\tiny, align=left, fill=white, fill opacity=0.6, text opacity=1, inner sep=1.5pt] at ([shift={(2pt,2pt)}]I.south west) {#3};
    \node[anchor=north, font=\scriptsize] at ([yshift=-2pt]I.south) {#4};
\end{tikzpicture}%
}%
% Top row: MSAA comparisons
\filtcellmark{inputbg}{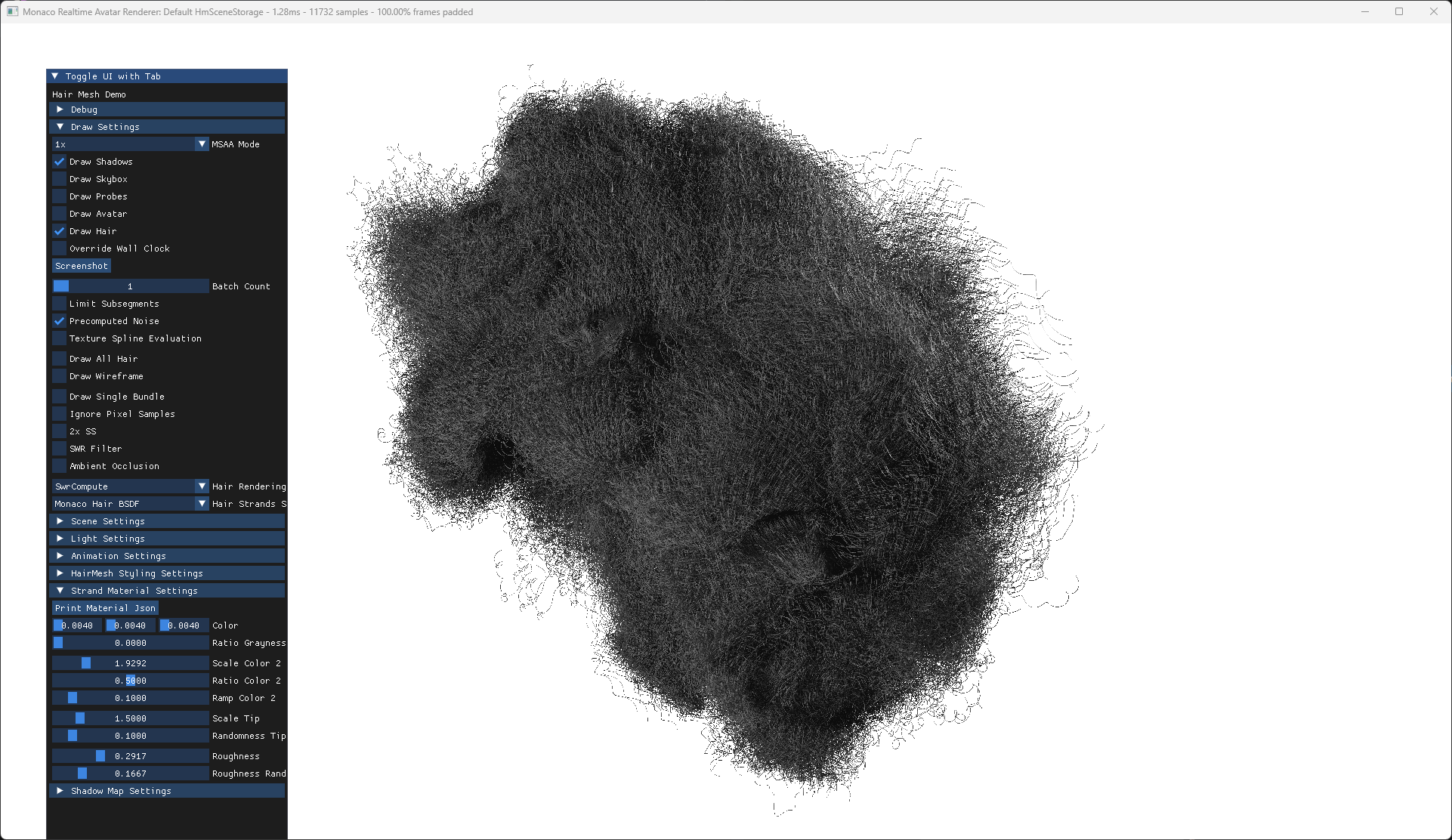}{Strands: 1.8ms\\Shading: 0.6ms\\Filter: 0.0ms\\Total: 2.4ms (35.2\%)}{MSAA 1 (Center Sample)}%
\hfill
\filtcell{white}{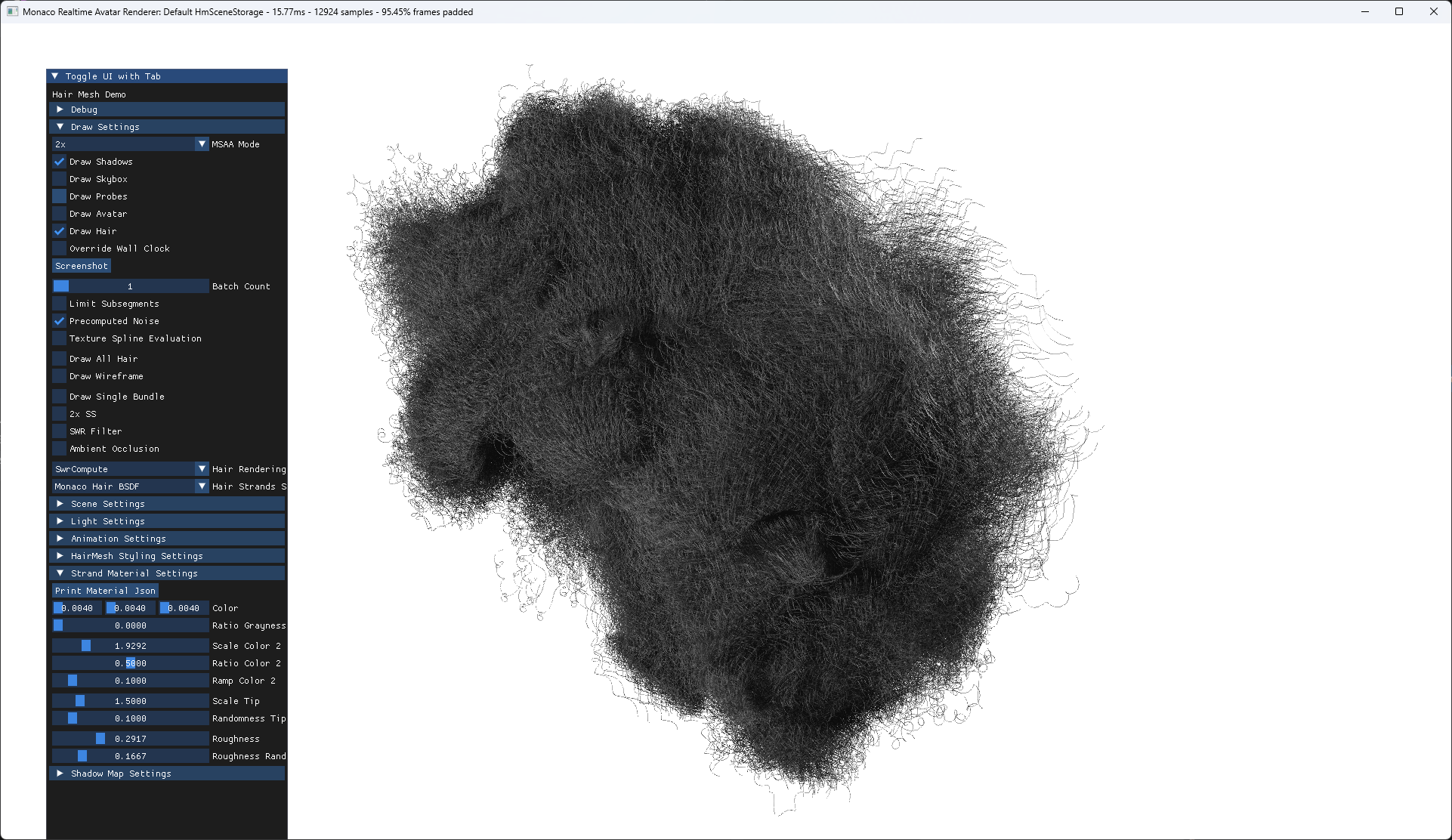}{Strands: 2.1ms\\Shading: 1.0ms\\Filter: 0.0ms\\Total: 3.1ms (45.5\%)}{MSAA 2}%
\hfill
\filtcell{white}{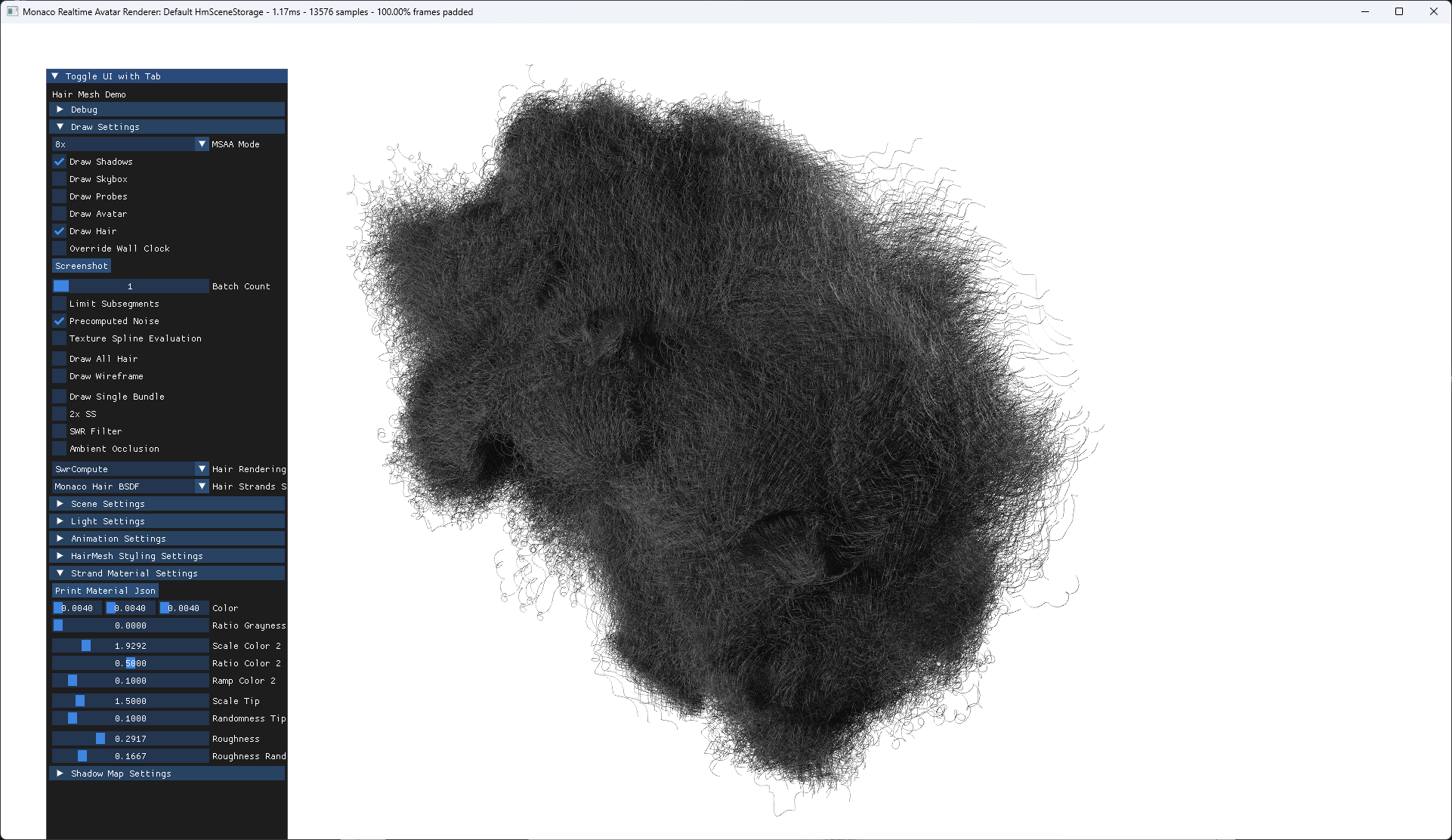}{Strands: 3.5ms\\Shading: 3.3ms\\Filter: 0.0ms\\Total: 6.8ms (100\%)}{MSAA 8 (Reference)}%
\\[1pt]
% Bottom row: Filter pipeline stages
\filtcell{inputbg}{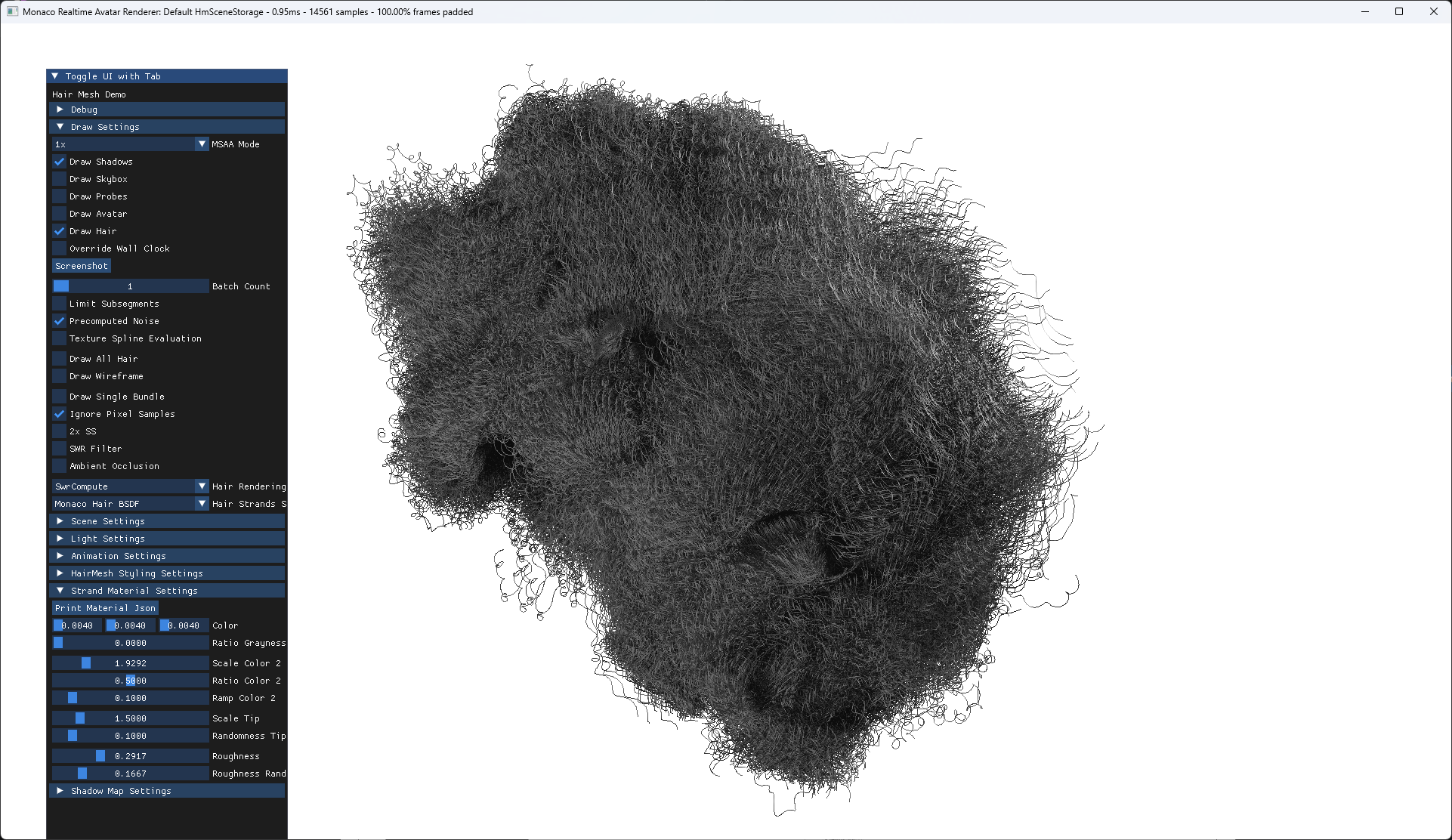}{Strands: 2.1ms\\Shading: 0.6ms\\Filter: 0.0ms\\Total: 2.7ms (39.7\%)}{Conservative Rasterization}%
\hfill
\filtcell{finalbg}{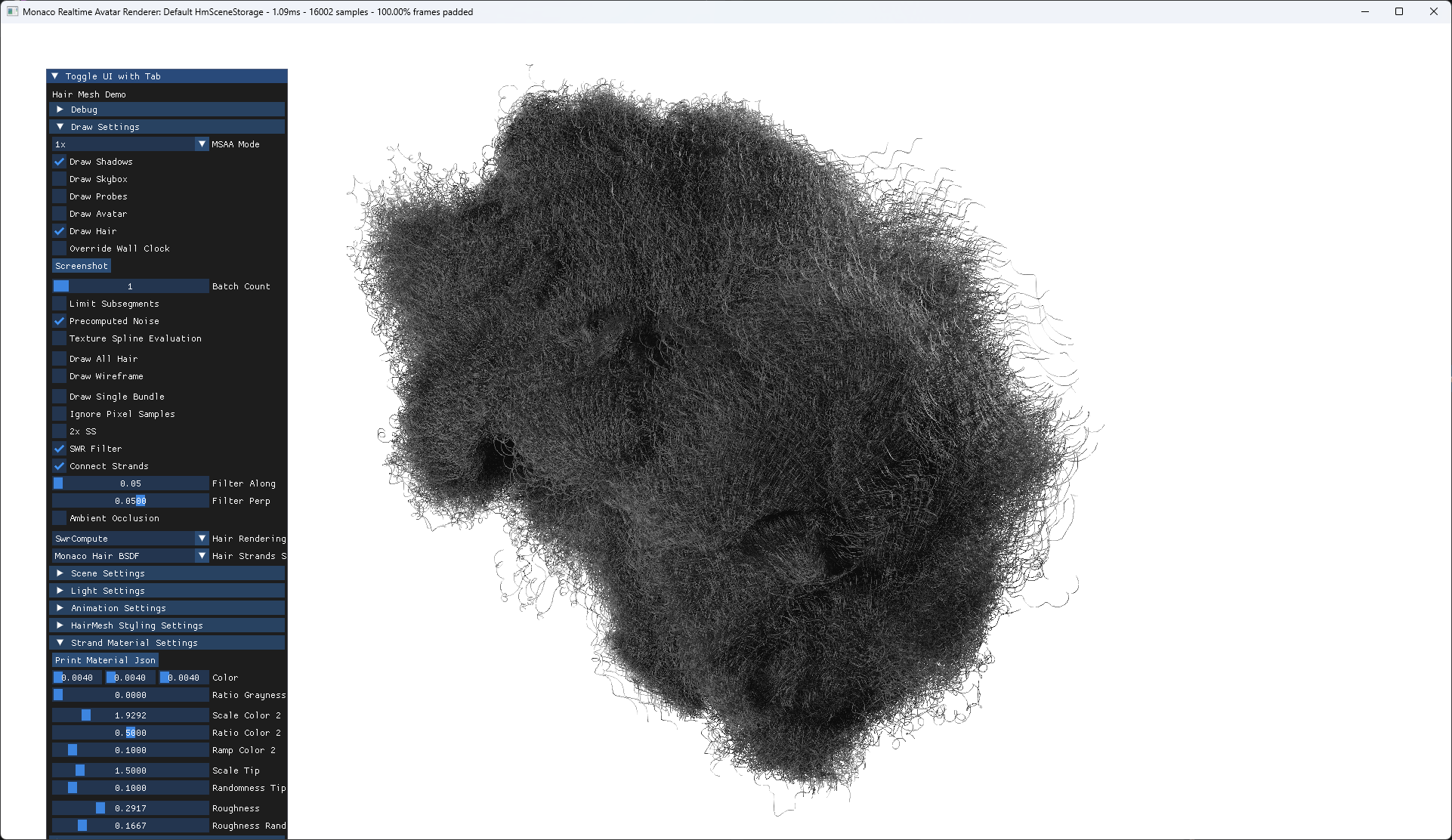}{Strands: 1.8ms\\Shading: 0.6ms\\Filter: 0.6ms\\Total: 3.0ms (44.1\%)}{Intermediate (Strand Reconnect)}%
\hfill
\filtcell{finalbg}{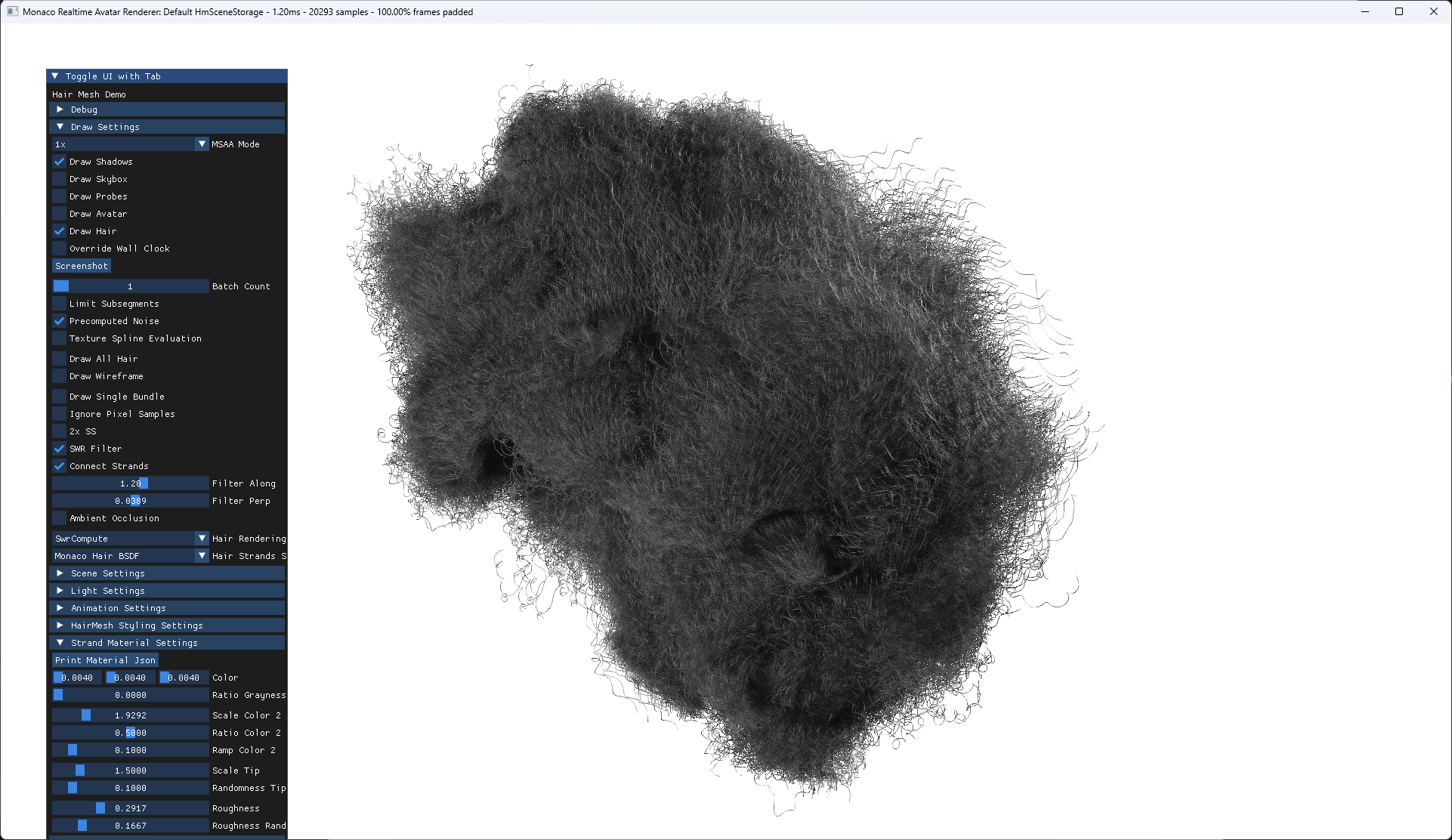}{Strands: 1.8ms\\Shading: 0.6ms\\Filter: 0.6ms\\Total: 3.0ms (44.1\%)}{Final (Strand Reconnect + Bilateral Blur)}%
\caption[Filter reconstruction pipeline overview]{The filter takes \colorbox{inputbg}{two inputs}: a center-hit render at MSAA\,1 and a \rev{conservative}{ignore-pixel-samples} render, which captures all strand-touched pixels regardless of center hits and serves purely as spatial guidance. These feed into a \colorbox{finalbg}{two-stage pipeline}: a reconnection pass that promotes isolated strand pixels by borrowing shading from nearby center-hit neighbors, followed by a directional blur that smooths along the strand tangent. Both stages are executed within a single $7{\times}7$ kernel pass. The final result closely matches the quality of high-sample references at a fraction of the cost.}
\label{fg:hm_filter}
\end{figure*}

In order to achieve low frame render times, \revlegal{the number of pixels available}{our budget for pixels} to write via atomic operations is very limited (only 64 bits per pixel, with 24 of those being used for depth), which limits us to a single sample per pixel in high-resolution renders. This results in significant aliasing and disconnected strands, since only strands passing through the pixel's center sample point are taken into account (see \cref{fg:hm_filter}, \emph{MSAA 1}). Using \rev{conservative}{line-to-pixel} rasterization on the other hand results in thick lines, specially for isolated strands (see \cref{fg:hm_filter} \emph{Conservative Rasterization}).

%At the same time we can only store 64\,bits per pixel, of which 24 are used for depth, limiting us to essentially one sample per pixel. This restricts the rendering to only strands that pass through the center sample, resulting in disconnected lines (see Fig. \ref{fg:hm_filter} 'MSAA 1'); ignoring the center test and outputting the strand whenever it intersects the pixel yields overly thick lines, especially for isolated strands (see Fig. \ref{fg:hm_filter} 'Conservative Rasterization').

We use an approach inspired by the work of \citet{huangDetailPreservingRealTimeHair2025}. However, they tackle these problems using a multi-pass approach that includes an explicit strand-connection step relying on per-strand indices which we cannot provide given our tight framebuffer. 
Instead, our solution stores two layers per pixel: A \emph{center-sample} layer that records only the strands passing through the pixel center, and a \emph{conservative} layer which \rev{records the atomic-min G-buffer value over all strands intersecting the pixel.}{records every strand intersecting the pixel}. Combined, these layers provide the filter with sufficient context about strand connectivity to reconnect disconnected strands, eliminating the need for the explicit connection pass proposed by \citet{huangDetailPreservingRealTimeHair2025}. We also extend this approach by supporting multisampling for the software rasterizer, recording samples into multiple layers of a \texttt{uint64} image and resolving them later (see \cref{fg:hm_filter}, \emph{MSAA 8}). However, we only use this approach as a reference, since it is not practical in terms of memory and \revlegal{computational}{} cost. 

\subsubsection{Elliptical Bilateral Filter}

We denoise the center-sample frame buffer with an orientation-aware bilateral filter similar to \citet{huangDetailPreservingRealTimeHair2025}. We apply the filter over a square window of radius~$r{=}5$ (i.e., a $(2r+1){\times}(2r+1)$ kernel). For every \rev{neighbor}{} pixel $\mathbf{Q}$ \rev{of center pixel $\mathbf{P}$}{} at offset~$\mathbf{d}$ the displacement is decomposed into tangential and perpendicular components using the screen-space tangent~$\mathbf{T}_P$:
\begin{equation}\label{eq:offset-decomp}
  \rev{\mathbf{d} = \mathbf{Q}-\mathbf{P}, \qquad}{}
  d_{\parallel} = \rev{\mathbf{d}}{Q-P} \cdot \mathbf{T}_P, \qquad
  d_{\perp}     = \bigl\|\mathbf{d} - d_{\parallel}\,\mathbf{T}_P\bigr\|.
\end{equation}
The bilateral weight \cite[Eq.~11]{huangDetailPreservingRealTimeHair2025} is the product of an elliptical spatial term and a color-similarity term, following
\begin{equation}\label{eq:bilateral}
  w_{PQ}
  = \underbrace{
      \exp\left(
        -\frac{d_{\parallel}^{2}}{\sigma_{\parallel}^{2}}
        -\frac{d_{\perp}^{2}}{\sigma_{\perp}^{2}}
      \right)
    }_{w_{\text{spatial}}}
  \;\cdot\;
  \underbrace{
      \exp\left(
        -\frac{\|\mathbf{C}_P - \mathbf{C}_Q\|^{2}}{\sigma_c^{2}}
      \right)
    }_{w_{\text{color}}},
\end{equation}
where $\sigma_{\parallel} = r \cdot s_{\parallel}$ and
$\sigma_{\perp} = r \cdot s_{\perp}$ are configurable major- and
minor-axis standard deviations, and \rev{$\sigma_c$ the color-similarity standard deviation (which we set to $\sigma_c = 0.9$)}{$\sigma_c = 0.9$}. \rev{}{In isotropic fallback mode the spatial term simplifies to $w_{\text{spatial}} = \exp(-\|\mathbf{d}\|^{2} / \sigma_{\text{iso}}^{2})$ with $\sigma_{\text{iso}} = 0.4\,r$.} \rev{Neighbors whose depth differs from~$P$ by more than a threshold of $1.45 \times 10^{-3}$ are rejected before weighting.}{Neighbors whose depth differs from~$P$ by more than $1.45 \times 10^{-3}$ are rejected before weighting.} 
The filtered color $\hat{\mathbf{C}}_P$ is thus the result of the convolution of the kernel following $\hat{\mathbf{C}}_P = \frac{\sum_{Q} \mathbf{C}_Q \cdot w_{PQ}} {\sum_{Q} w_{PQ}}.$

\subsubsection{Strand Connection}
Pixels that appear in the conservative frame buffer but have a missing sample in the center-sample frame buffer are \emph{promoted}: They borrow shading from nearby center-hit neighbors. Because the conservative rasterization layer already captures every strand that touches a pixel, this promotion mechanism implicitly reconnects visually disconnected strand segments without the explicit connection pass of \citet{huangDetailPreservingRealTimeHair2025}. When a closer strand exists in the conservative rasterization layer in front of the center-hit strand (depth mismatch~$\ge 5 \times 10^{-4}$), the reconstruction is blended on top of the existing center-hit frame-buffer. For promoted pixels the color-similarity term is disabled ($w_{\text{color}}=1$) because the pixel has no shaded color to compare against.

% \begin{figure}[t]
%     \includegraphics[width=\columnwidth]{resources/filter_new.pdf}
%   \caption{The filter takes two inputs (blue): a center-hit render at MSAA\,1 and an ignore-pixel-samples render, which captures all strand-touched pixels regardless of center hits and serves purely as spatial guidance — it is never directly shaded. These feed into a two-stage pipeline (orange): a reconnection pass that promotes isolated strand pixels by borrowing shading from nearby center-hit neighbors, followed by a directional blur that smooths along the strand tangent. Both stages are executed within a single $7{\times}7$ kernel pass. The final result closely matches the quality of high-sample references at a fraction of the cost. \todo{draft, maybe better images} \adrian{Add total time, and speed up wrt reference (I guess center sample? or 8 samples?). Also label better 2/8 samples (this is MSAA, no?) and ``ignore samples'', which I have no idea what it is.}} 
%   \label{fg:hm_filter}
% \end{figure}

\subsubsection{Alpha Estimation}
Per-pixel opacity is derived from the ratio of hits in the center-sample frame buffer versus hits in the conservative frame buffer along a tangent-aligned strip of \rev{adjustable width $w_\text{strip}$ (we choose $0.5$\,px), whose length is bounded by the bilateral filter's kernel diameter}{half-width $0.5$\,px}. Within this strip we count center hits $H_\text{center}$ and conservative hits pixels $H_\text{cons}$ and estimate the transparency of the pixel $\mathbf{P}$ as $\alpha_P = \frac{H_\text{center}}{H_\text{cons}}$. For center-sample hit pixels with a conservative layer depth mismatch the output opacity is \rev{1}{the sample's own alpha} as blending already happened \rev{(important for depth separated flyaway strands in front of groom)}{}; otherwise it equals~$\alpha_P$ (see \cref{cod:filter_code} for a complexity-reduced version).

\subsection{Shading Model}
\label{sec:shading}

\begin{figure}[t]
\centering
\setlength{\fboxsep}{0pt}
\setlength{\fboxrule}{0.3pt}
\begin{subfigure}[t]{0.49\columnwidth}
    \fcolorbox{imgbordercolor}{white}{\includegraphics[width=\linewidth, trim=180 10 110 100px, clip]{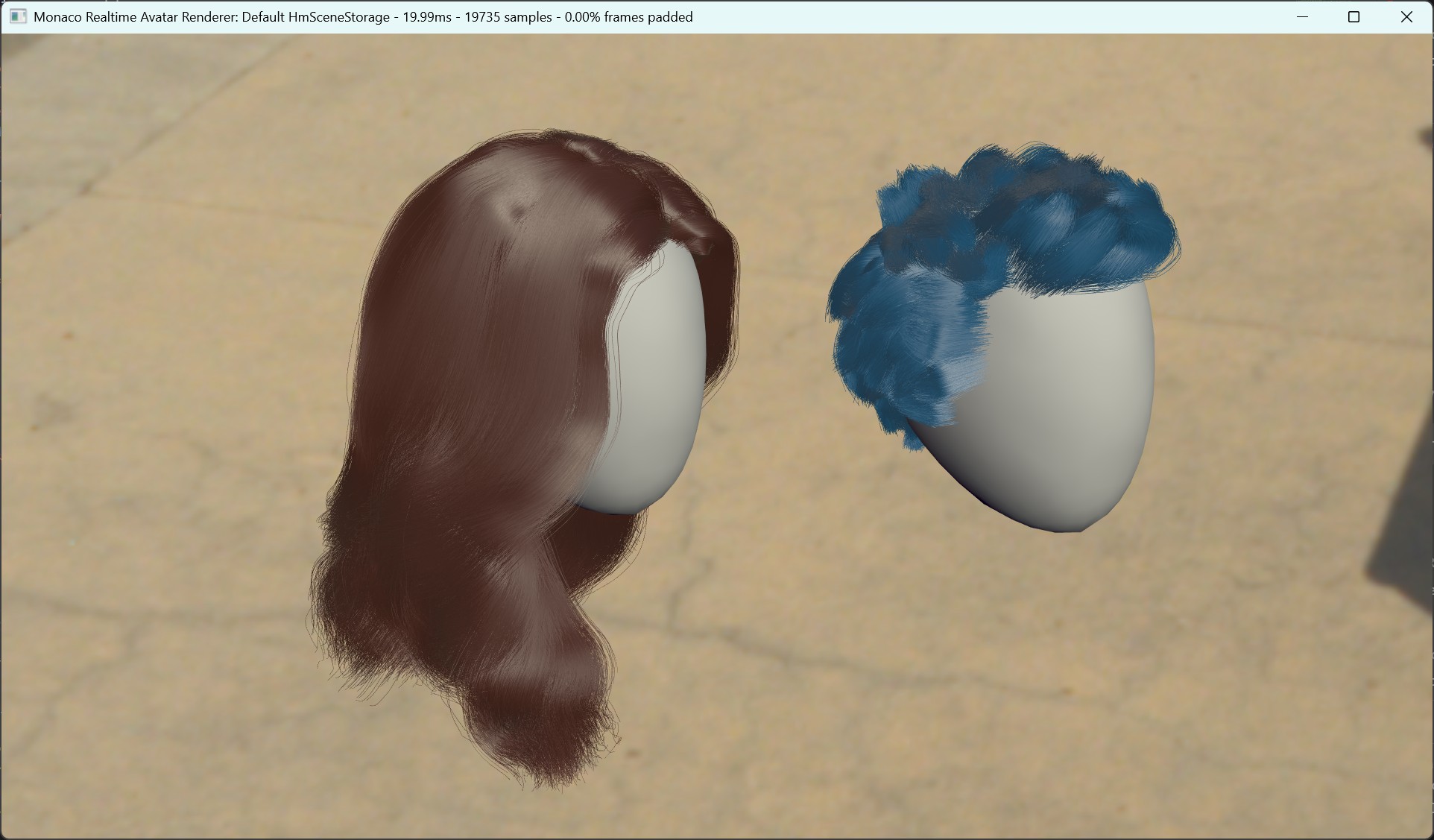}}
    \caption{Without Ambient Occlusion}
\end{subfigure}
\hfill
\begin{subfigure}[t]{0.49\columnwidth}
    \fcolorbox{imgbordercolor}{white}{\includegraphics[width=\linewidth, trim=180 10 110 100px, clip]{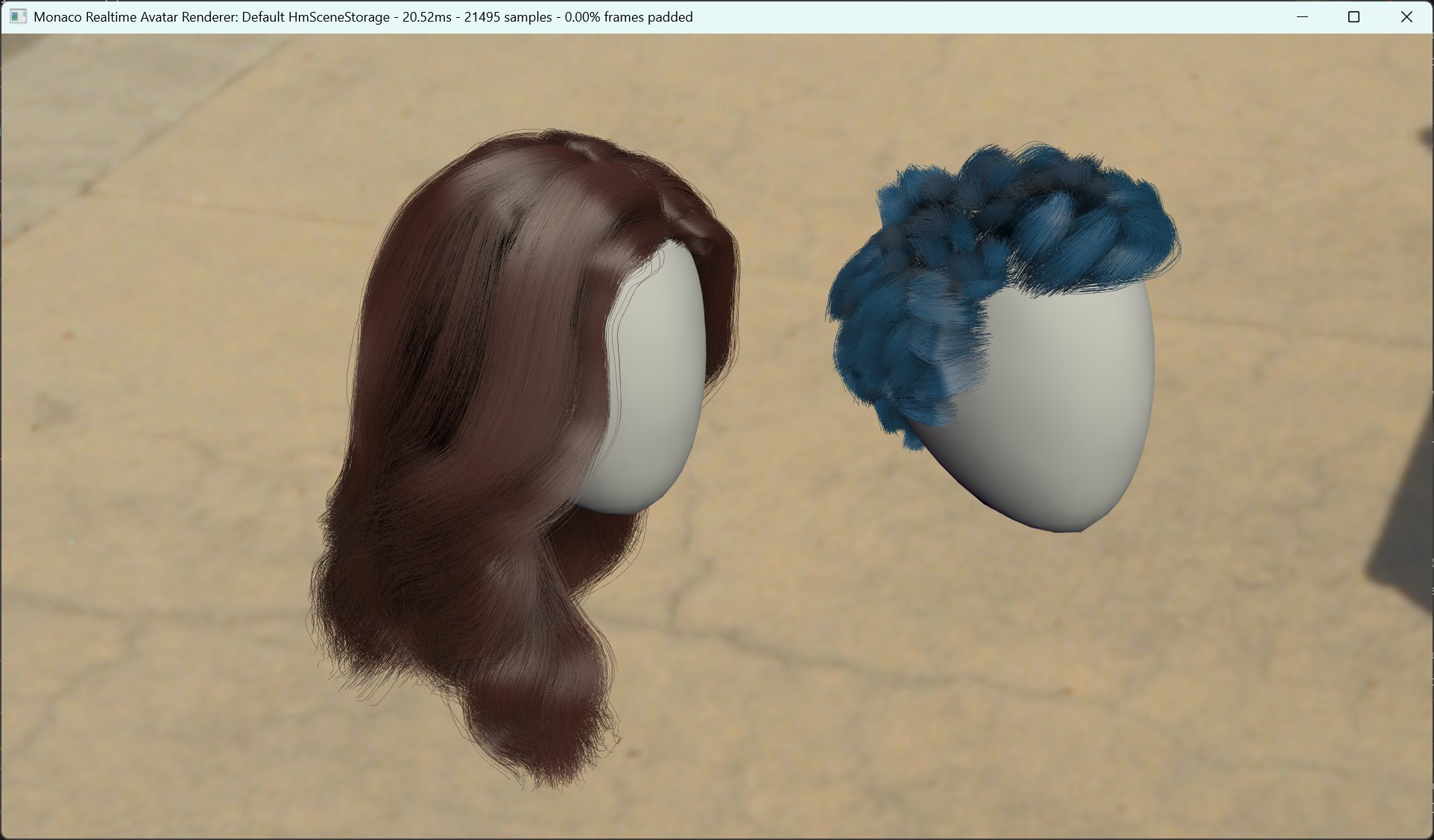}}
    \caption{With Ambient Occlusion}
\end{subfigure}
\caption{Comparison of the effect of pre-baked ambient occlusion. Both images are rendered using our software rasterizer with LOD and reconstruction filter (SWR\,1+F+L) without (a) and with (b) baked ambient occlusion. The ambient occlusion term adds depth and volume to the groom by darkening occluded regions near the scalp and within dense strand clusters. }
\label{fg:ao_comparison}
\end{figure}

As described in \cref{sec:swr}, we defer shading to a second step that operates in screen space over the G-Buffer with the information stored in that buffer. We use a physics-based hair shading model similar to the one proposed by \citet{pekelis2015data}, but fitted to Chiang's \cite{chiang2016practical} model instead of Marschner's. 
We obtain the parameters of the model on a per-pixel level, by querying the shading parameters using the uvw-coordinates stored in the G-Buffer.

\paragraph*{Point Light Sources }
We render point light sources by evaluating the shading model directly, plus a Lambertian term similar to Kajiya-Kay's~\shortcite{kajiya1989rendering} to simulate the contribution of multiple scattering.
Volumetric shadows are computed using deep-opacity maps~\cite{yuksel2008deep}, which we query using the depth-correction described in \cref{sec:lod}.

\paragraph*{Probe Lighting and Baked \rev{Ambient}{} Occlusion}
For probe-based lighting, we use the standard approach of pre-filtered environment maps for specular components (the single scattering hair shading model), and spherical harmonics expansion for more diffuse contributions (the diffuse multiple scattering). For the specular components, we approximate the R and TRT lobes with a single GGX lobe, which we use for querying the pre-filtered probe. 
For the diffuse part, we simply use Ramamoorthi's \cite{ramamoorthi2002frequency}{} irradiance expansion centered at a normal of the fiber oriented towards the view direction. This introduces a subtle view-dependent effect, but enables accounting for more directional incoming light than tangent-based projections \cite{mehta2012analytic}. 

A key element for improving shading quality from probes is our baked occlusion approach. In the context of hair, screen-space approaches \cite{jimenez2016practical} might fail, since these assume surfaces and not volumetric-like assets. Instead, we leverage the hair mesh structure and pre-bake an \rev{ambient}{} occlusion value in each vertex of the hair mesh proxy. This value is interpolated in the same way as the position or tangents during the strand assembly, and rasterized into the G-Buffer, which is very efficient. \Cref{fg:ao_comparison} shows the effect of accounting for this pre-baked ambient occlusion term.

\section{Results}

We evaluate our method on an NVIDIA RTX\,5080 \rev{(SG size = 32)}{} at 1920$\times$1080 resolution using a hair scene with 127k strands. 
Figure~\ref{fig:benchmark} and Table~\ref{tab:benchmark} report rendering quality and GPU frame times across three camera distances (close, mid, far) for all pipeline configurations for two different grooms. 
We compare our deferred software rasterizer against the mesh-shader-based approach from \citet{bhokareRealTimeHairRendering2024a}, with and without our proposed LOD schemes, and with the baked ambient occlusion. In \cref{fg:teaser} we render the scene with a high-frequency environment map and three fill point light sources, while for \cref{fig:benchmark} we only use three point light sources.

\paragraph{Quality}
Hair rendering at low sample counts suffers from strong aliasing: Individual strands are thinner than a pixel and their sub-pixel coverage changes rapidly with camera distance, producing flickering and loss of fine detail. This is especially pronounced at MSAA\,1, where Figure~\ref{fig:benchmark} reveals severe strand dropout and fragmentation across all distances. Our reconstruction filter (SWR\,1+F) directly addresses this by aggregating coverage information across neighboring pixels, recovering strand detail comparable to MSAA\,4--8 while maintaining performance close to MSAA\,2. Adding LOD (SWR\,1+F+L, highlighted in yellow) further reduces \revlegal{computational}{} cost at distance without sacrificing this quality level, making it our recommended configuration. \rev{Depending on the hairstyle it can help to adjust LOD $\lambda$ or the filters $\sigma_{\perp}$, $\sigma_{\parallel}$ or radius. Overall we found no cases where the filter did not work at all.}{}

\paragraph{Performance}
Table~\ref{tab:benchmark} shows that our software rasterizer outperforms the mesh shader baseline across all distances and MSAA levels. At MSAA\,1 without LOD, the software rasterizer takes 3.1\,ms vs.\ 13.6\,ms for mesh shaders — a $4.4\times$ speedup. The reconstruction filter adds moderate \revlegal{computational}{} overhead, except at far distance where heavy atomic contention in the conservative layer causes a spike to 4.5\,ms (shown in \textcolor{orange}{orange}). Enabling LOD as we will discuss next or disabling the strand connection pass at far distance are possible solutions to address this issue. \revlegal{Computational cost}{Cost} is dominated by the strand generation and software rasterization phase, which takes up about $96\%$ of the rendering \revlegal{computational}{} cost even in the presence of complex shading, which has a very small \revlegal{computational}{} cost thanks to the deferred approach. Similarly, the filtering \revlegal{computational}{} cost is negligible. 

\paragraph{LOD}
LOD brings the largest absolute gains at far distance, where strand counts drop from 127k to 21k.
For the software rasterizer with filter (our recommended configuration), the far frame time falls from 4.5\,ms to 0.4\,ms — an $11\times$ reduction — while close and mid views performance improves by 12\% and 28\% respectively.
In contrast, the mesh shader LOD variant is \emph{slower} than its non-LOD counterpart at close and mid distances (shown in \textcolor{red}{red}).
This counter-intuitive result stems from a fundamental limitation of the task shader model: a task shader can only launch mesh-shader workloads at the granularity of an entire workgroup.
Our compute pre-pass, in contrast, evaluates this per thread — one thread per bundle — and feeds the result into an indirect dispatch, incurring nearly zero \revlegal{computational}{} cost due to better parallelization, and LOD is strictly beneficial at every distance.
The pre-pass for determining the LOD value adds a very small \revlegal{computational}{} cost ($3\%$ of the total frame), but significantly accelerates the strans generation and software rasterization phases.

\paragraph{Multiple Grooms}
Our pipeline scales naturally to scenes containing multiple grooms. Each groom is processed sequentially through the same rasterization and filtering pipeline, with its result blended into the render target before the next groom is processed. This allows an arbitrary number of grooms to be composed without any modifications to the core pipeline. By preserving the depth buffer between groom passes, strands occluded by previously rendered grooms are rejected early, reducing atomic write contention and improving performance. Figure~\ref{fg:teaser} demonstrates a scene with multiple grooms rendered this way.

\begin{table}[H]
\caption{GPU frame time benchmark (ms) on an RTX\,5080 at 1920$\times$1080. The number in parentheses is the raw strand rasterization time. The MSAA column denotes samples per pixel. \textcolor{red}{Red} text reveals increased mesh shader frame times due to culling \revlegal{computational}{} overhead. \textcolor{orange}{Orange} indicates performance degradation from atomic contention in the filter layer. The \textcolor{yellow!70!black}{yellow}-highlighted row represents our best quality-to-performance ratio.}
\label{tab:benchmark}
\centering
\scriptsize
\begin{tabular}{l c c c c c c c c}
    \toprule
    & & & \multicolumn{3}{c}{\textbf{Groom 1}} & \multicolumn{3}{c}{\textbf{Groom 2}} \\
    \cmidrule(lr){4-6} \cmidrule(lr){7-9}
    \textbf{Rasterizer} & \textbf{Filter} & \textbf{MSAA} & {\textbf{Close}} & {\textbf{Mid}} & {\textbf{Far}} & {\textbf{Close}} & {\textbf{Mid}} & {\textbf{Far}} \\
    \midrule
    \multicolumn{9}{l}{\textit{Software Rasterizer -- Compute}} \\
    \addlinespace[2pt]
                              &            & 1 & 3.08\,\tiny(2.68) & 2.35\,\tiny(2.17) & 1.81\,\tiny(1.76) & 3.40\,\tiny(2.87) & 2.74\,\tiny(2.50) & 2.40\,\tiny(2.33) \\
                              & \checkmark & 1 & 3.72\,\tiny(2.89) & 2.70\,\tiny(2.37) & \color{orange} 4.50\,\tiny(4.38) & 4.23\,\tiny(3.03) & 3.00\,\tiny(2.60) & \color{orange} 4.23\,\tiny(4.09) \\
                              &            & 2 & 4.10\,\tiny(3.37) & 2.84\,\tiny(2.56) & 2.03\,\tiny(1.94) & 4.37\,\tiny(3.41) & 3.12\,\tiny(2.76) & 2.58\,\tiny(2.47) \\
                              &            & 4 & 5.97\,\tiny(4.61) & 3.68\,\tiny(3.21) & 2.37\,\tiny(2.21) & 6.24\,\tiny(4.45) & 3.96\,\tiny(3.33) & 2.87\,\tiny(2.68) \\
                              &            & 8 & 8.27\,\tiny(5.67) & 4.71\,\tiny(3.77) & 2.77\,\tiny(2.44) & 9.10\,\tiny(5.61) & 4.95\,\tiny(3.85) & 3.18\,\tiny(2.84) \\
    \addlinespace[3pt]
    \multicolumn{9}{l}{\textit{Software Rasterizer -- Compute + LOD}} \\
    \addlinespace[2pt]
                              &            & 1 & 2.73\,\tiny(2.30) & 1.69\,\tiny(1.49) & 0.25\,\tiny(0.18) & 3.07\,\tiny(2.52) & 2.24\,\tiny(1.95) & 0.35\,\tiny(0.27) \\
    \rowcolor{yellow!20}
                              & \checkmark & 1 & 3.35\,\tiny(2.48) & 1.94\,\tiny(1.56) & 0.36\,\tiny(0.22) & 3.79\,\tiny(2.59) & 2.40\,\tiny(1.97) & 0.45\,\tiny(0.30) \\
                              &            & 2 & 3.73\,\tiny(2.99) & 2.15\,\tiny(1.81) & 0.29\,\tiny(0.19) & 4.08\,\tiny(3.08) & 2.55\,\tiny(2.17) & 0.49\,\tiny(0.34) \\
                              &            & 4 & 5.54\,\tiny(4.16) & 2.87\,\tiny(2.36) & 0.42\,\tiny(0.22) & 5.95\,\tiny(4.14) & 3.27\,\tiny(2.65) & 0.52\,\tiny(0.32) \\
                              &            & 8 & 7.71\,\tiny(5.08) & 3.69\,\tiny(2.78) & 0.62\,\tiny(0.25) & 8.56\,\tiny(5.06) & 4.17\,\tiny(3.03) & 0.76\,\tiny(0.36) \\
    \midrule
    \multicolumn{9}{l}{\textit{Hardware Rasterizer -- Mesh Shader}} \\
    \addlinespace[2pt]
                              &  & 1 & 13.58 & 10.58 & 9.66 & 15.07 & 11.24 & 10.55 \\
                              &  & 2 & 16.42 & 12.14 & 9.75 & 18.47 & 12.83 & 10.67 \\
                              &  & 4 & 19.80 & 14.15 & 10.14 & 22.36 & 14.92 & 11.12 \\
                              &  & 8 & 23.14 & 16.47 & 10.76 & 26.38 & 18.44 & 12.39 \\
    \addlinespace[3pt]
    \multicolumn{9}{l}{\textit{Hardware Rasterizer -- Mesh Shader + LOD}} \\
    \addlinespace[2pt]
                              &  & 1 & \color{red} 19.45 & \color{red} 13.31 & 2.28 & \color{red} 20.47 & \color{red} 15.10 & 3.02 \\
                              &  & 2 & \color{red} 23.12 & \color{red} 16.54 & 2.40 & \color{red} 25.49 & \color{red} 19.51 & 3.15 \\
                              &  & 4 & \color{red} 26.18 & \color{red} 18.51 & 2.68 & \color{red} 28.60 & \color{red} 21.19 & 3.57 \\
                              &  & 8 & \color{red} 28.40 & \color{red} 20.34 & 3.16 & \color{red} 32.00 & \color{red} 23.70 & 4.60 \\
    \bottomrule
\end{tabular}
\end{table}

\def\groomIlabel{Groom 1}
\def\groomIIlabel{Groom 2}

\def\refimgcloseI{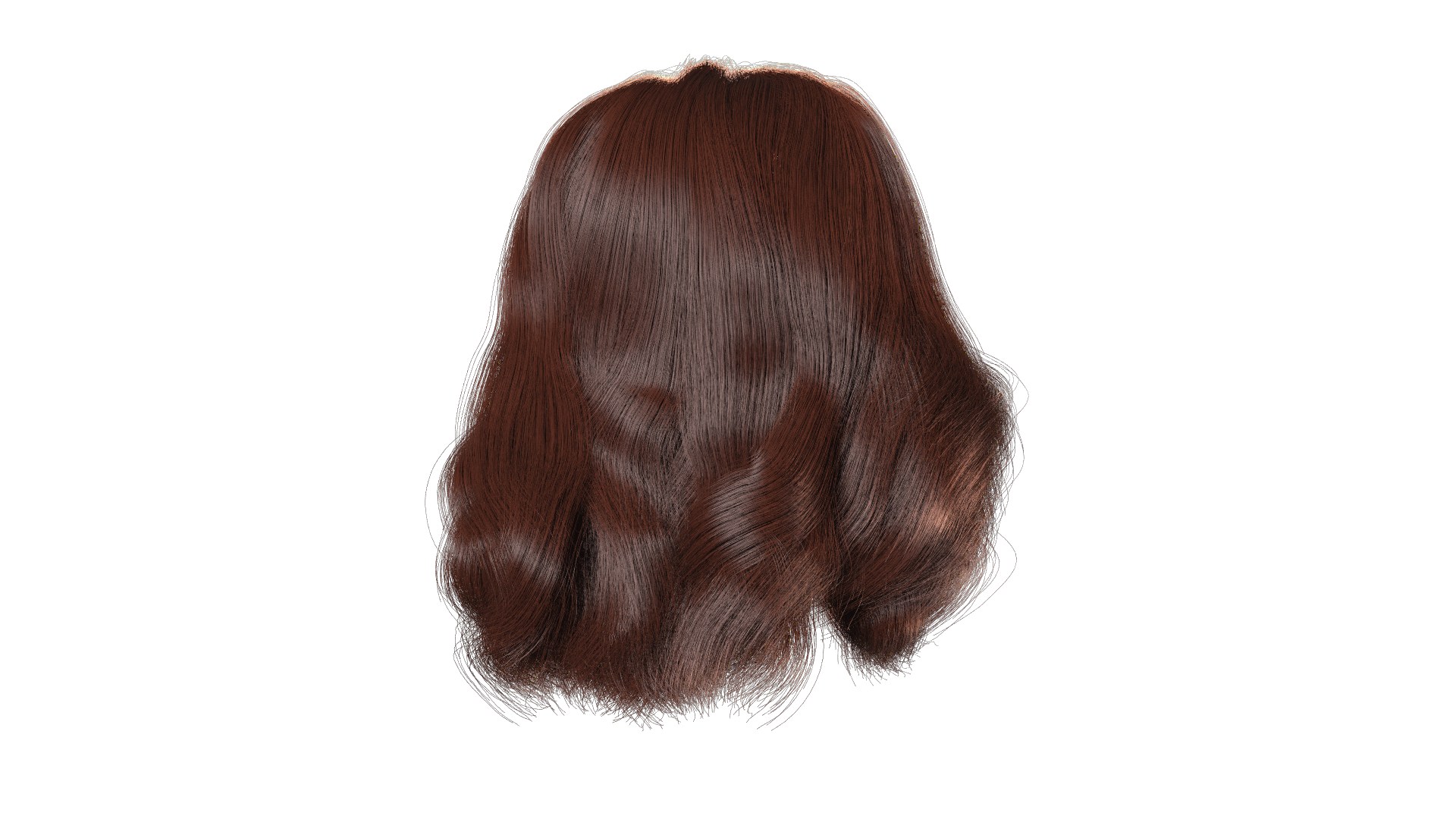}
\def\refimgmidI{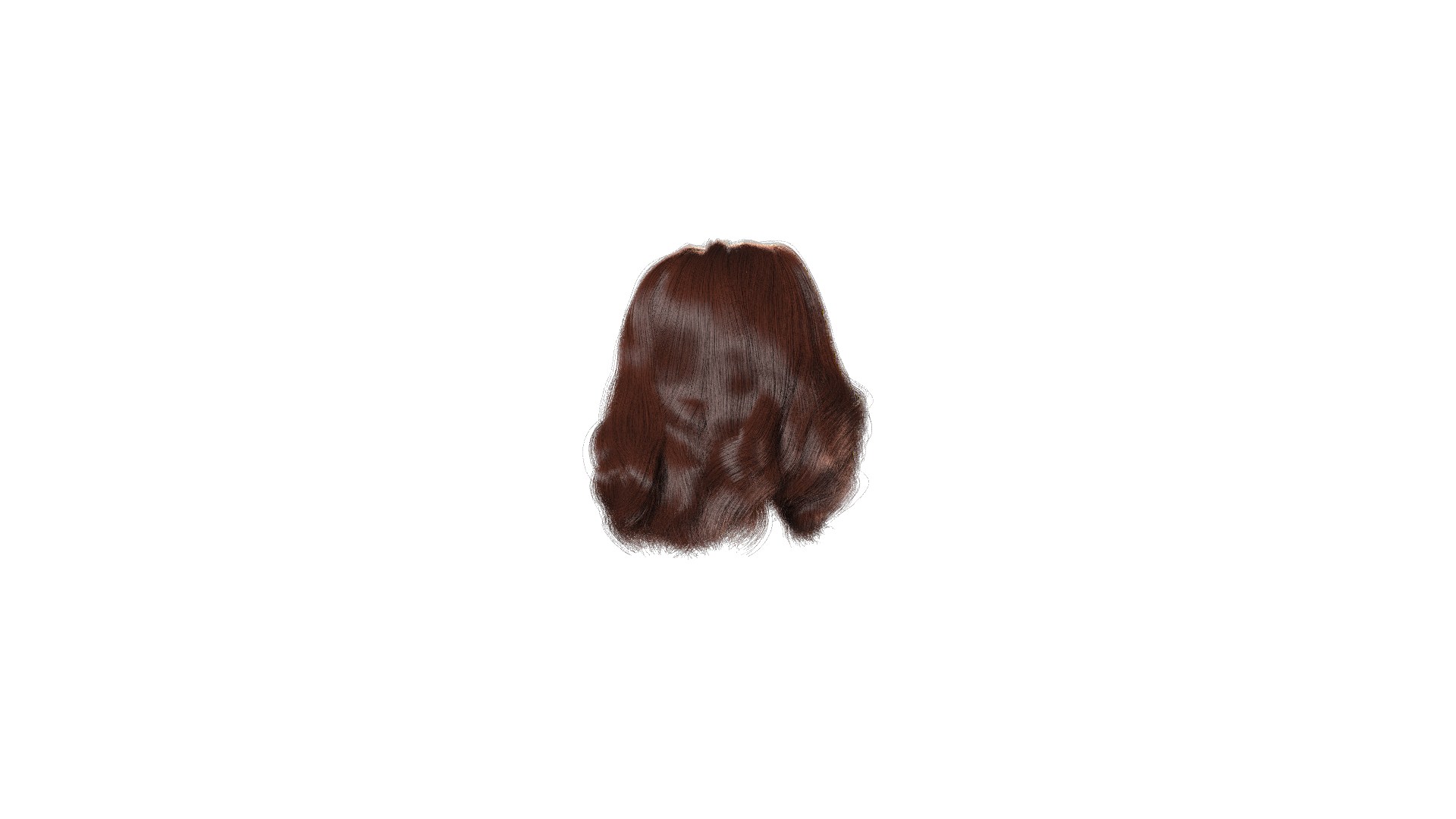}
\def\refimgfarI{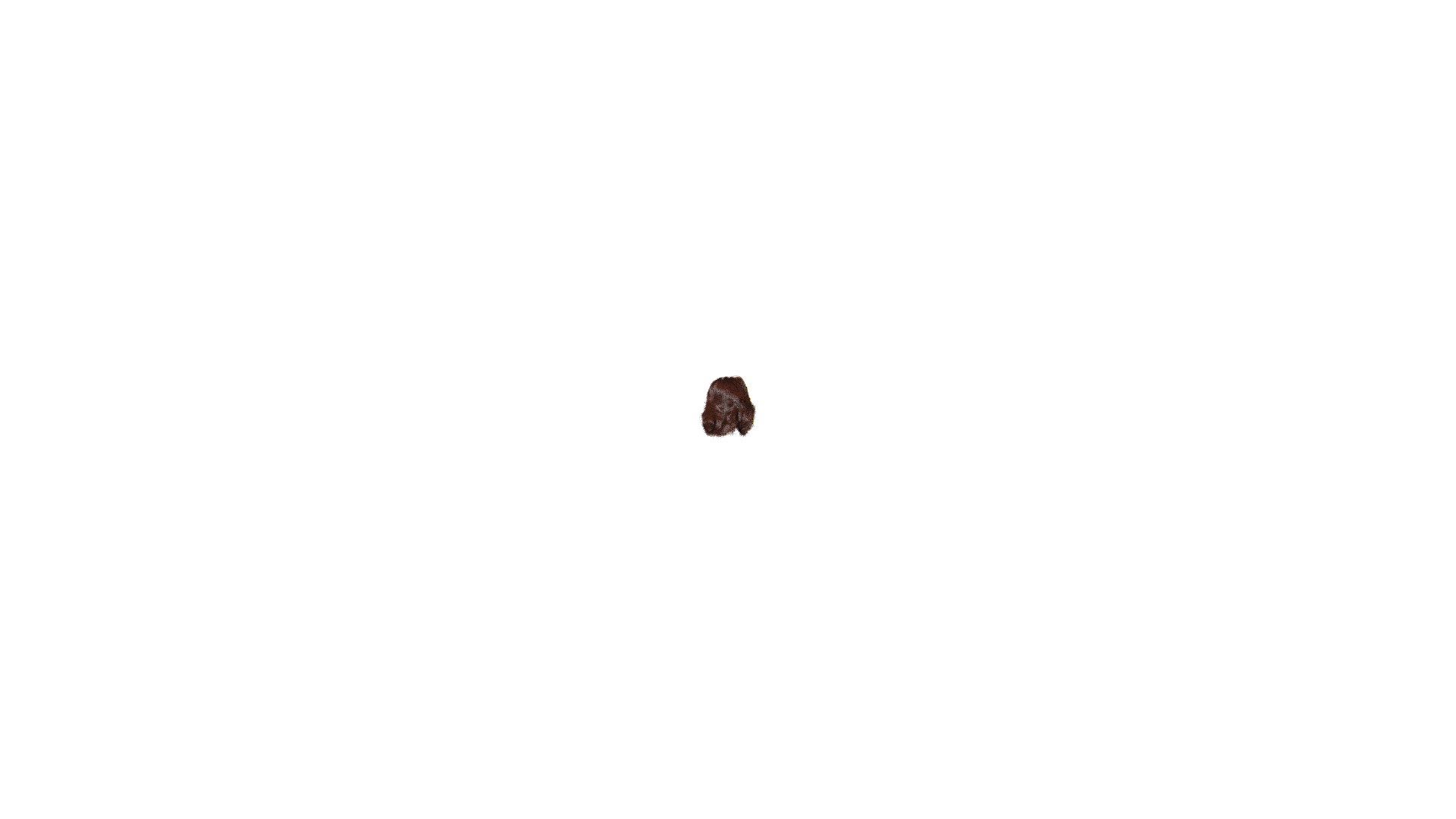}

\def\refimgcloseII{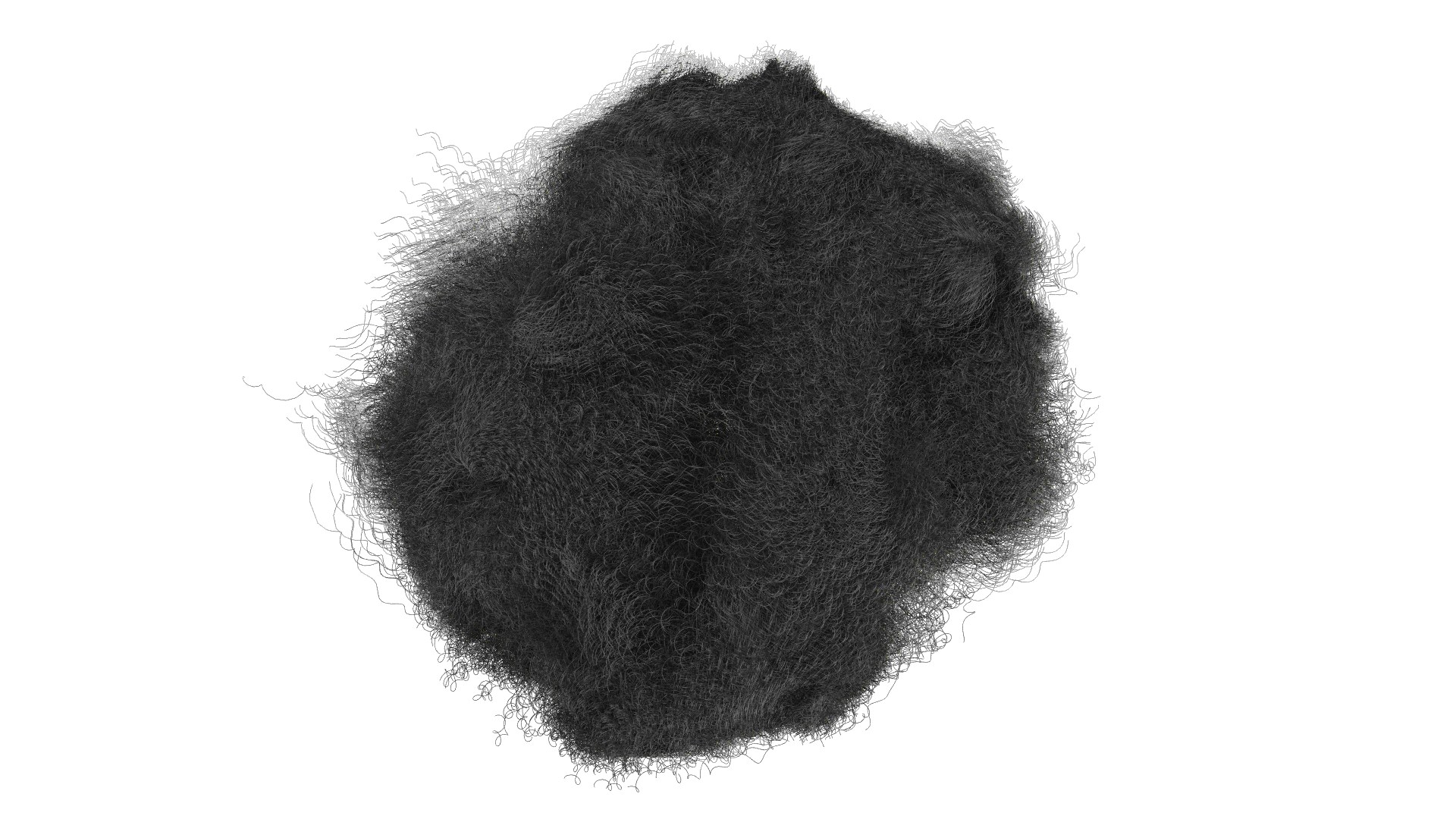}
\def\refimgmidII{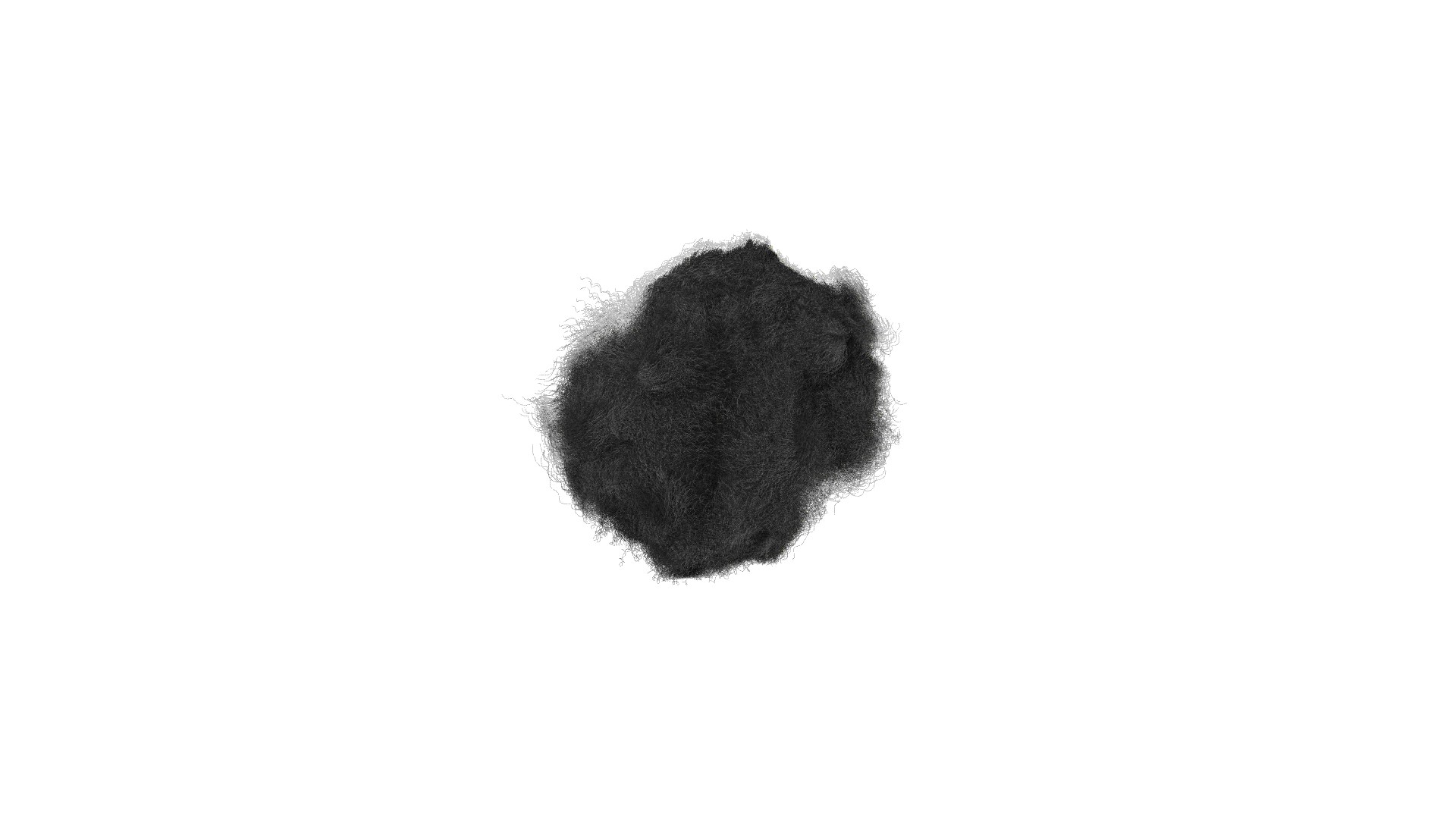}
\def\refimgfarII{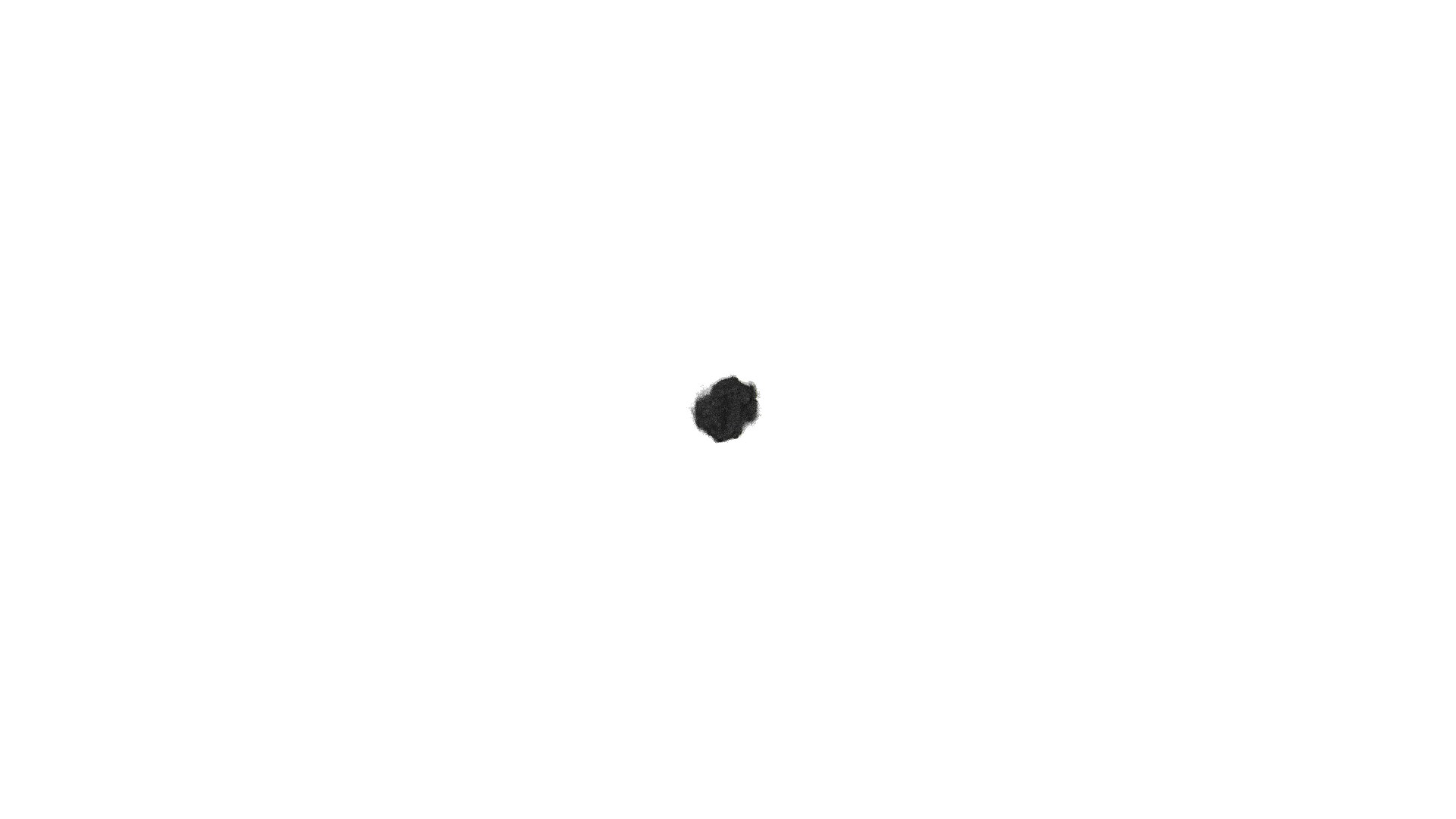}

% --- Groom 1: Placeholder data ---
\def\expAlabel{SWR~1}     \def\expAclose{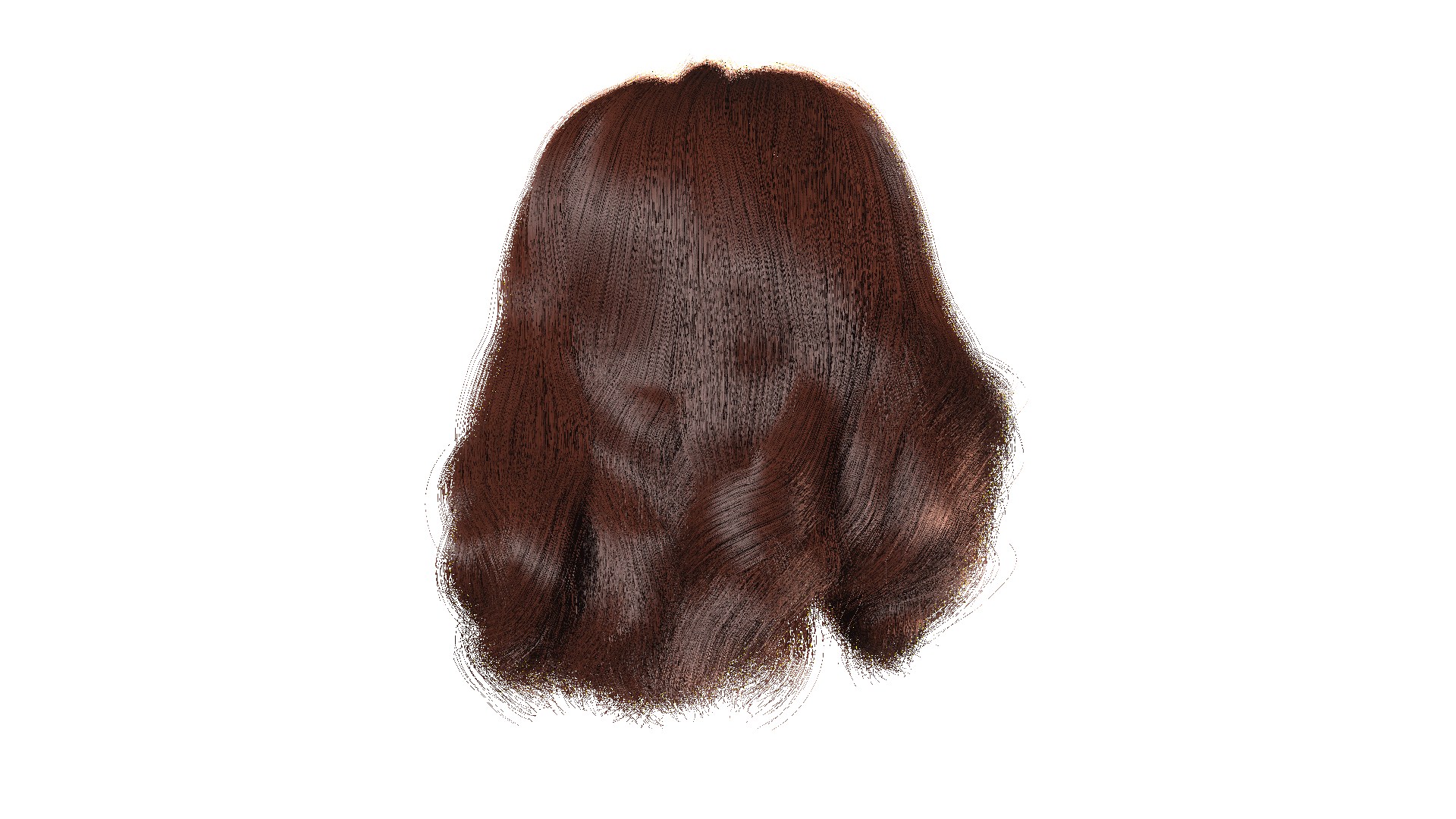}              \def\expAmid{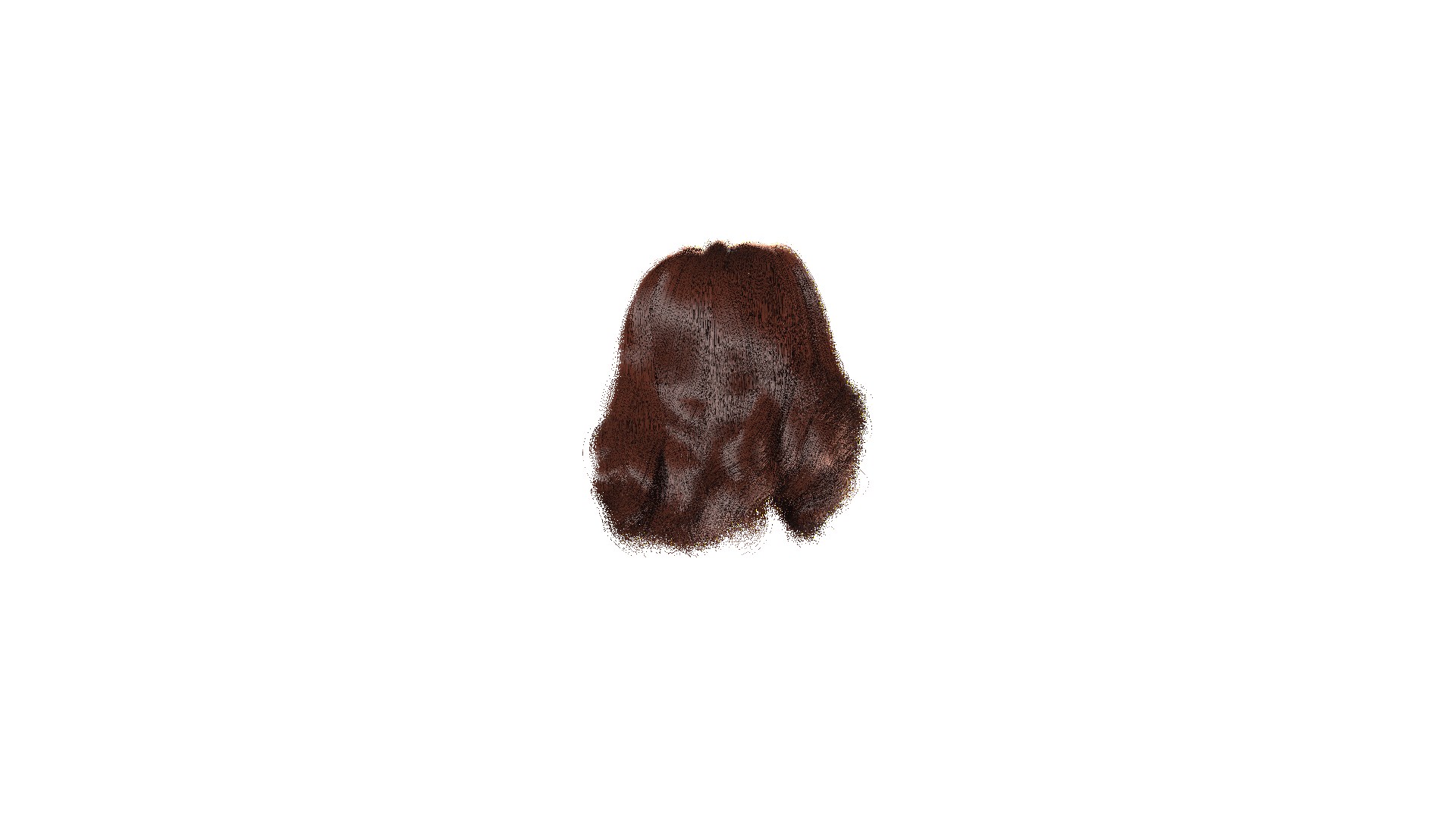}              \def\expAfar{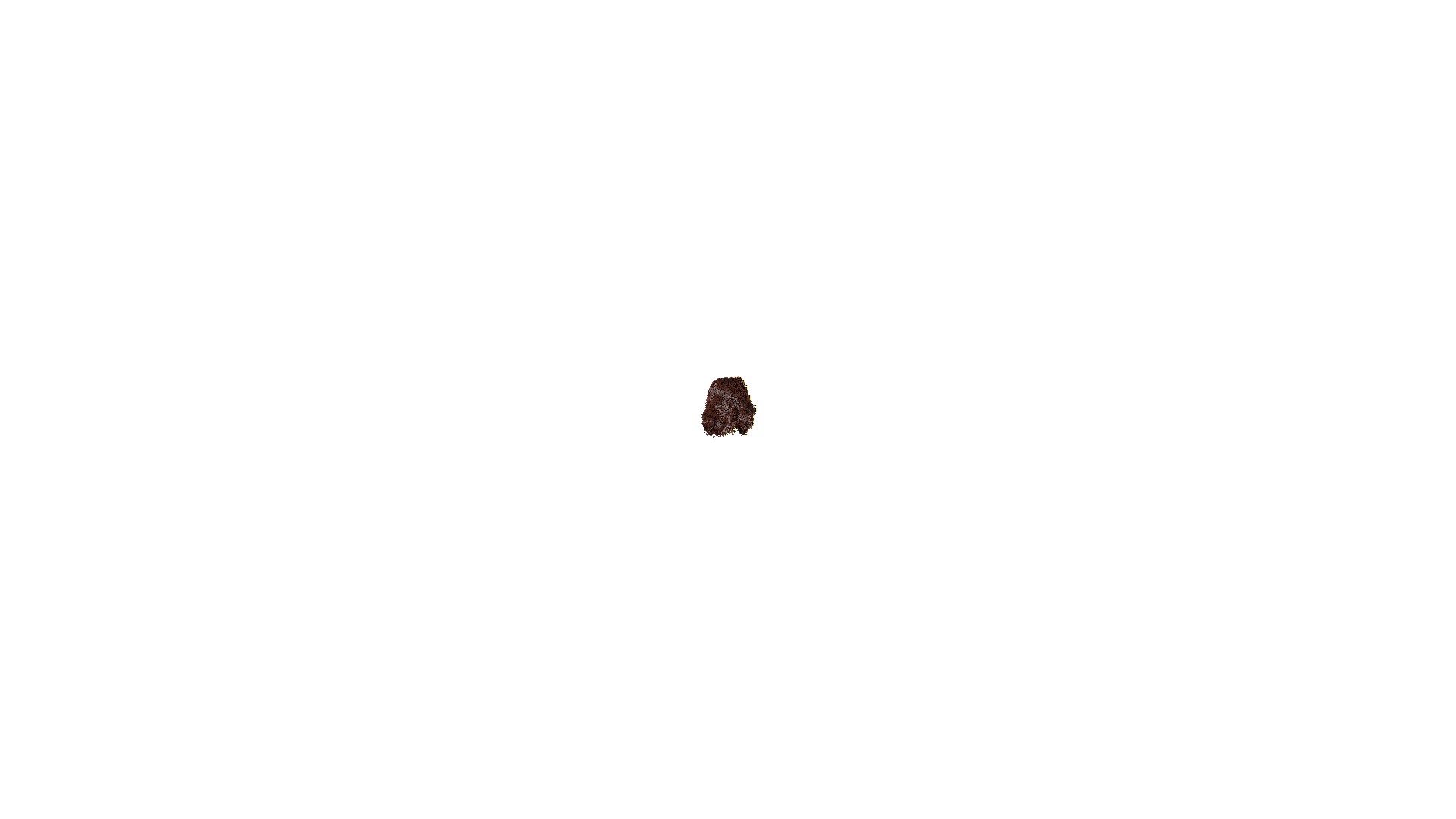}
\def\expBlabel{SWR~1+F}   \def\expBclose{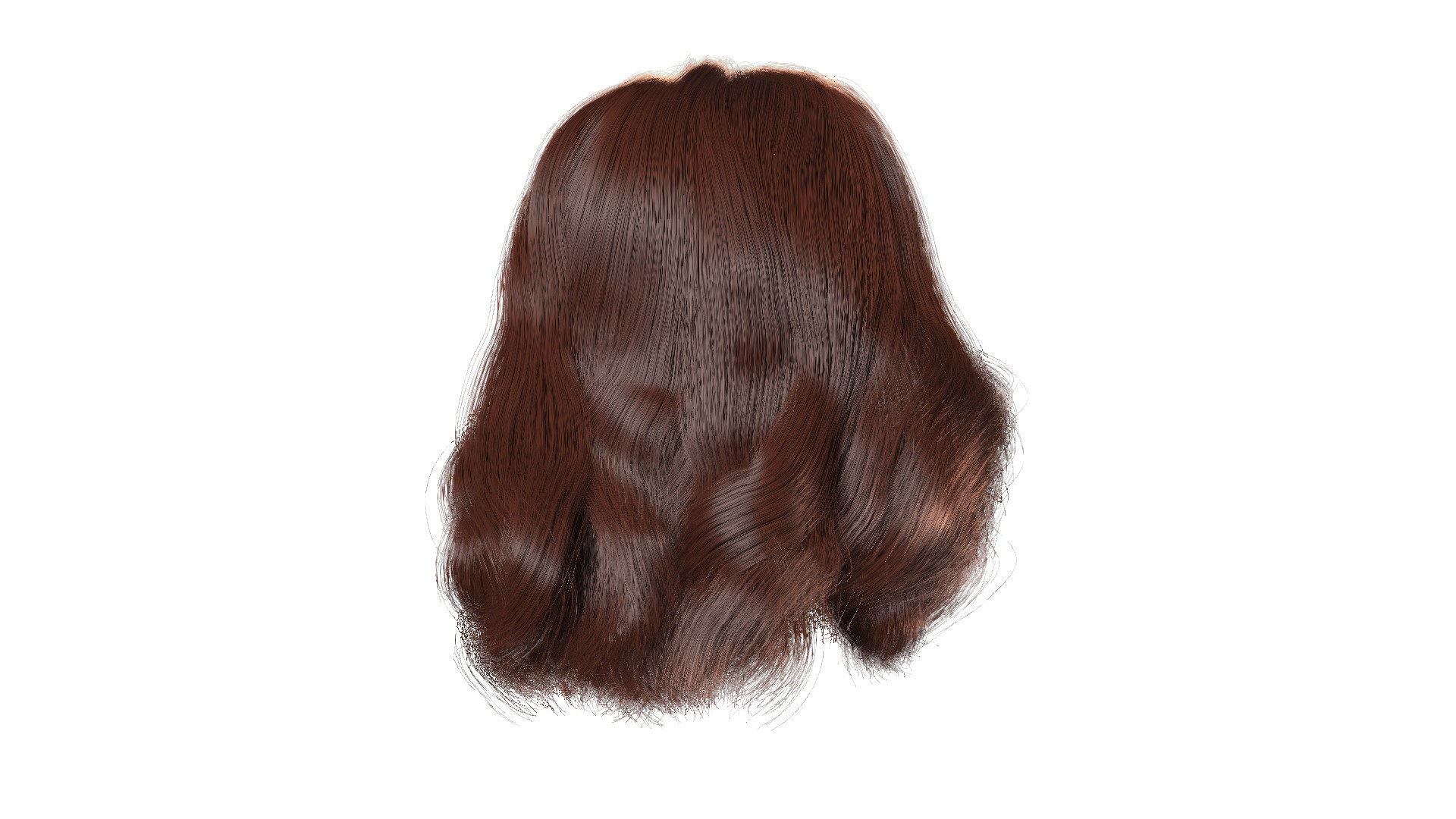}       \def\expBmid{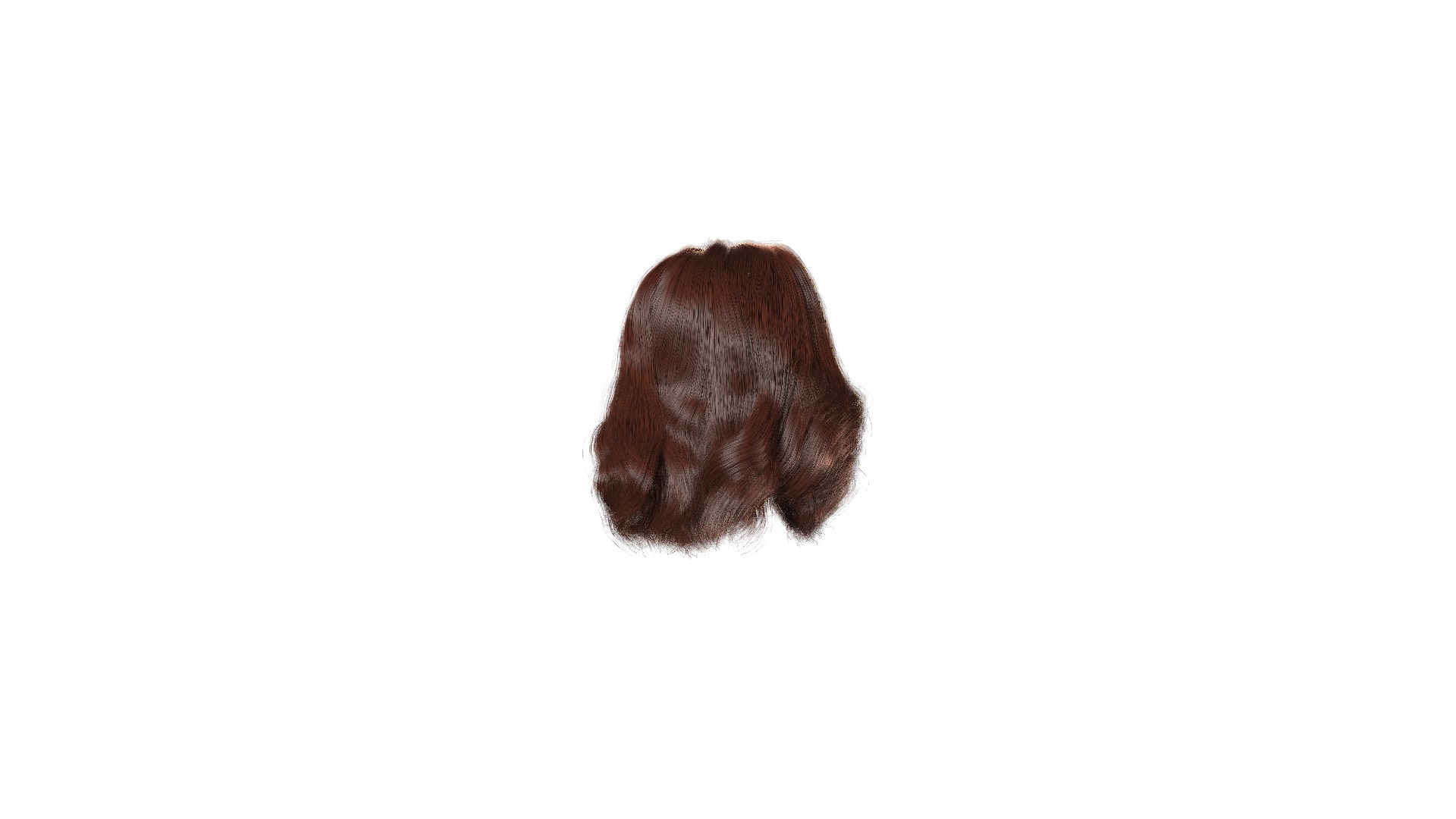}       \def\expBfar{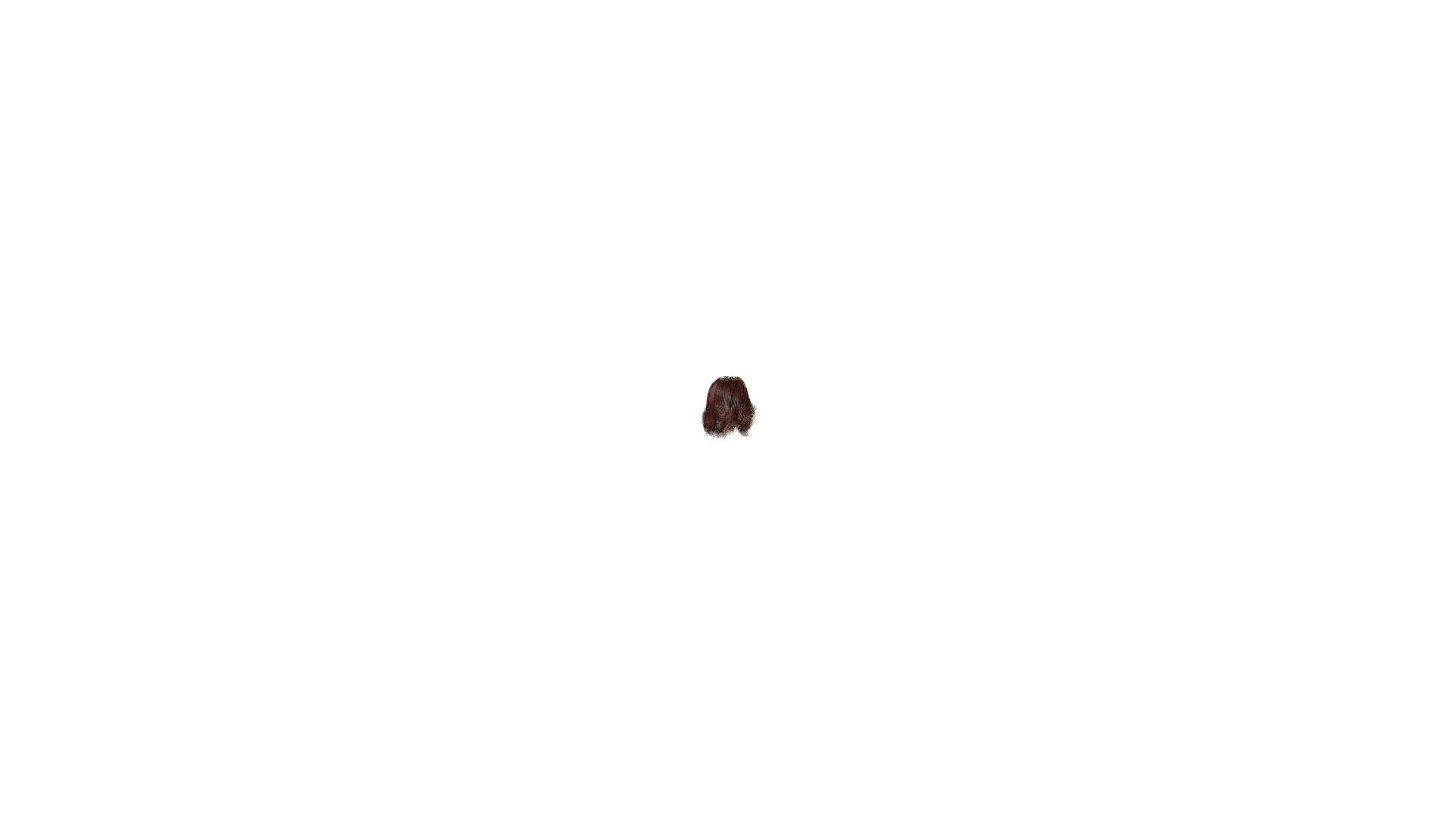}
\def\expClabel{SWR~1+F+L} \def\expCclose{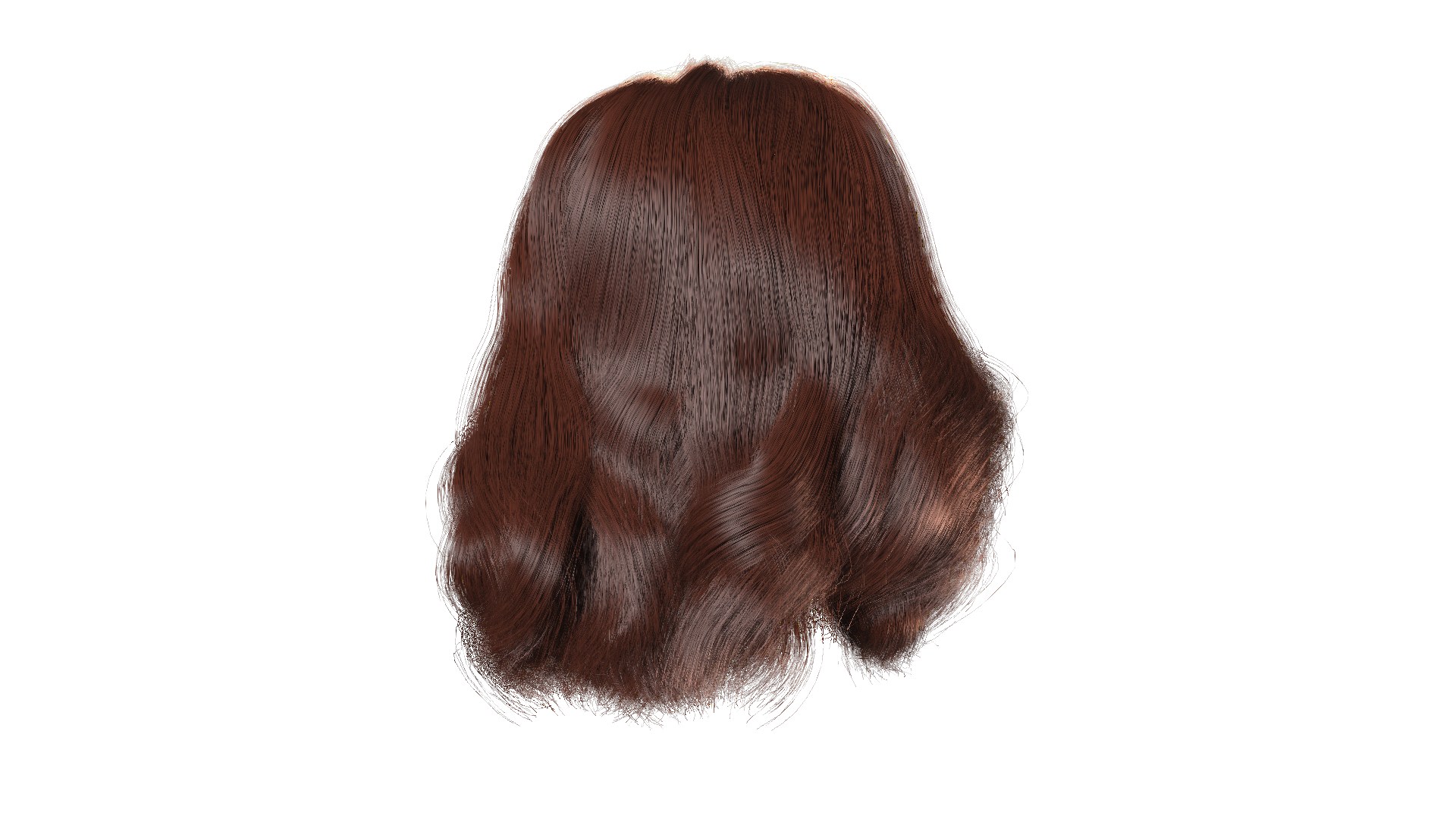}   \def\expCmid{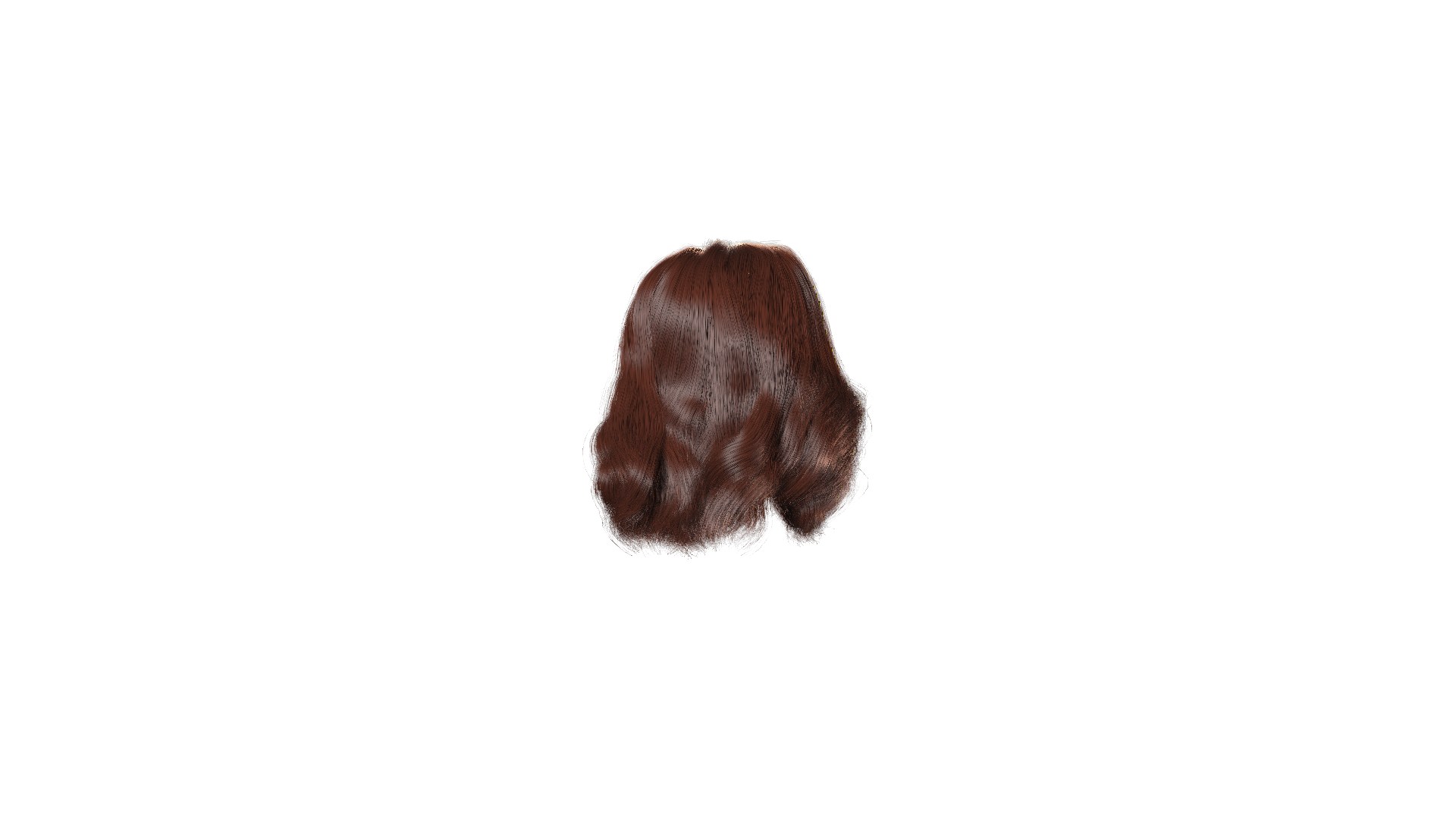}   \def\expCfar{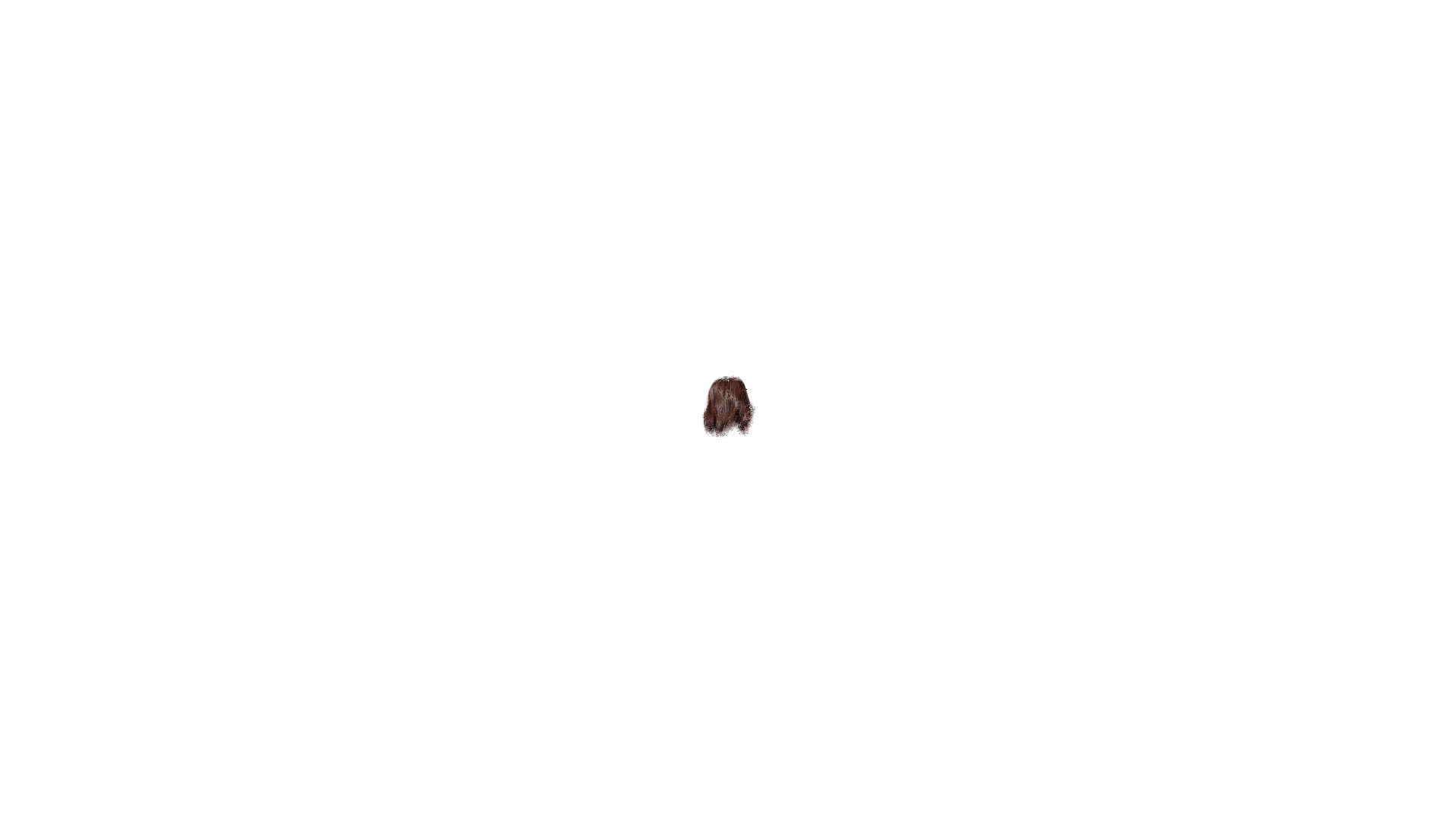}
\def\expDlabel{SWR~2}     \def\expDclose{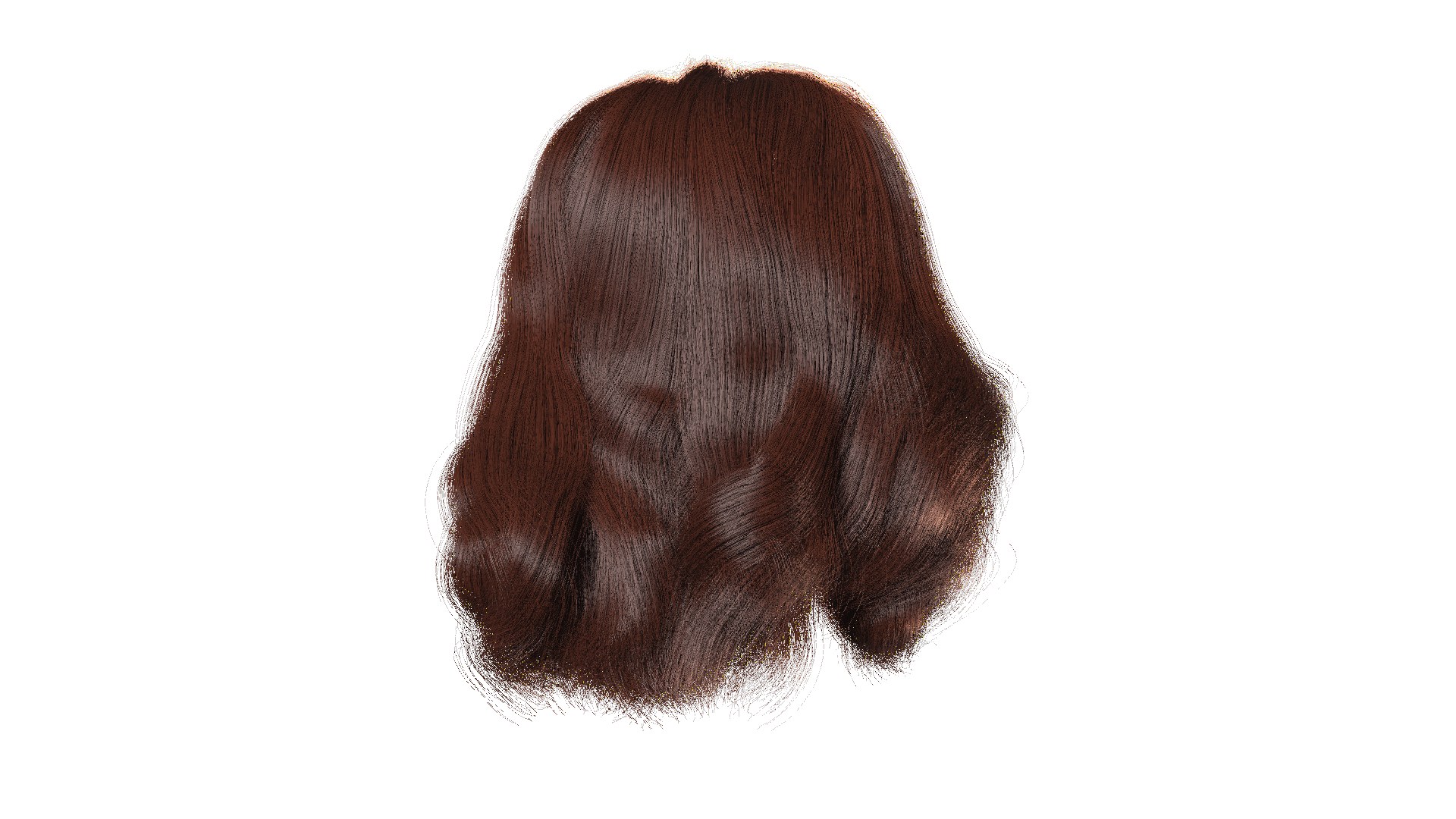}              \def\expDmid{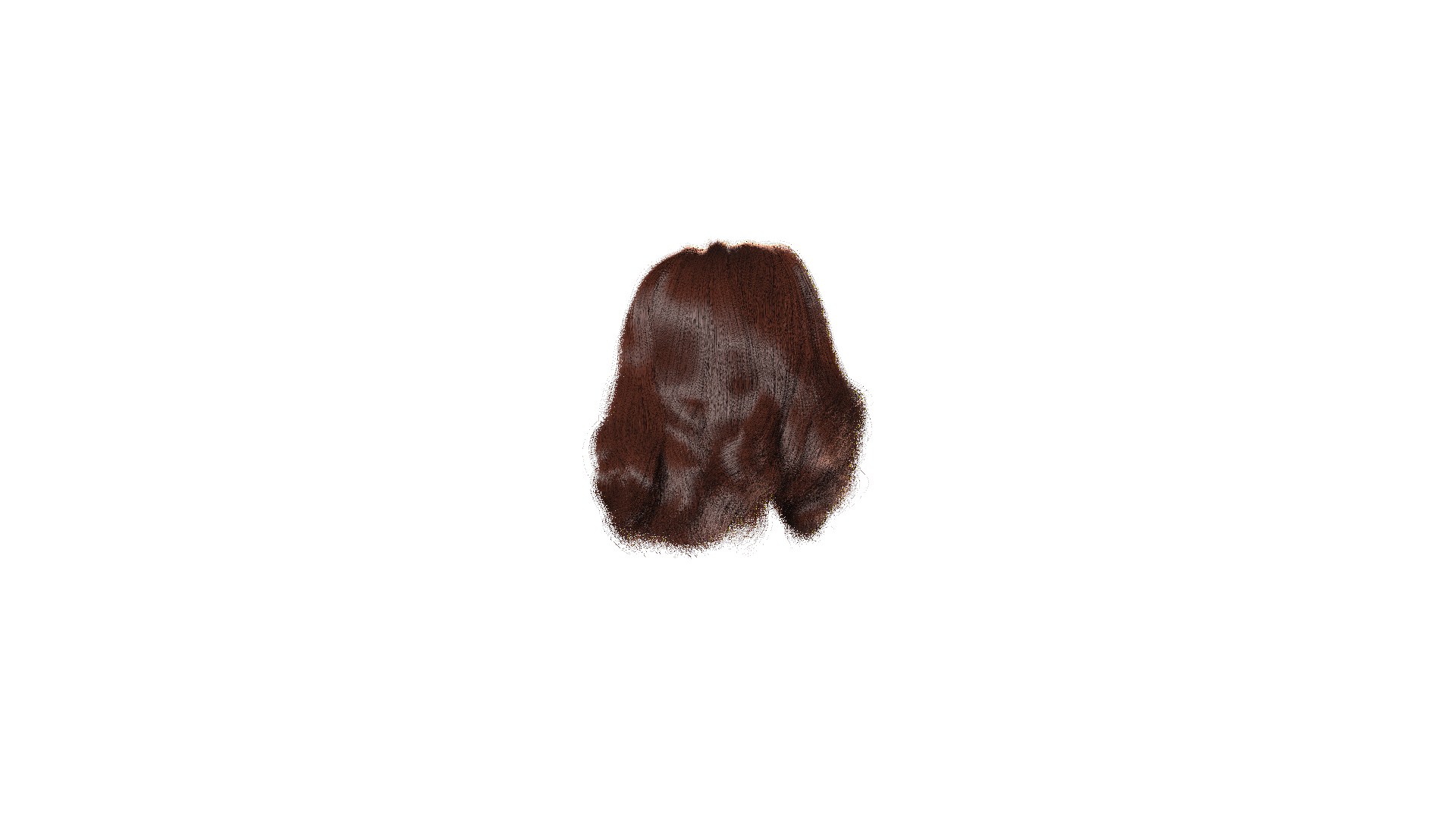}              \def\expDfar{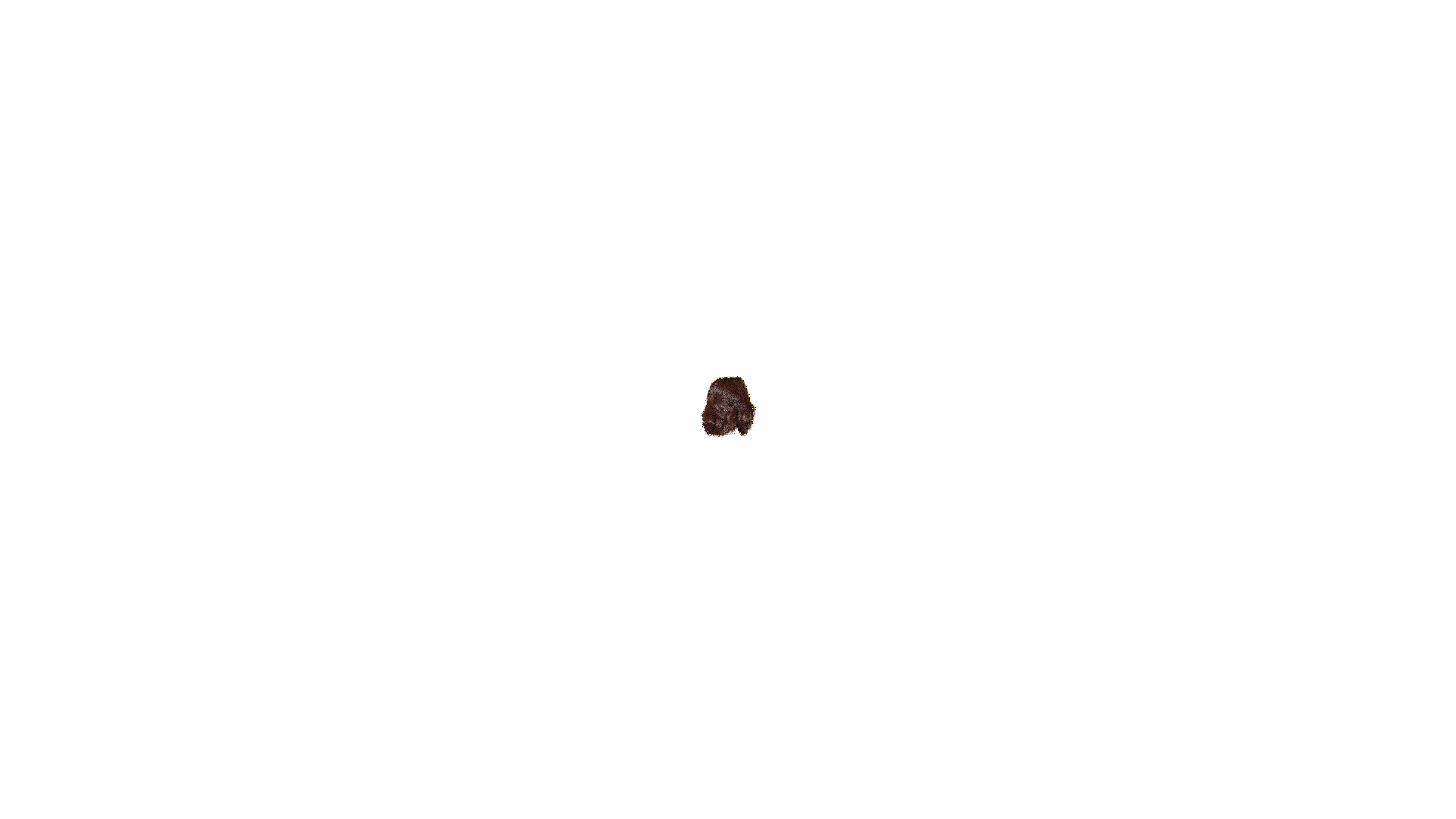}
\def\expElabel{SWR~2+L}   \def\expEclose{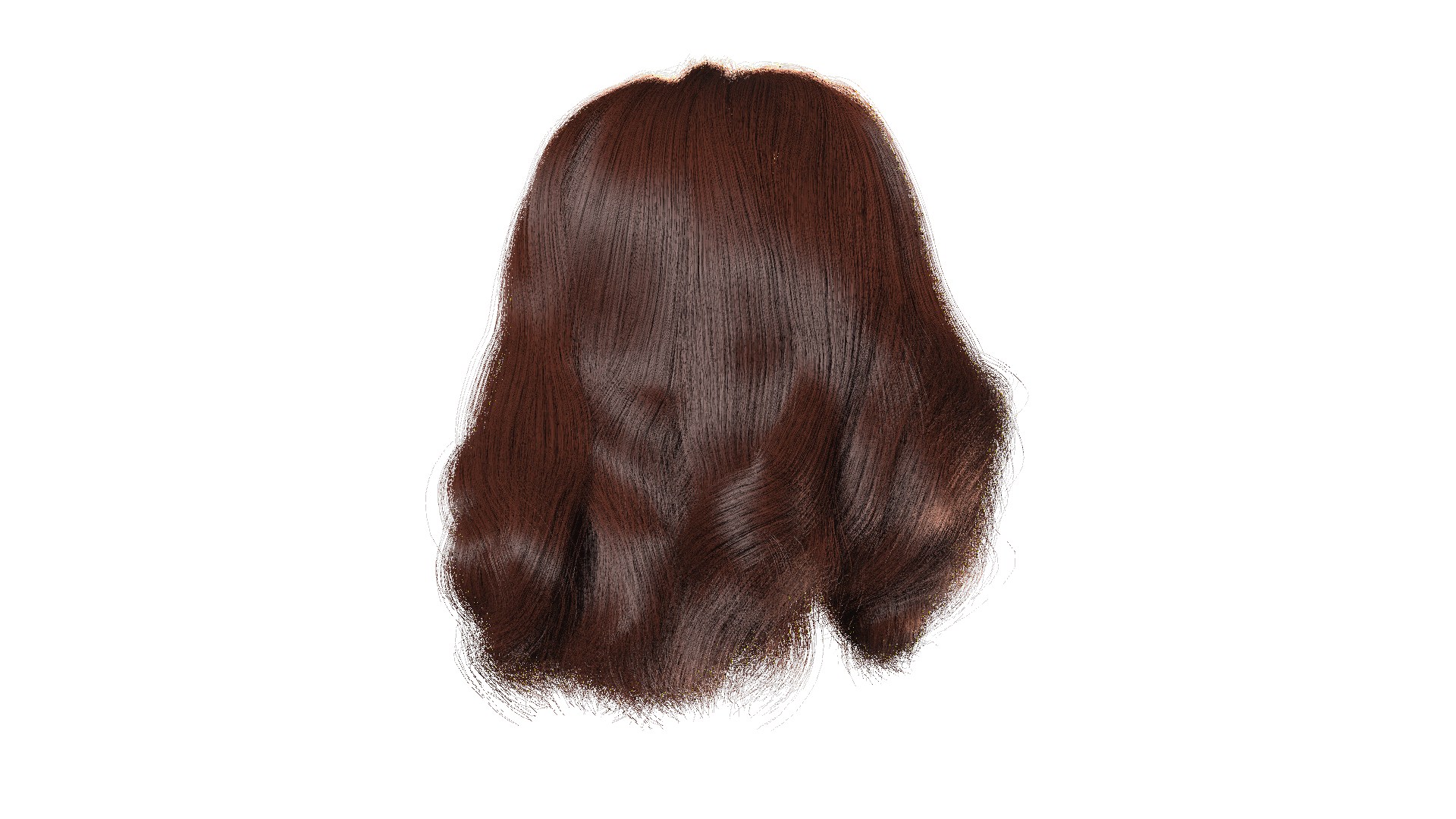}          \def\expEmid{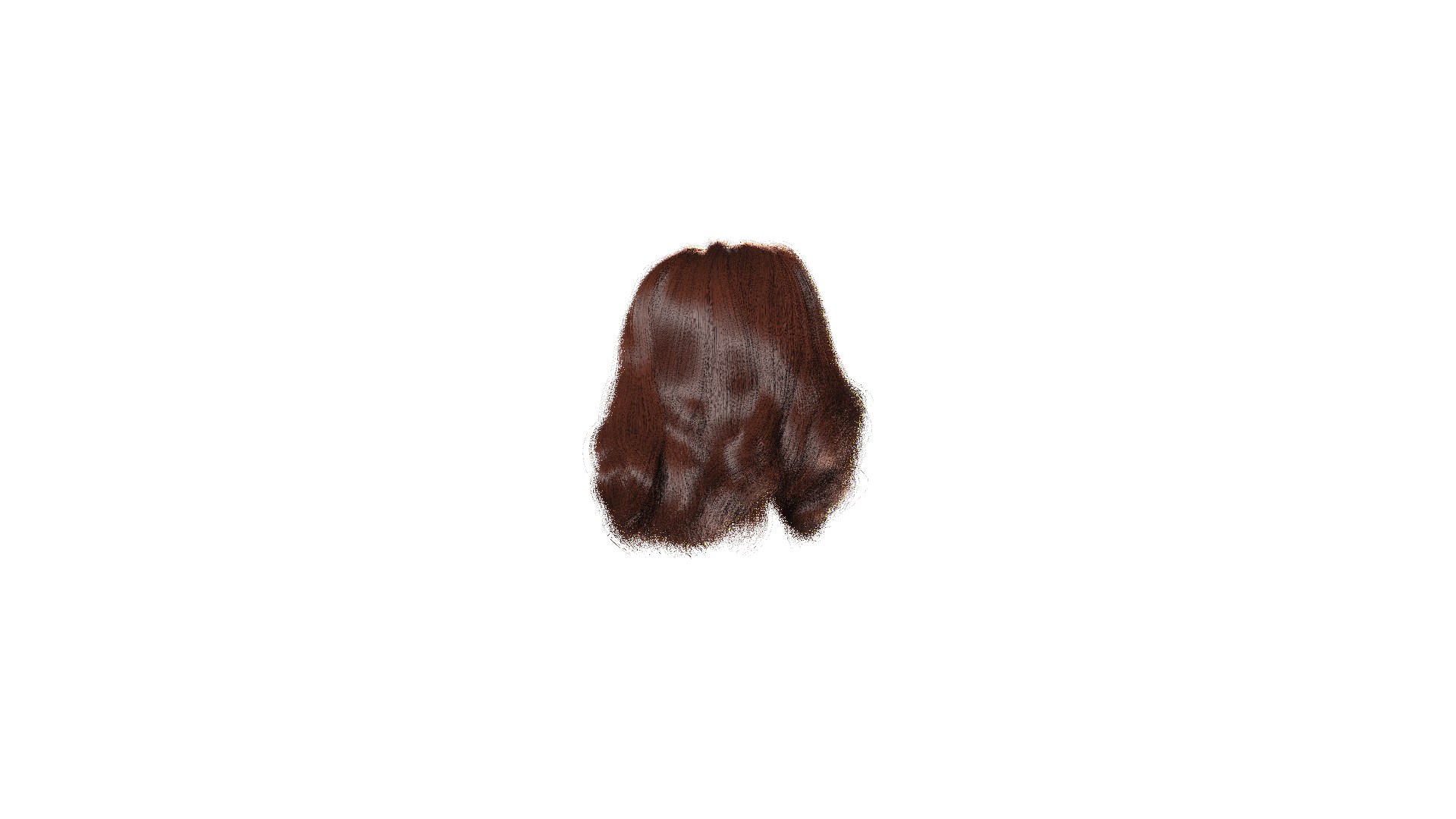}          \def\expEfar{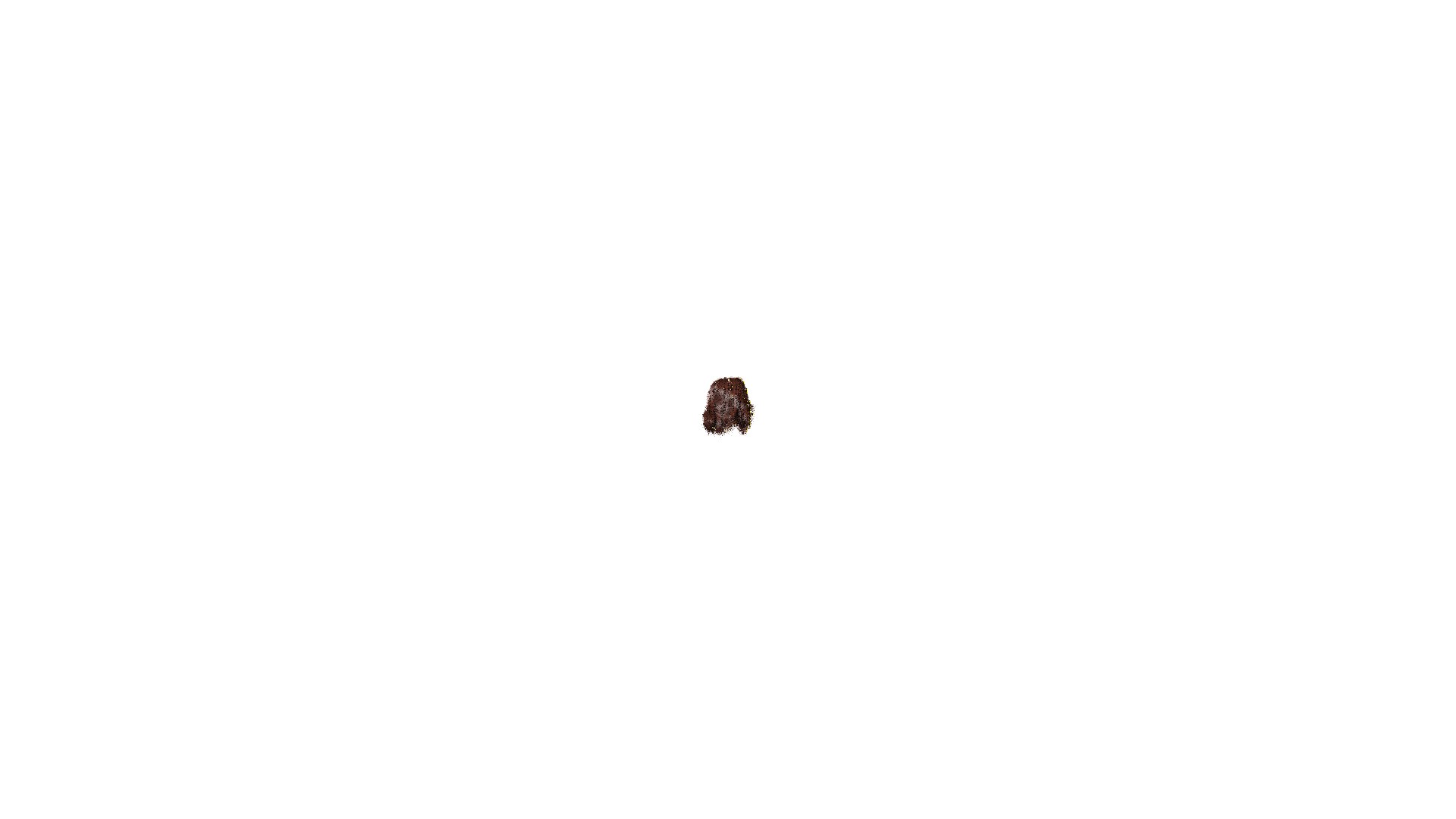}
\def\expFlabel{SWR~8}     \def\expFclose{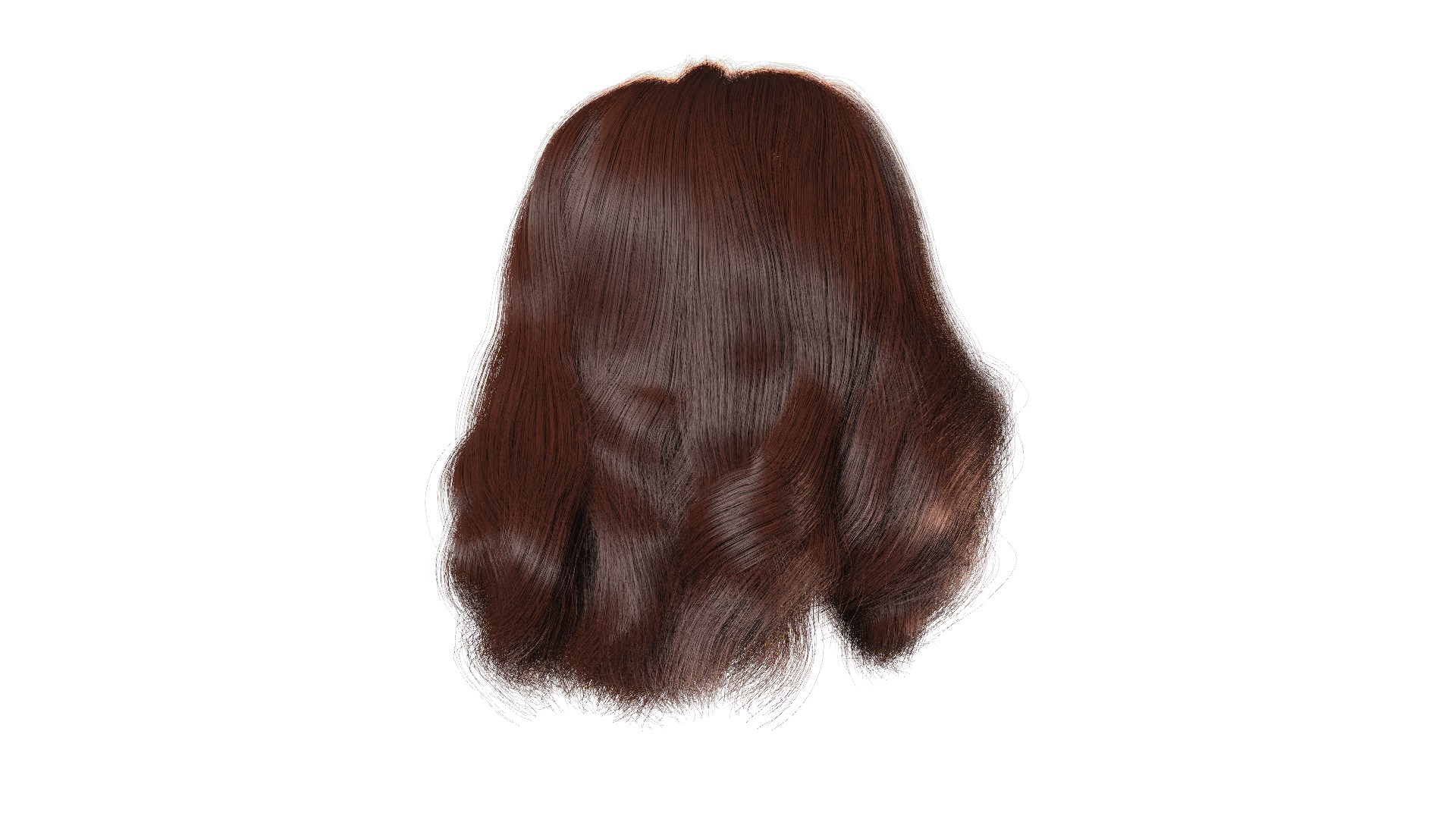}              \def\expFmid{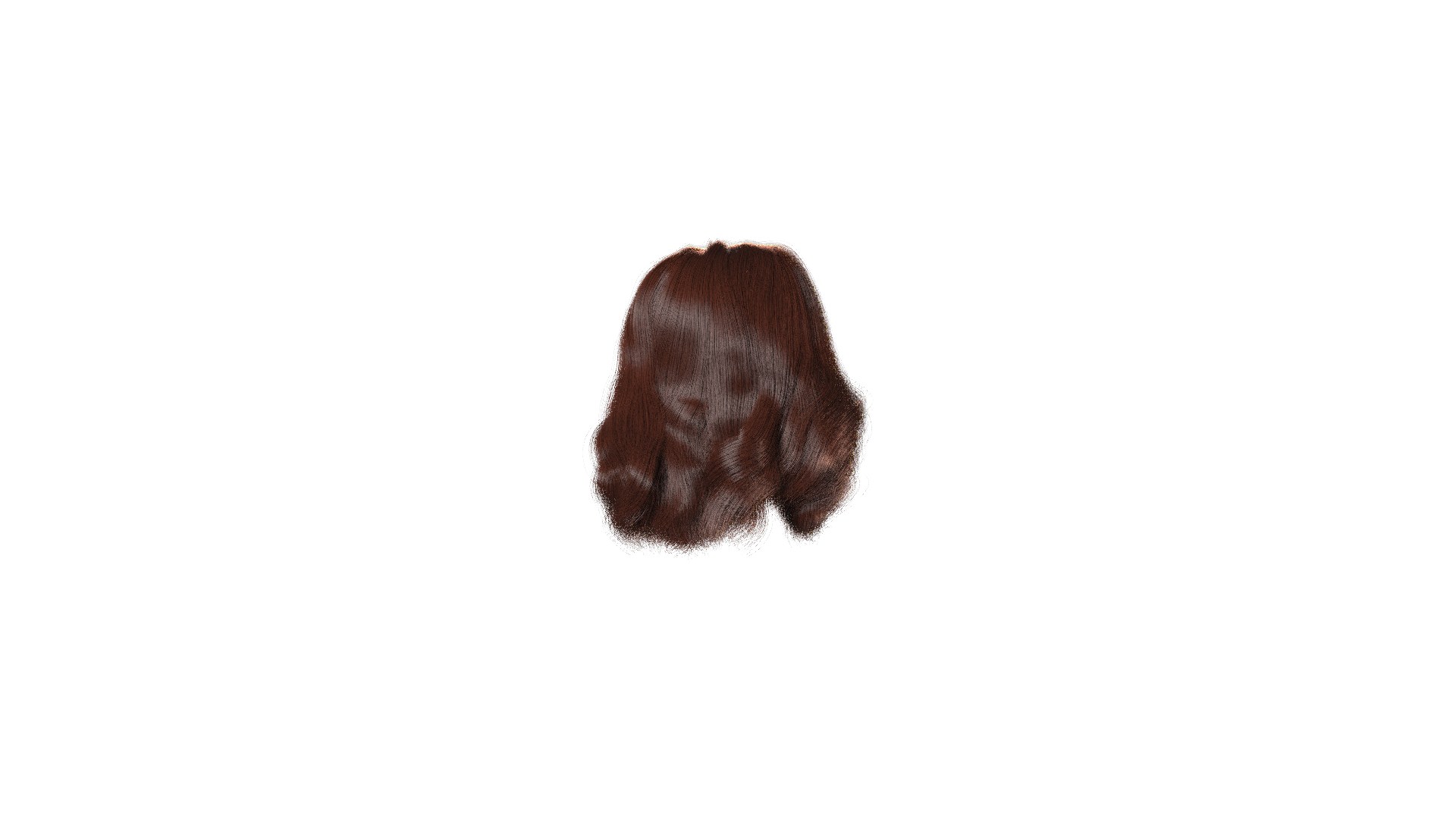}              \def\expFfar{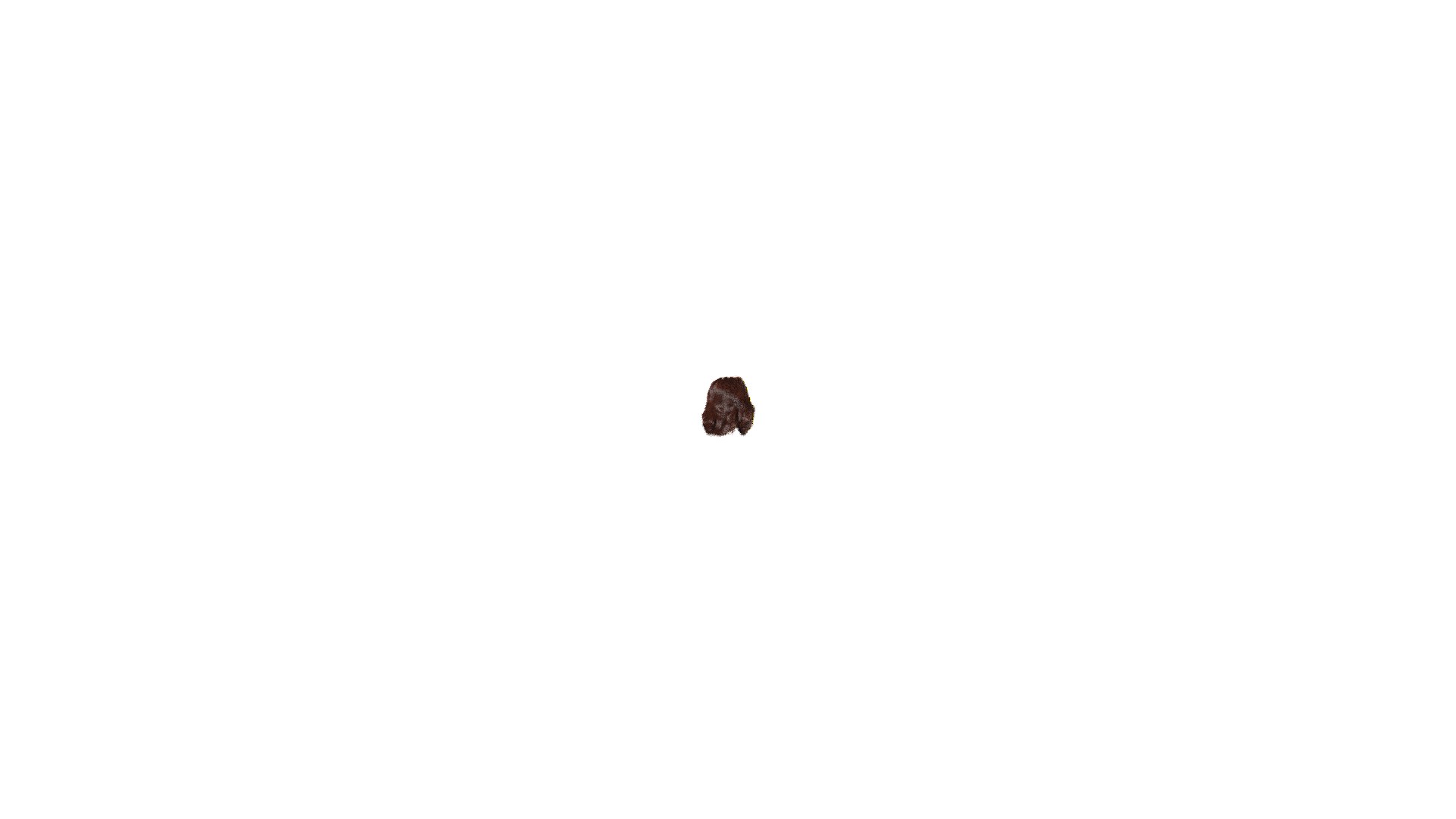}
\def\expGlabel{Mesh~8}    \def\expGclose{figure/benchmark/mesh8_offset0.jpg}             \def\expGmid{figure/benchmark/mesh8_offset100.jpg}             \def\expGfar{figure/benchmark/mesh8_offset900.jpg}

% --- Groom 2: Placeholder data (7 experiments) ---
\def\expIIAlabel{SWR~1}     \def\expIIAclose{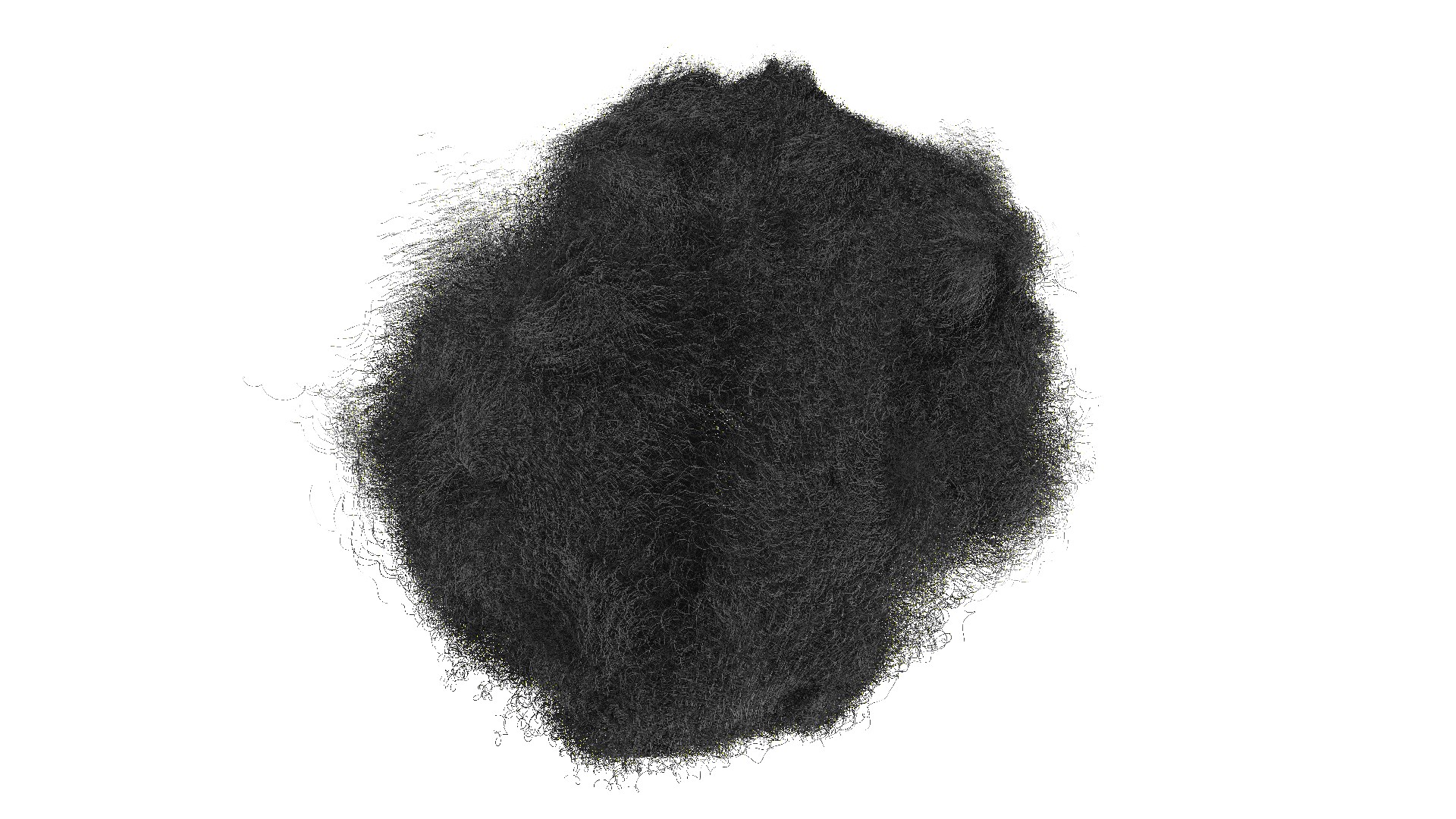}              \def\expIIAmid{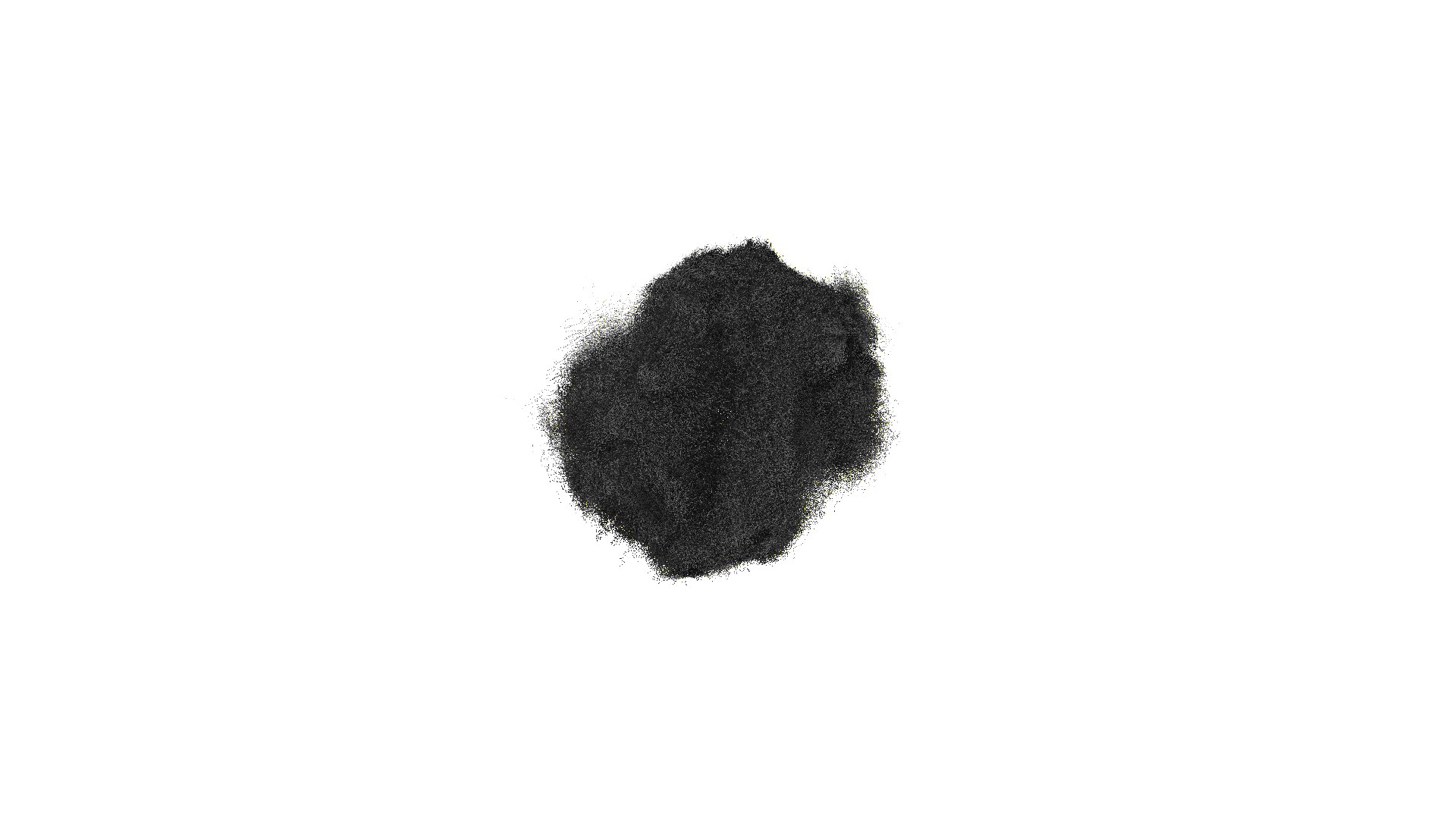}              \def\expIIAfar{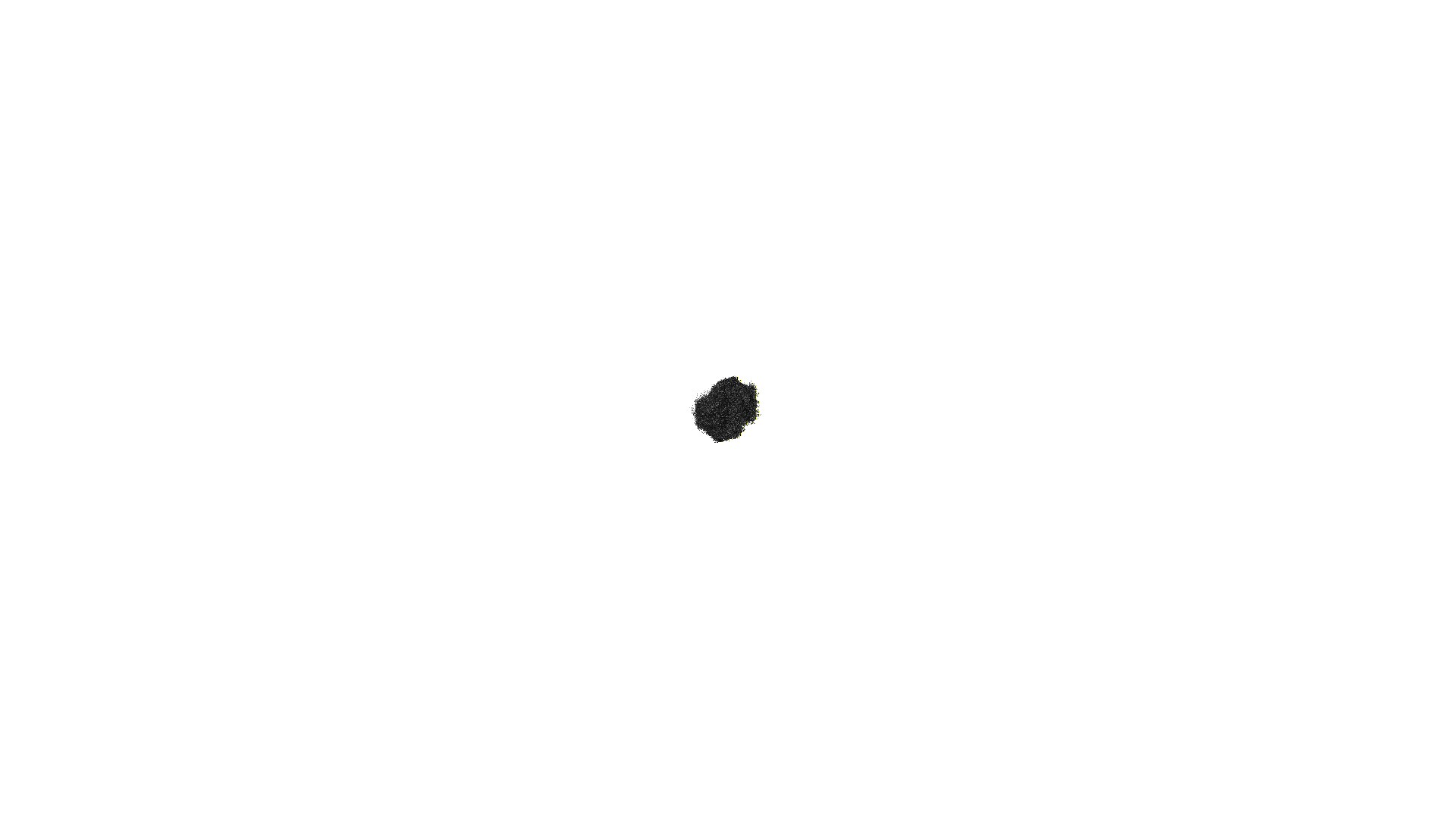}
\def\expIIBlabel{SWR~1+F}   \def\expIIBclose{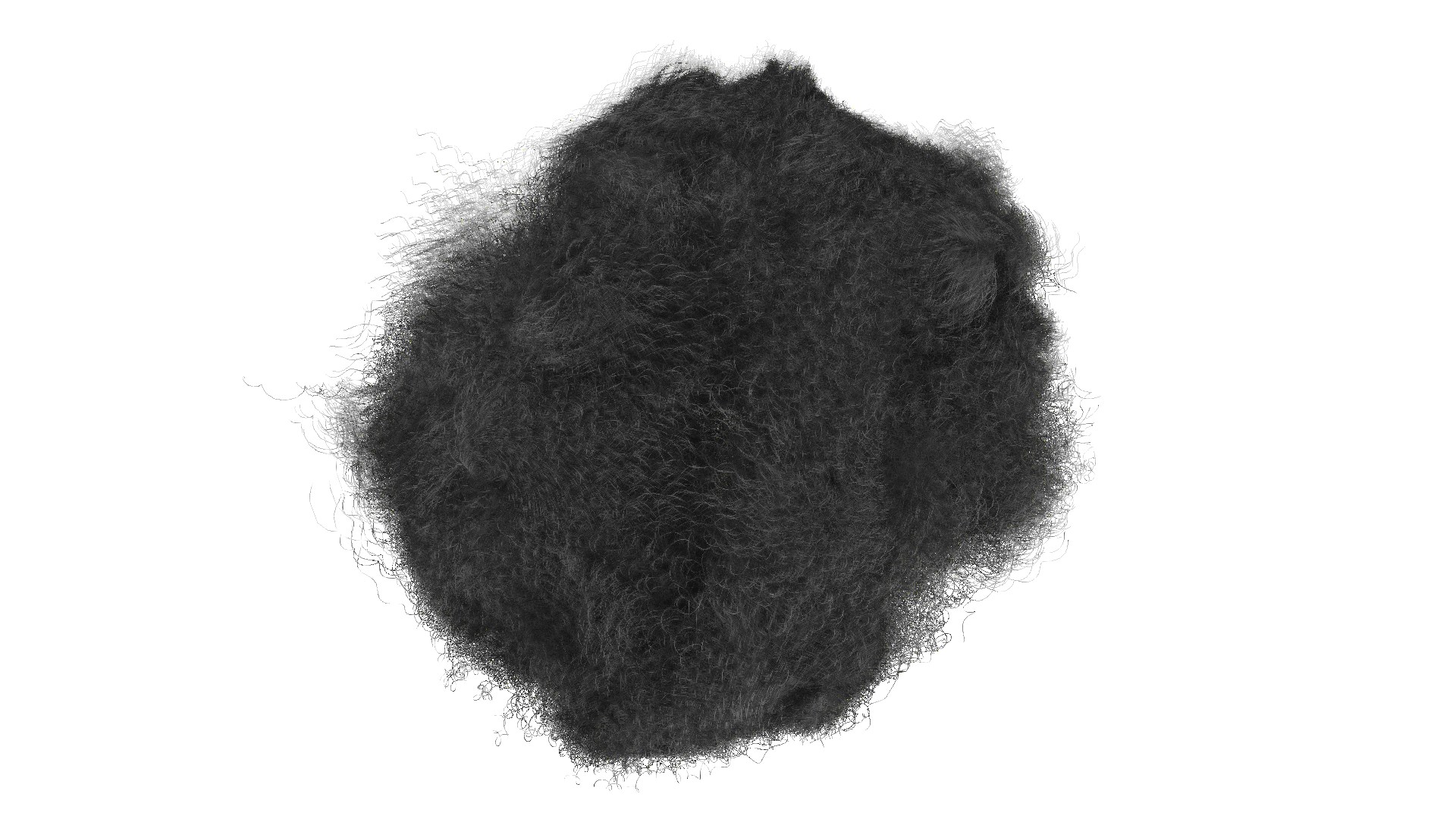}       \def\expIIBmid{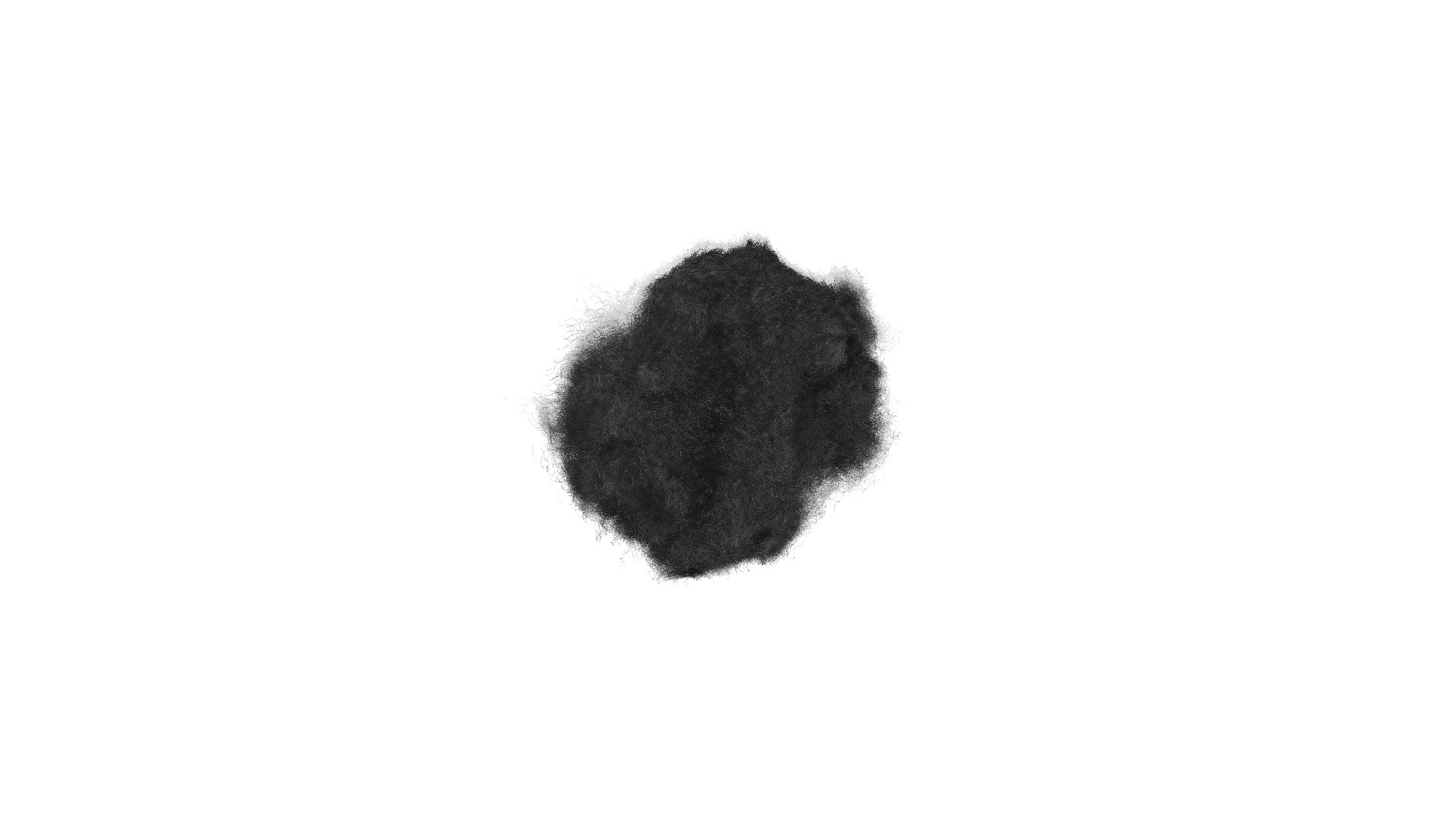}       \def\expIIBfar{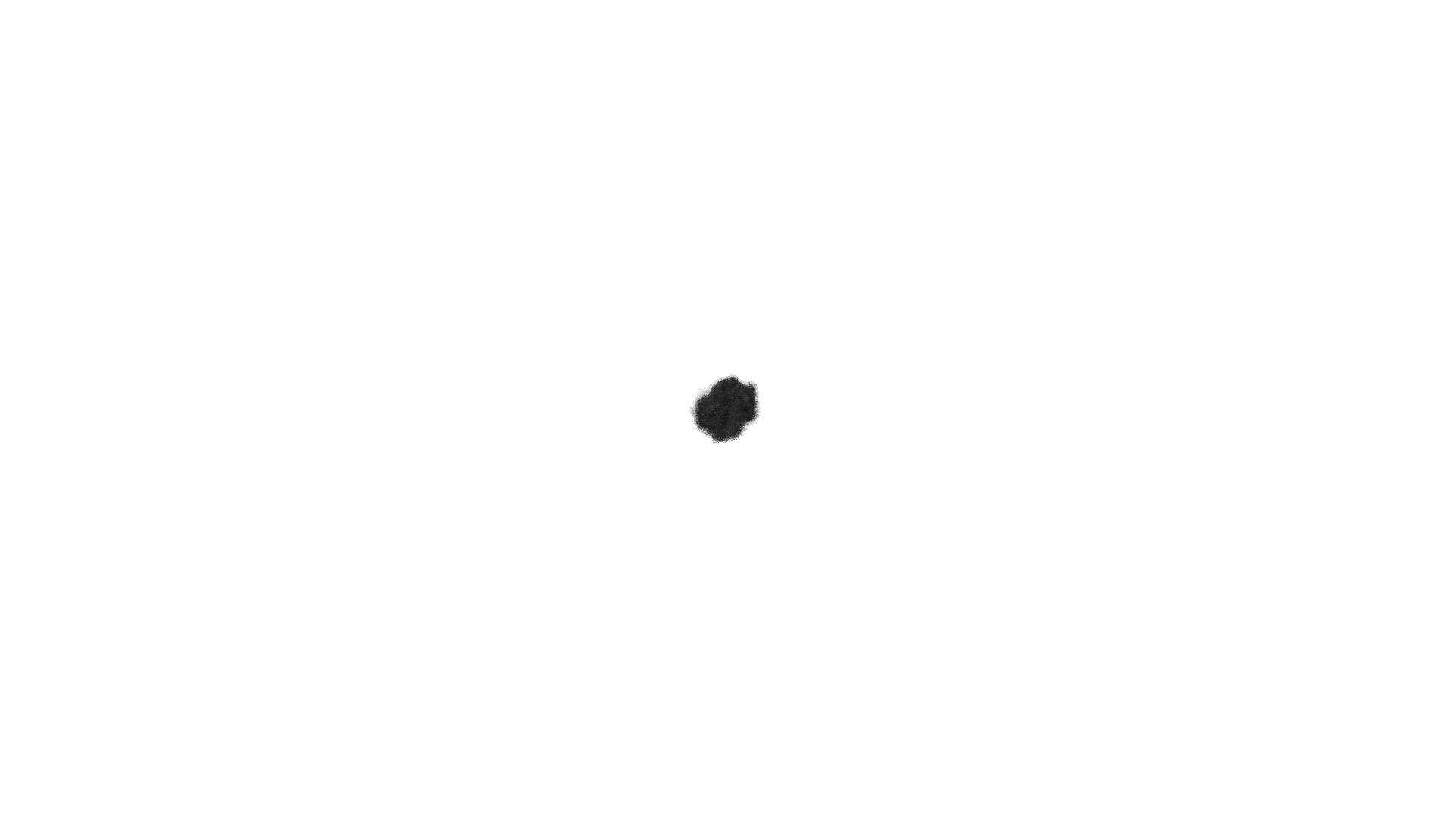}
\def\expIIClabel{SWR~1+F+L} \def\expIICclose{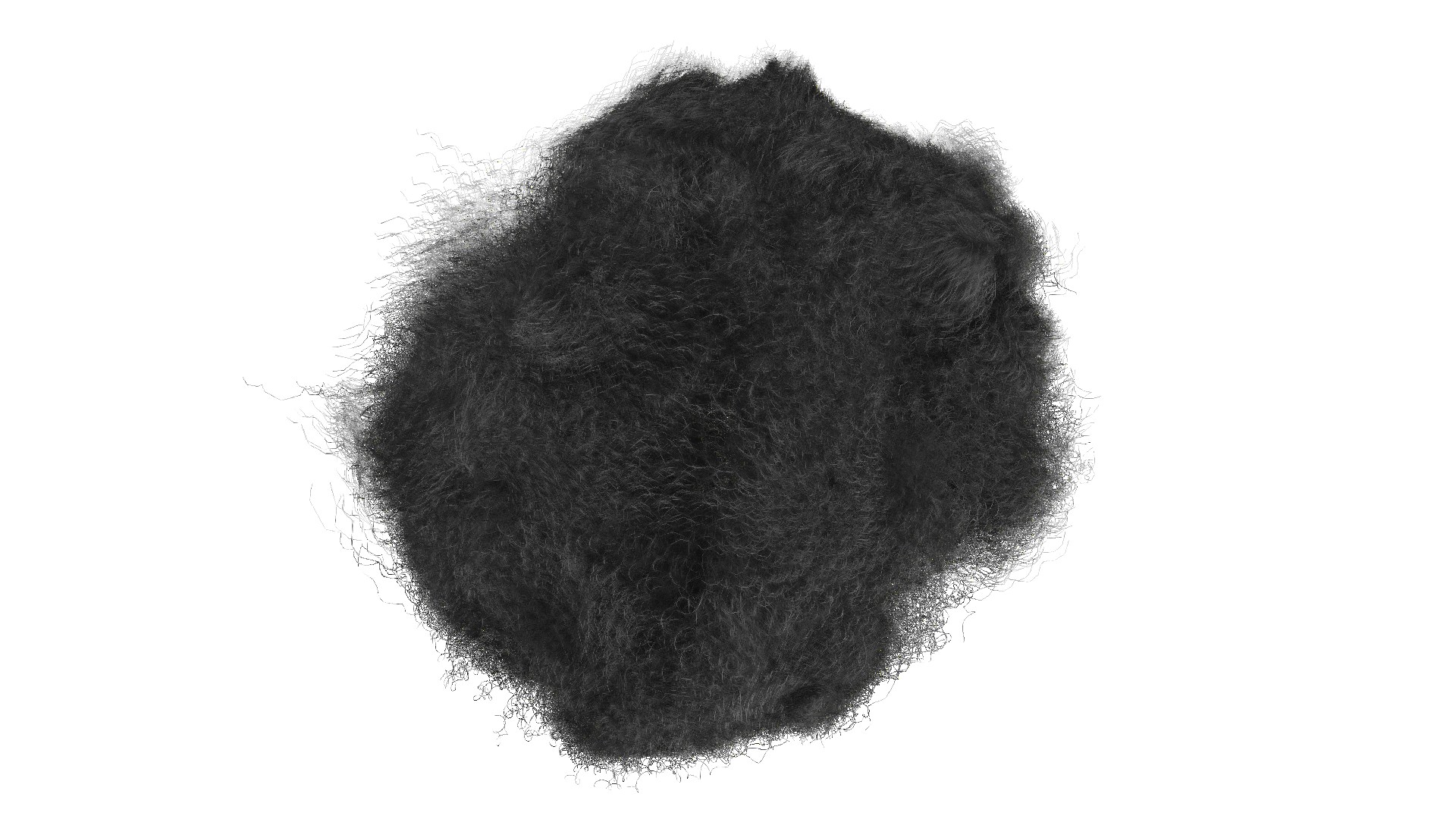}   \def\expIICmid{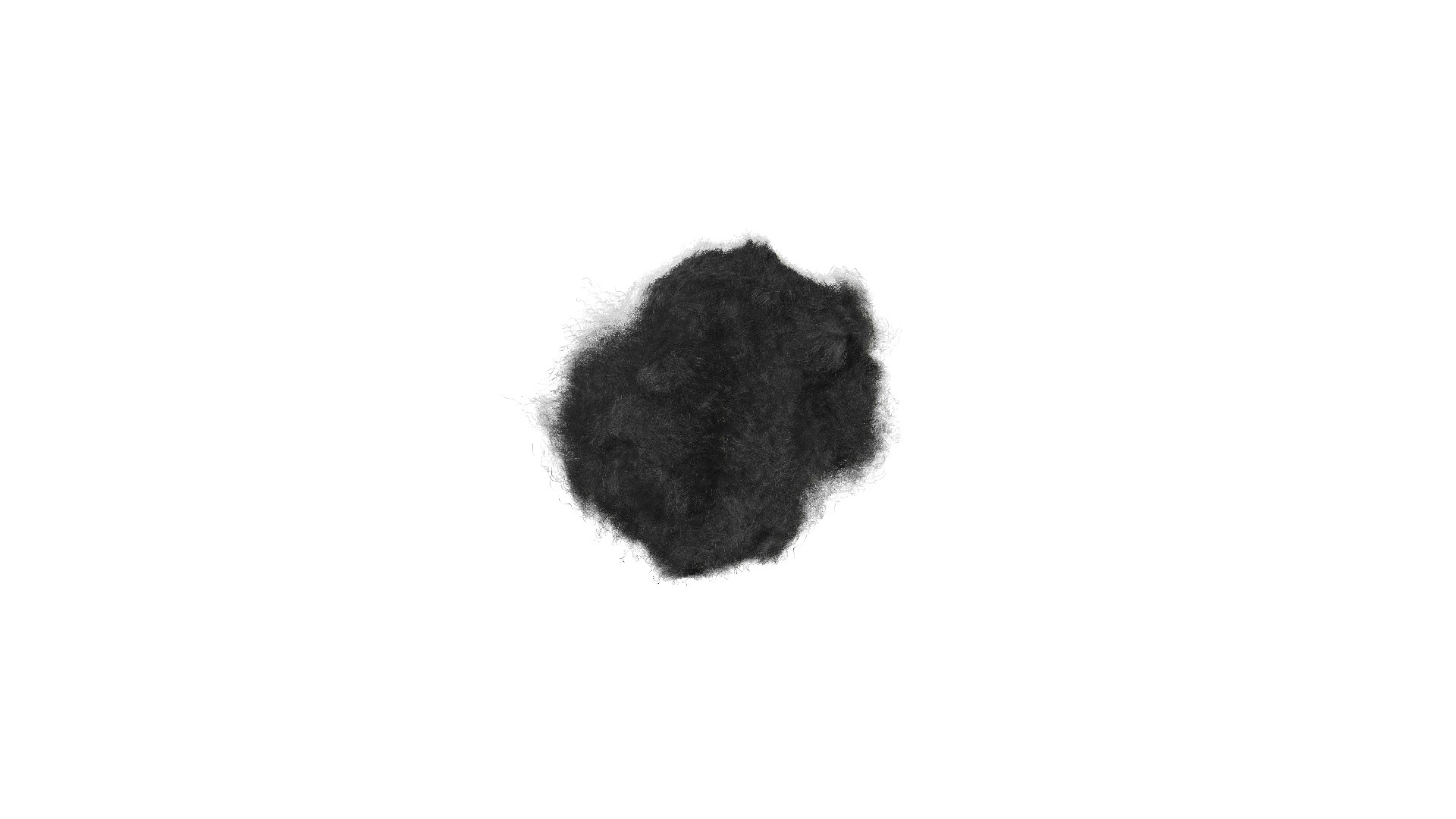}   \def\expIICfar{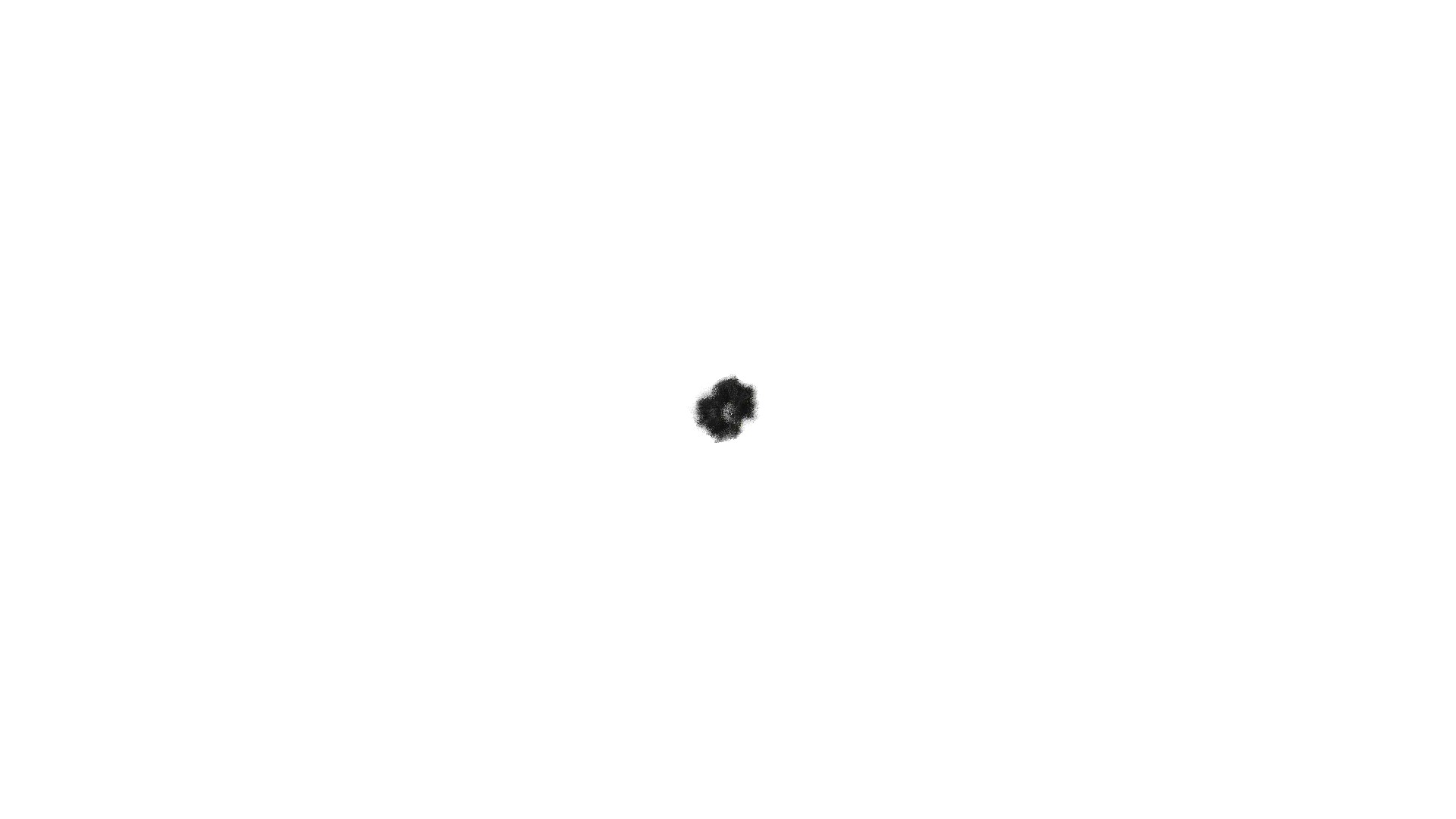}
\def\expIIDlabel{SWR~2}     \def\expIIDclose{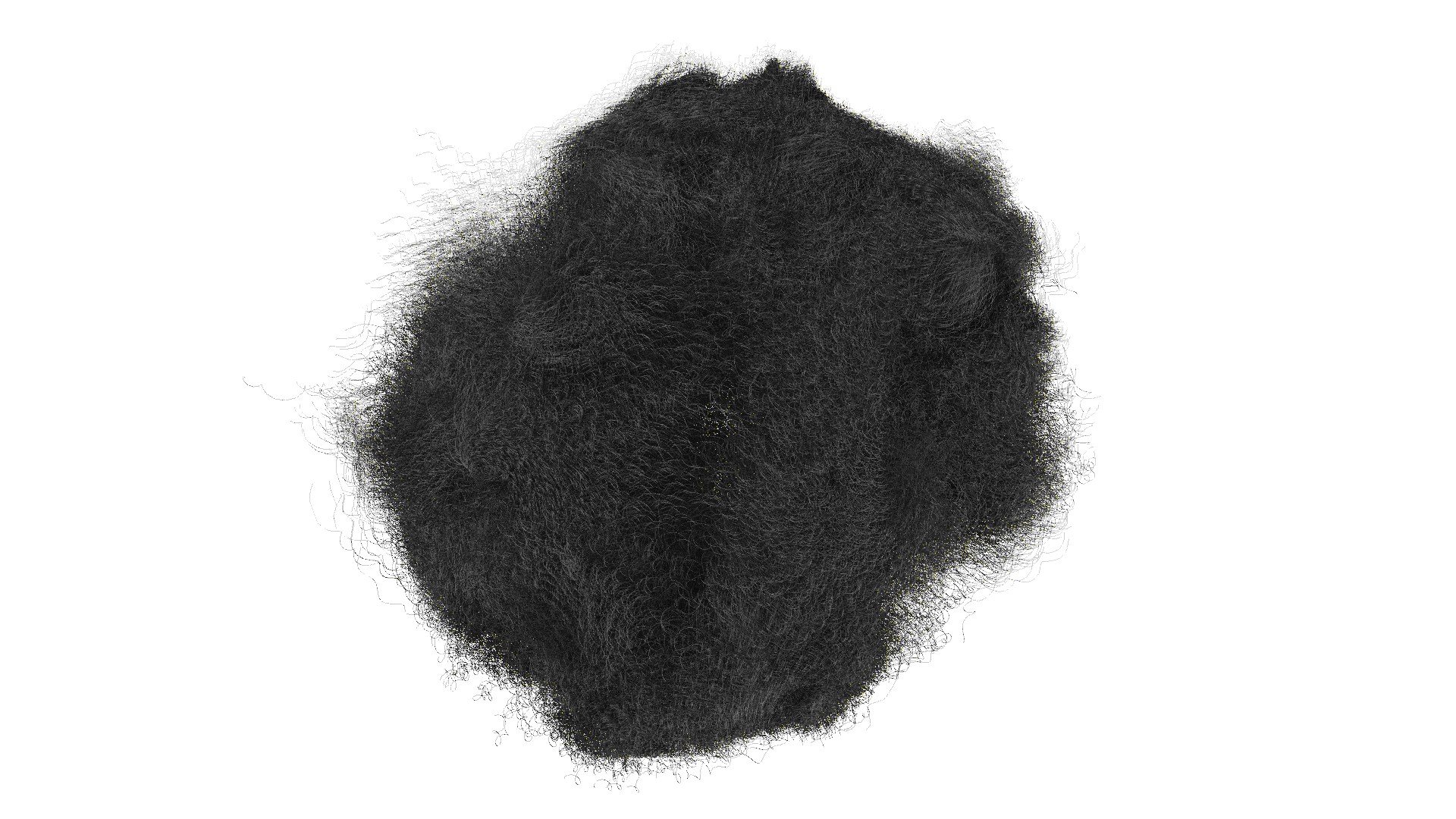}              \def\expIIDmid{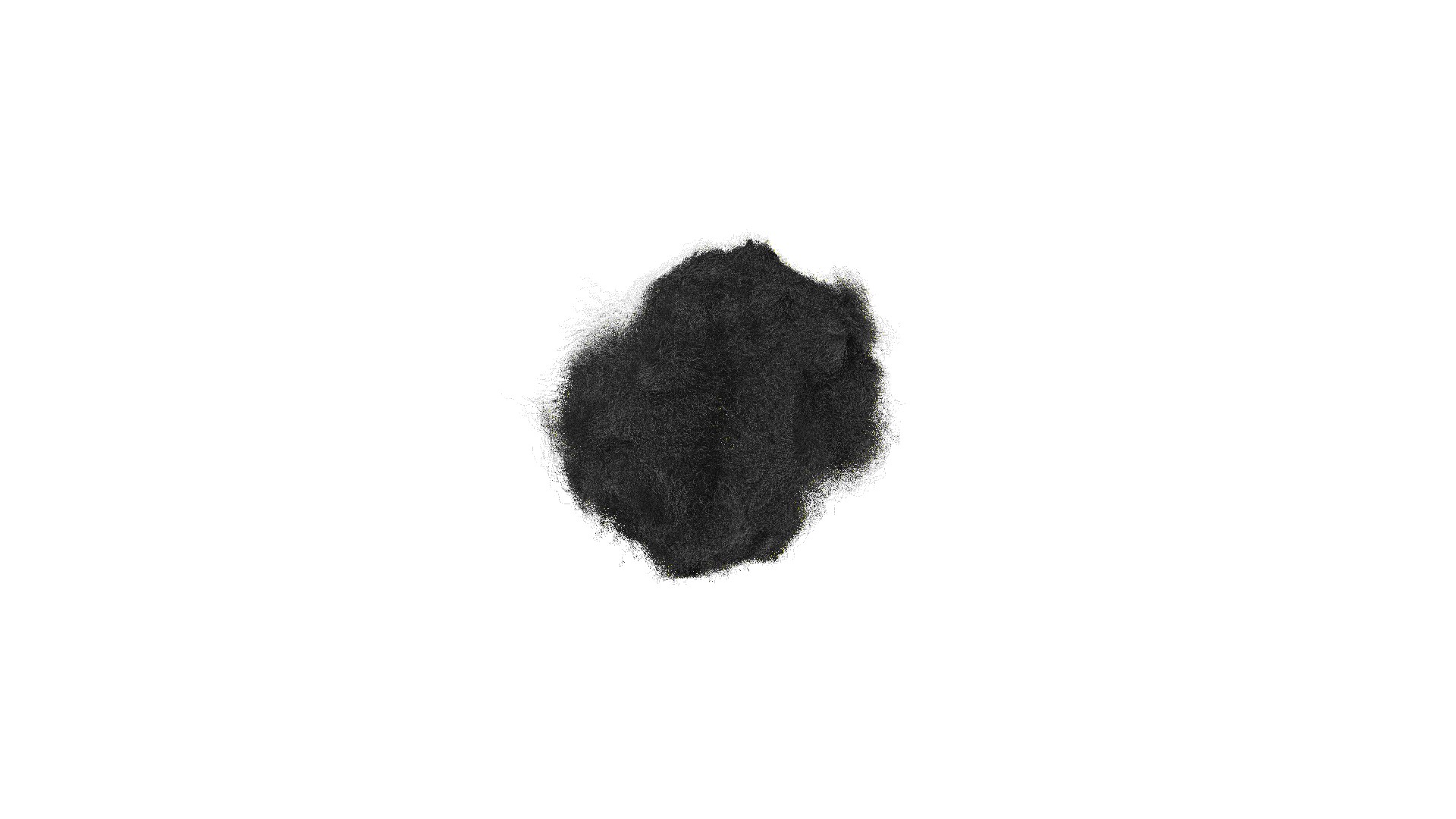}              \def\expIIDfar{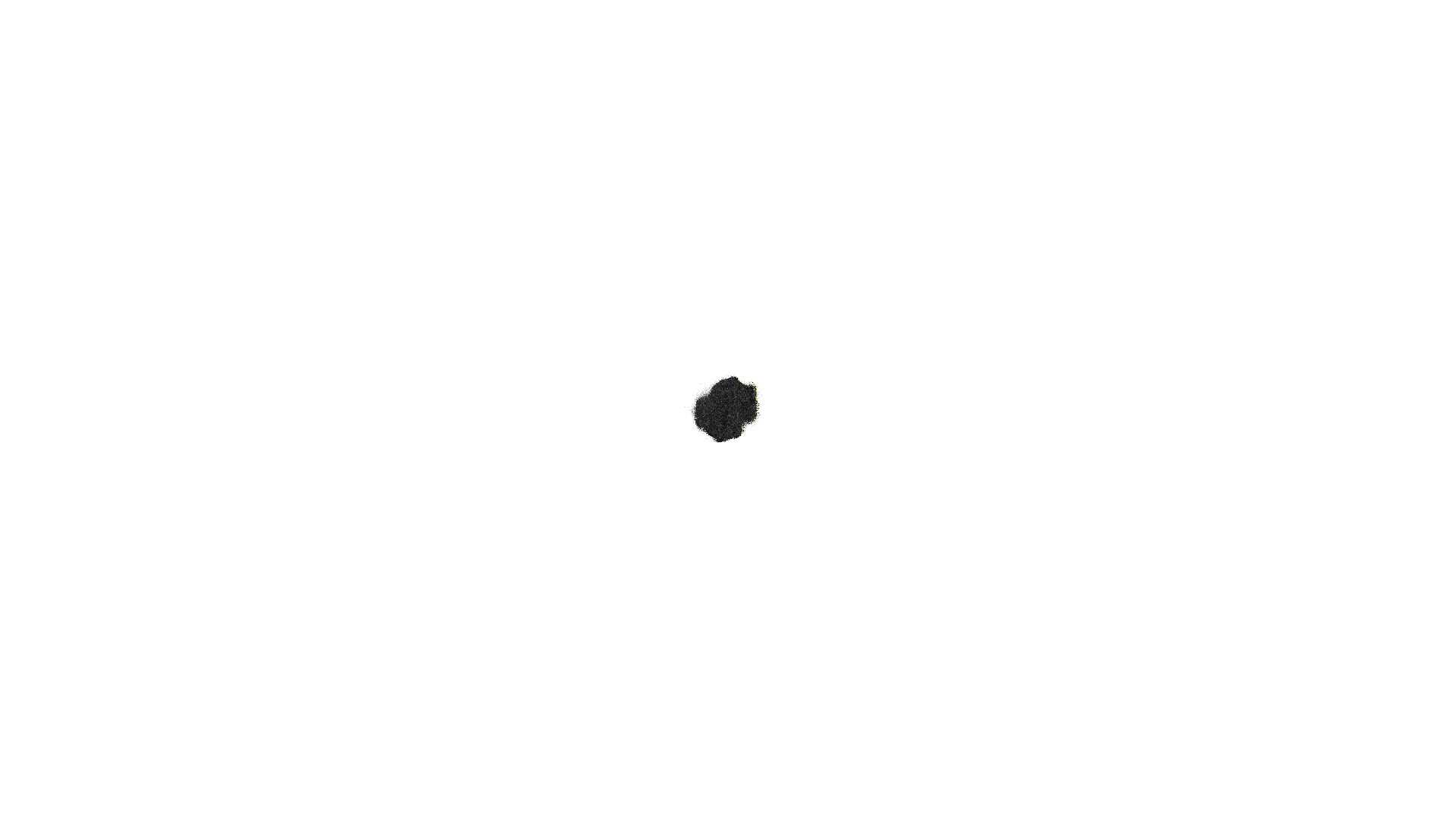}
\def\expIIElabel{SWR~2+L}   \def\expIIEclose{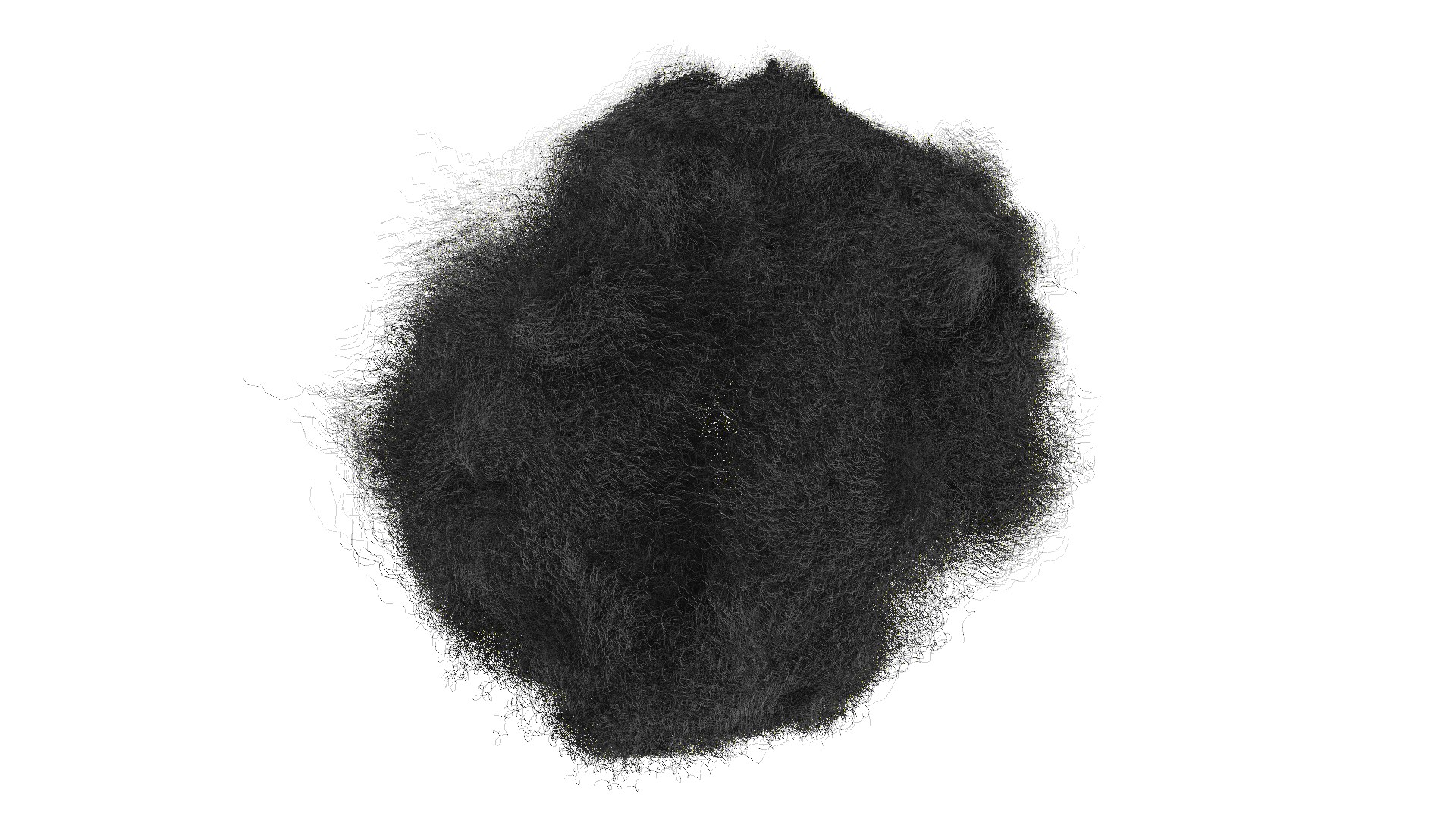}          \def\expIIEmid{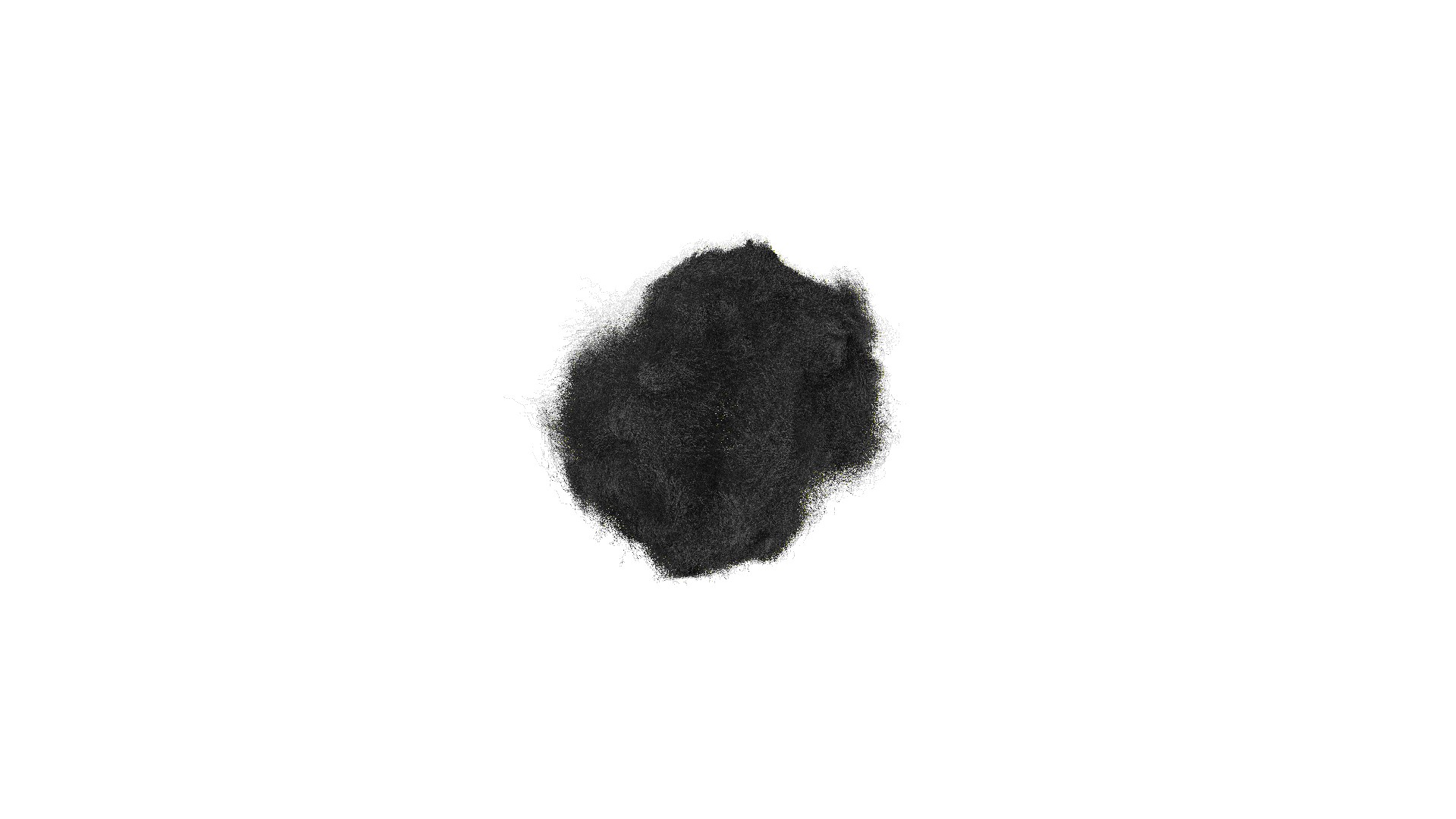}          \def\expIIEfar{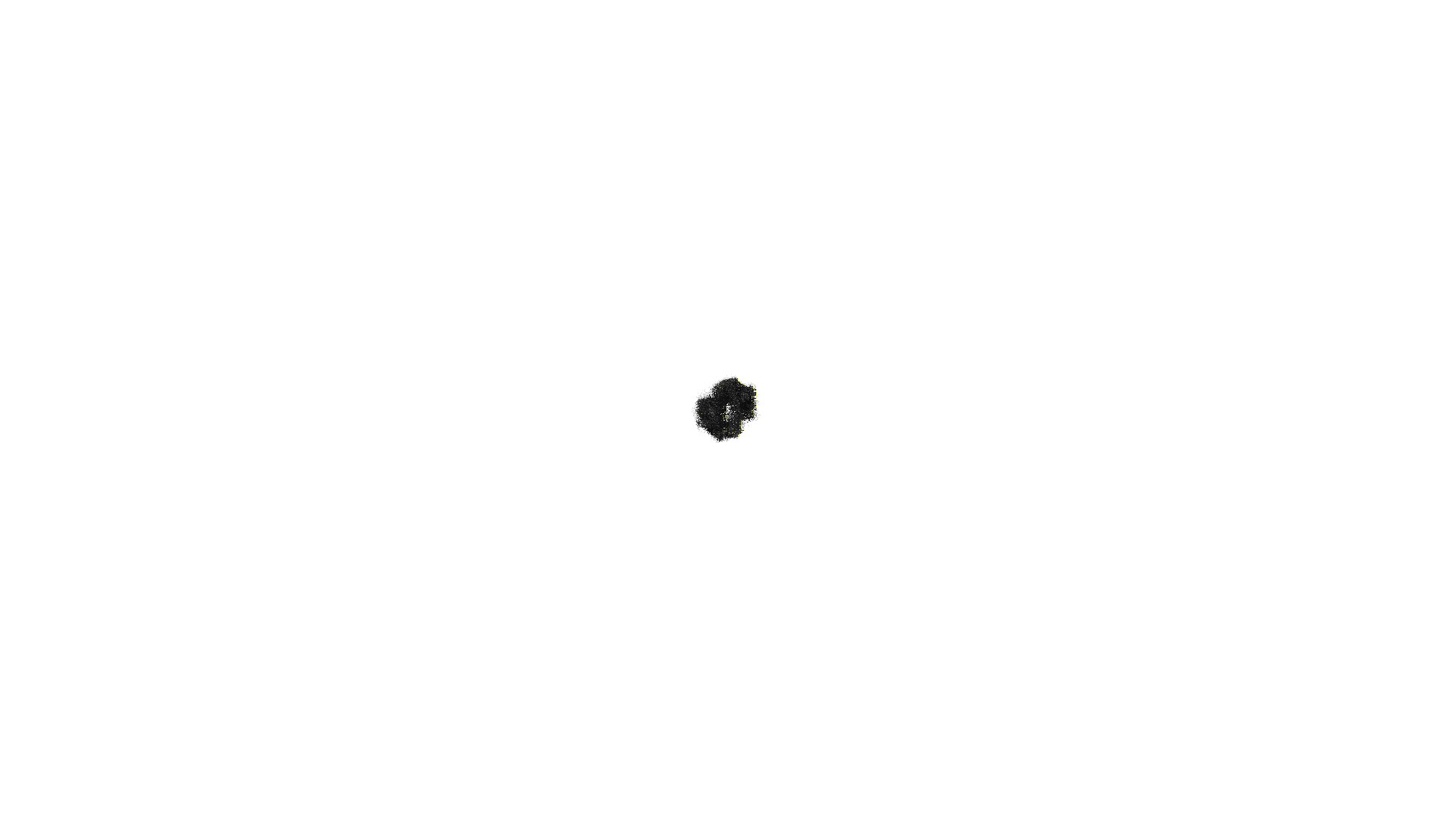}
\def\expIIFlabel{SWR~8}     \def\expIIFclose{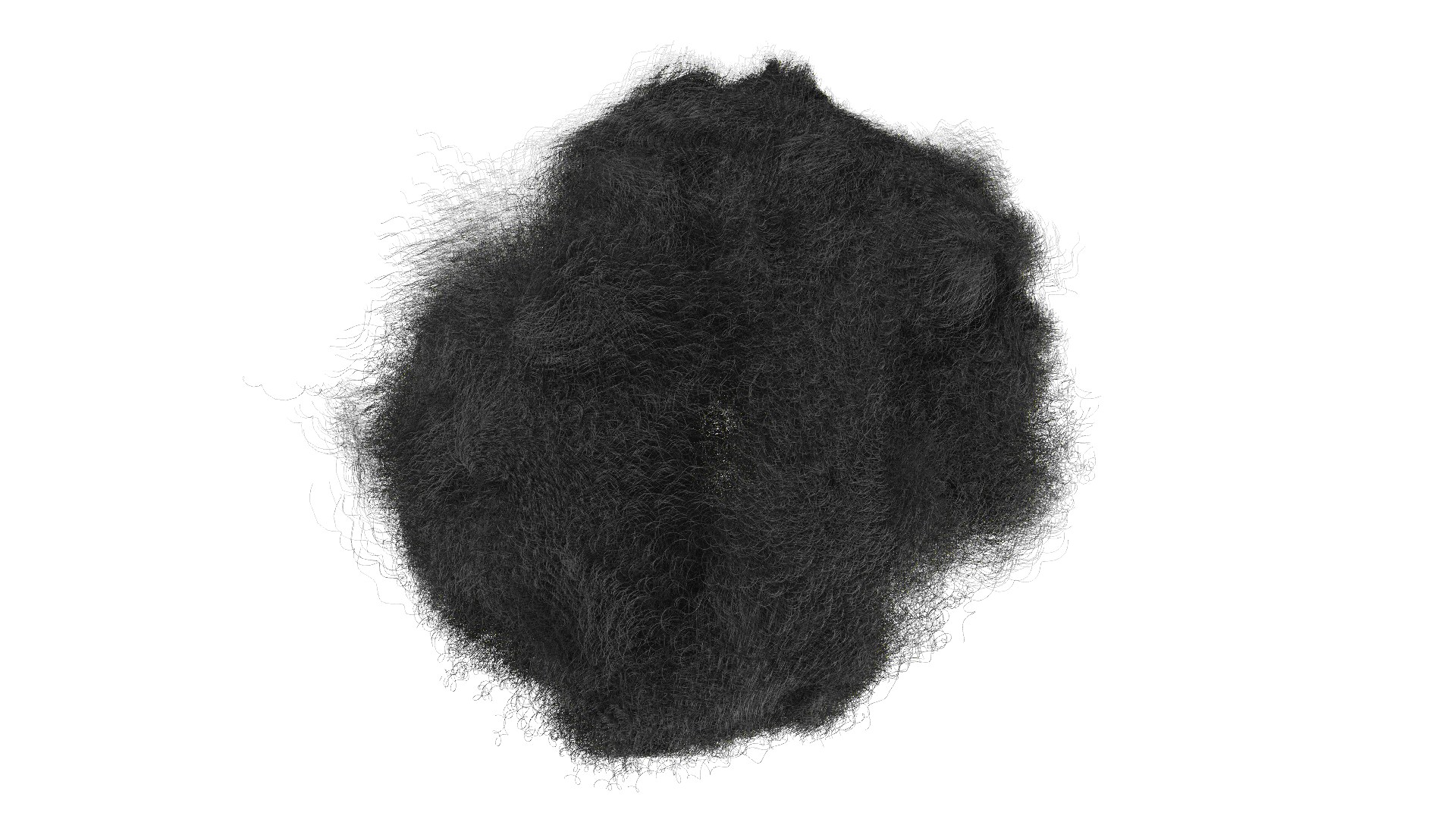}              \def\expIIFmid{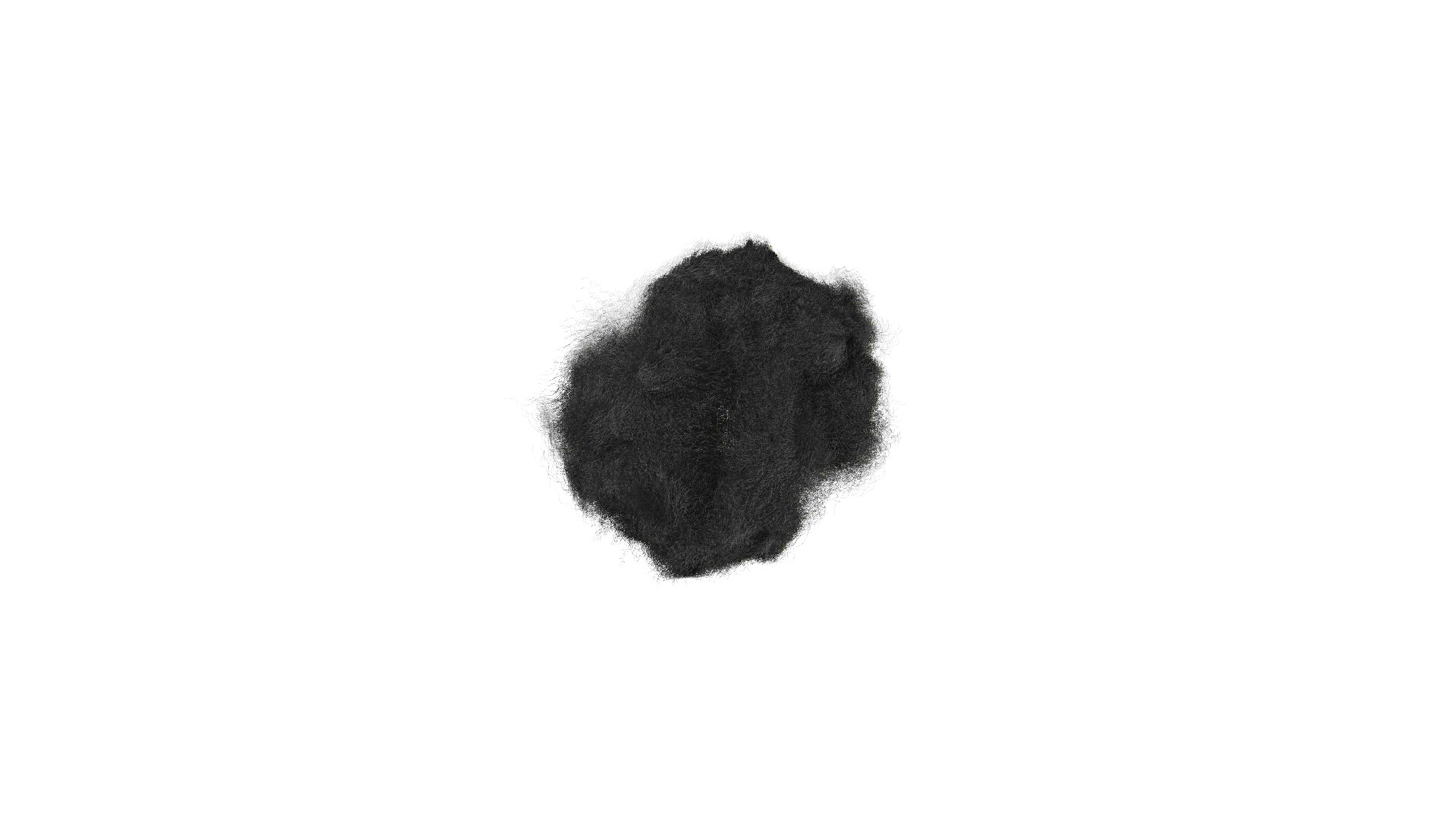}              \def\expIIFfar{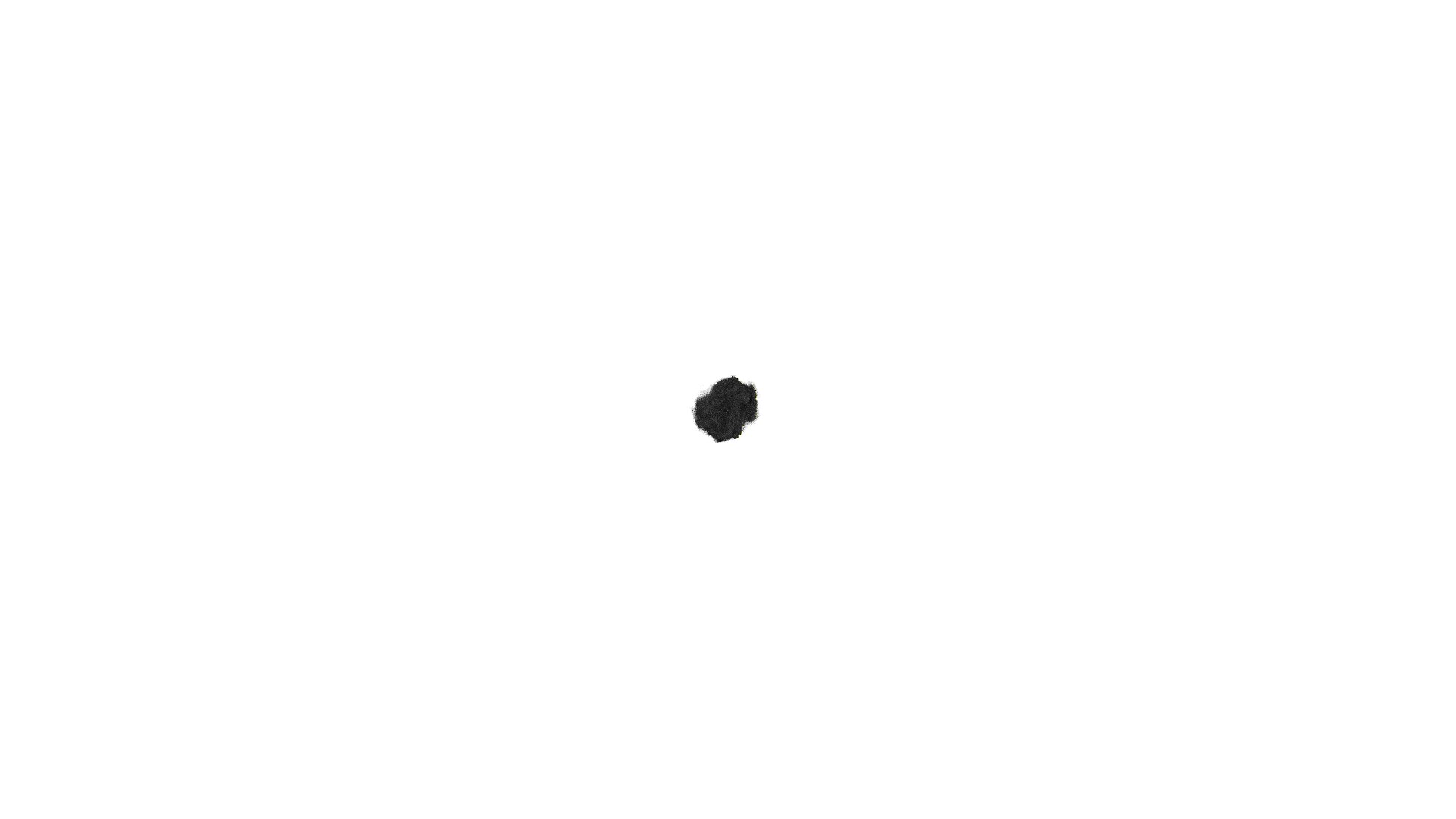}
\def\expIIGlabel{Mesh~8}    \def\expIIGclose{figure/benchmark_afro/mesh8_offset0.jpg}             \def\expIIGmid{figure/benchmark_afro/mesh8_offset100.jpg}             \def\expIIGfar{figure/benchmark_afro/mesh8_offset900.jpg}

% =====================================================================
% END OF EXPERIMENT DATA — everything below is layout/rendering logic.
% =====================================================================

% === Distance colors ===
\definecolor{closecolor}{HTML}{388E3C}
\definecolor{midcolor}{HTML}{F44336}
\definecolor{farcolor}{HTML}{1E88E5}

% === Display sizes ===
\newcommand{\refW}{2.88cm}   % Reference image display width (3.2 * 0.9)
\newcommand{\sliceS}{2.25cm} % Slice / composite image size (square, 2.5 * 0.9)
\newcommand{\borderW}{0.8pt} % Colored border width

% === Cropped slice commands ===
% Groom bounding boxes (from pixel analysis):
%   offset0:   center=(957,519), size=794x886
%   offset100: center=(959,524), size=382x420
%   offset900: center=(960,536), size=71x80
% Each crop = 1.15 * max(w,h), square, centered on groom center.
% Image natural size at 96 DPI: 20in x 11.25in. trim = {left} {bottom} {right} {top}
%
% Close crop: 1018x1018 centered at (957,519) -> pixels x=448..1466, y=10..1028
\newcommand{\closeslice}[1]{%
  \includegraphics[trim=4.6667in 0.5417in 4.7292in 0.1042in, clip, width=\sliceS, height=\sliceS]{#1}%
}
% Mid crop: 482x482 centered at (959,524) -> pixels x=718..1200, y=283..765
\newcommand{\midslice}[1]{%
  \includegraphics[trim=7.4792in 3.2812in 7.5000in 2.9479in, clip, width=\sliceS, height=\sliceS]{#1}%
}
% Far crop: 92x92 centered at (960,536) -> pixels x=914..1006, y=490..582
\newcommand{\farslice}[1]{%
  \includegraphics[trim=9.5208in 5.1875in 9.5208in 5.1042in, clip, width=\sliceS, height=\sliceS]{#1}%
}

% === Experiment column macro (composite with angled bands) ===
% #1=x position, #2=y position, #3=label, #4=close img, #5=mid img, #6=far img
\newcommand{\experimentcolumn}[6]{%
  \def\compY{#2}%
  \def\h{1.125}%       % half of composite size (2.25cm)
  \def\da{-0.45}%      % left divider bottom x offset
  \def\db{-0.07}%      % left divider top x offset
  \def\dc{0.07}%       % right divider bottom x offset
  \def\dd{0.45}%       % right divider top x offset
  \node[anchor=south, font=\footnotesize\sffamily] at (#1, {#2+\h+0.1}) {#3};
  %
  % --- Close band (left) ---
  \begin{scope}
    \clip ({#1-\h}, \compY-\h) --
          ({#1+\da}, \compY-\h) --
          ({#1+\db}, \compY+\h) --
          ({#1-\h}, \compY+\h) -- cycle;
    \node[inner sep=0] at (#1, \compY) {\closeslice{#4}};
  \end{scope}
  %
  % --- Mid band (middle) ---
  \begin{scope}
    \clip ({#1+\da}, \compY-\h) --
          ({#1+\dc}, \compY-\h) --
          ({#1+\dd}, \compY+\h) --
          ({#1+\db}, \compY+\h) -- cycle;
    \node[inner sep=0] at (#1, \compY) {\midslice{#5}};
  \end{scope}
  %
  % --- Far band (right) ---
  \begin{scope}
    \clip ({#1+\dc}, \compY-\h) --
          ({#1+\h}, \compY-\h) --
          ({#1+\h}, \compY+\h) --
          ({#1+\dd}, \compY+\h) -- cycle;
    \node[inner sep=0] at (#1, \compY) {\farslice{#6}};
  \end{scope}
  %
  % --- Borders ---
  % Close: left side + top-left + bottom-left
  \draw[closecolor, line width=0.4pt, line cap=round, dash pattern=on 2pt off 2pt, dash phase=0.5pt]
    ({#1-\h}, \compY-\h) -- ({#1-\h}, \compY+\h);
  \draw[closecolor, line width=0.4pt, line cap=round, dash pattern=on 2pt off 2pt, dash phase=0.5pt]
    ({#1-\h}, \compY+\h) -- ({#1+\db}, \compY+\h);
  \draw[closecolor, line width=0.4pt, line cap=round, dash pattern=on 2pt off 2pt, dash phase=0.5pt]
    ({#1-\h}, \compY-\h) -- ({#1+\da}, \compY-\h);
  % Far: right side + top-right + bottom-right
  \draw[farcolor, line width=0.4pt, line cap=round, dash pattern=on 2pt off 2pt, dash phase=0.5pt]
    ({#1+\h}, \compY-\h) -- ({#1+\h}, \compY+\h);
  \draw[farcolor, line width=0.4pt, line cap=round, dash pattern=on 2pt off 2pt, dash phase=0.5pt]
    ({#1+\dd}, \compY+\h) -- ({#1+\h}, \compY+\h);
  \draw[farcolor, line width=0.4pt, line cap=round, dash pattern=on 2pt off 2pt, dash phase=0.5pt]
    ({#1+\dc}, \compY-\h) -- ({#1+\h}, \compY-\h);
  % Mid: top + bottom between dividers
  \draw[midcolor, line width=0.4pt, line cap=round, dash pattern=on 2pt off 2pt, dash phase=0.5pt]
    ({#1+\db}, \compY+\h) -- ({#1+\dd}, \compY+\h);
  \draw[midcolor, line width=0.4pt, line cap=round, dash pattern=on 2pt off 2pt, dash phase=0.5pt]
    ({#1+\da}, \compY-\h) -- ({#1+\dc}, \compY-\h);
  % Angled dividers (orange, dotted)
  \draw[midcolor, line width=0.4pt, line cap=round, dash pattern=on 2pt off 2pt, dash phase=0.5pt]
    ({#1+\da}, \compY-\h) -- ({#1+\db}, \compY+\h);
  \draw[midcolor, line width=0.4pt, line cap=round, dash pattern=on 2pt off 2pt, dash phase=0.5pt]
    ({#1+\dc}, \compY-\h) -- ({#1+\dd}, \compY+\h);
}

% === Reference column macro ===
% #1=y offset, #2=close img, #3=mid img, #4=far img
% refH = refW * 9/16 = 2.88 * 0.5625 = 1.62cm
\newcommand{\referencecolumn}[4]{%
  % Close reference image + dotted crop rectangle
  \node[inner sep=0, draw=imgbordercolor, line width=0.8pt] (rc) at (\refX, {#1}) {%
      \includegraphics[width=\refW]{#2}%
    };
    \coordinate (rcBL) at ($(rc.south west) + (0.672, 0.058)$);
    \coordinate (rcTR) at ($(rc.south west) + (2.199, 1.585)$);
    \coordinate (rcTL) at (rcBL |- rcTR);
    \coordinate (rcBR) at (rcBL -| rcTR);
    \draw[closecolor, line width=0.4pt, line cap=round, dash pattern=on 2pt off 2pt, dash phase=1.2pt] (rcBL) -- (rcBR);
    \draw[closecolor, line width=0.4pt, line cap=round, dash pattern=on 2pt off 2pt, dash phase=1.2pt] (rcBR) -- (rcTR);
    \draw[closecolor, line width=0.4pt, line cap=round, dash pattern=on 2pt off 2pt, dash phase=1.2pt] (rcTR) -- (rcTL);
    \draw[closecolor, line width=0.4pt, line cap=round, dash pattern=on 2pt off 2pt, dash phase=1.2pt] (rcTL) -- (rcBL);
  % Mid reference image + dotted crop rectangle
  \node[inner sep=0, draw=imgbordercolor, line width=0.8pt] (rm) at (\refX, {#1-1.62-\rowgap}) {%
    \includegraphics[width=\refW]{#3}%
  };
    \coordinate (rmBL) at ($(rm.south west) + (1.077, 0.463)$);
    \coordinate (rmTR) at ($(rm.south west) + (1.800, 1.185)$);
    \coordinate (rmTL) at (rmBL |- rmTR);
    \coordinate (rmBR) at (rmBL -| rmTR);
    \draw[midcolor, line width=0.4pt, line cap=round, dash pattern=on 2pt off 2pt, dash phase=0.5pt] (rmBL) -- (rmBR);
    \draw[midcolor, line width=0.4pt, line cap=round, dash pattern=on 2pt off 2pt, dash phase=0.5pt] (rmBR) -- (rmTR);
    \draw[midcolor, line width=0.4pt, line cap=round, dash pattern=on 2pt off 2pt, dash phase=0.5pt] (rmTR) -- (rmTL);
    \draw[midcolor, line width=0.4pt, line cap=round, dash pattern=on 2pt off 2pt, dash phase=0.5pt] (rmTL) -- (rmBL);
  % Far reference image + dotted crop rectangle
  \node[inner sep=0, draw=imgbordercolor, line width=0.8pt] (rf) at (\refX, {#1-2*(1.62+\rowgap)}) {%
    \includegraphics[width=\refW]{#4}%
  };
    \coordinate (rfBL) at ($(rf.south west) + (1.371, 0.747)$);
    \coordinate (rfTR) at ($(rf.south west) + (1.509, 0.885)$);
    \coordinate (rfTL) at (rfBL |- rfTR);
    \coordinate (rfBR) at (rfBL -| rfTR);
    \draw[farcolor, line width=0.4pt, line cap=round, dash pattern=on 2pt off 2pt, dash phase=1.0pt] (rfBL) -- (rfBR);
    \draw[farcolor, line width=0.4pt, line cap=round, dash pattern=on 2pt off 2pt, dash phase=1.0pt] (rfBR) -- (rfTR);
    \draw[farcolor, line width=0.4pt, line cap=round, dash pattern=on 2pt off 2pt, dash phase=1.0pt] (rfTR) -- (rfTL);
    \draw[farcolor, line width=0.4pt, line cap=round, dash pattern=on 2pt off 2pt, dash phase=1.0pt] (rfTL) -- (rfBL);
}

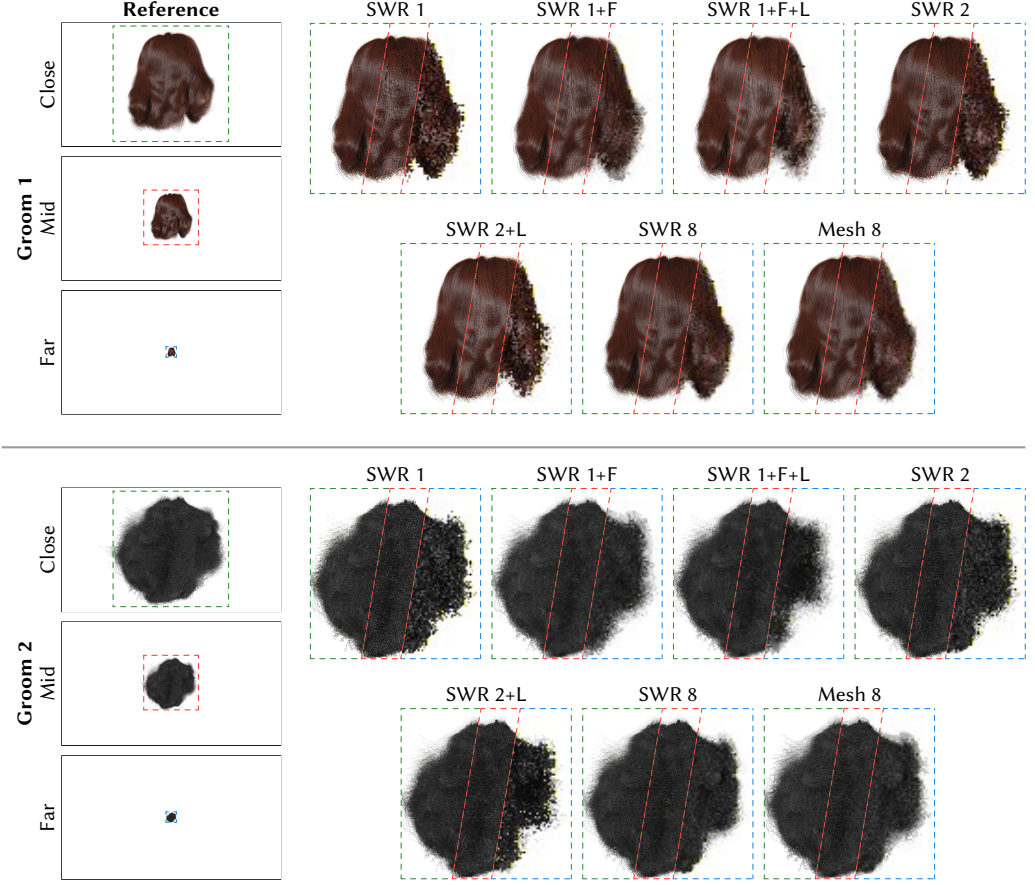
\begin{figure*}[htbp]
\centering
\begin{tikzpicture}[every node/.style={inner sep=0, outer sep=0}]

% === Layout coordinates ===
\def\rowgap{0.15}   % cm between rows
\def\colgap{0.15}   % cm between experiment columns
\def\refexgap{0.4}  % cm gap between reference and experiments
\def\groomsep{1.8}  % cm vertical separation between groom sections

% Derived sizes (from refW and sliceS)
\def\refH{1.62}        % refW * 9/16 = 2.88 * 0.5625
\def\halfRefH{0.81}    % refH / 2
\def\halfSlice{1.125}  % sliceS / 2 = 2.25 / 2
\def\sliceStep{2.4}    % sliceS + colgap = 2.25 + 0.15

% Row Y centers (top to bottom) — for groom 1 reference column
\pgfmathsetmacro{\rowAI}{0}
\pgfmathsetmacro{\rowBI}{-(\refH + \rowgap)}
\pgfmathsetmacro{\rowCI}{-2*(\refH + \rowgap)}

% Groom 1 experiment grid
\pgfmathsetmacro{\compRowTopI}{\rowAI + \halfRefH - \halfSlice}
\pgfmathsetmacro{\compRowBotI}{\rowCI - \halfRefH + \halfSlice}

% Groom 2 vertical offset
\pgfmathsetmacro{\groomIIoffset}{\rowCI - \halfRefH - \groomsep}
\pgfmathsetmacro{\rowAII}{\groomIIoffset}
\pgfmathsetmacro{\rowBII}{\rowAII - (\refH + \rowgap)}
\pgfmathsetmacro{\rowCII}{\rowAII - 2*(\refH + \rowgap)}

% Groom 2 experiment grid
\pgfmathsetmacro{\compRowTopII}{\rowAII + \halfRefH - \halfSlice}
\pgfmathsetmacro{\compRowBotII}{\rowCII - \halfRefH + \halfSlice}

\pgfmathsetmacro{\headerY}{\rowAI + \halfRefH + 0.1}

% Column X positions
\pgfmathsetmacro{\labelX}{-0.2}
\pgfmathsetmacro{\groomLabelX}{-0.5}
\pgfmathsetmacro{\refX}{1.44}                          % 2.88 / 2
\pgfmathsetmacro{\expColA}{2.88 + \refexgap + \halfSlice}
\pgfmathsetmacro{\expColB}{\expColA + \sliceStep}
\pgfmathsetmacro{\expColC}{\expColB + \sliceStep}
\pgfmathsetmacro{\expColD}{\expColC + \sliceStep}

% Bottom row centered: 3 items centered within the 4-column span
\pgfmathsetmacro{\botColCenter}{(\expColA + \expColD) / 2}
\pgfmathsetmacro{\botColA}{\botColCenter - \sliceStep}
\pgfmathsetmacro{\botColB}{\botColCenter}
\pgfmathsetmacro{\botColC}{\botColCenter + \sliceStep}

% Right edge of figure (for separator line)
\pgfmathsetmacro{\rightEdge}{\expColD + \halfSlice}

% === Column headers (shared, shown once at top) ===
\node[anchor=south, font=\footnotesize\sffamily\bfseries] at (\refX, \headerY) {Reference};

% === Row labels (rotated, like groom labels) ===
\node[anchor=center, rotate=90, font=\footnotesize\sffamily] at (\labelX, \rowAI) {\,Close\,};
\node[anchor=center, rotate=90, font=\footnotesize\sffamily] at (\labelX, \rowBI) {\,Mid\,};
\node[anchor=center, rotate=90, font=\footnotesize\sffamily] at (\labelX, \rowCI) {\,Far\,};

\node[anchor=center, rotate=90, font=\footnotesize\sffamily] at (\labelX, \rowAII) {\,Close\,};
\node[anchor=center, rotate=90, font=\footnotesize\sffamily] at (\labelX, \rowBII) {\,Mid\,};
\node[anchor=center, rotate=90, font=\footnotesize\sffamily] at (\labelX, \rowCII) {\,Far\,};

% === Groom labels (rotated, on far left) ===
\pgfmathsetmacro{\groomImidY}{(\rowAI + \rowCI) / 2}
\pgfmathsetmacro{\groomIImidY}{(\rowAII + \rowCII) / 2}
\node[anchor=center, rotate=90, font=\footnotesize\sffamily\bfseries]
  at (\groomLabelX, \groomImidY) {\groomIlabel};
\node[anchor=center, rotate=90, font=\footnotesize\sffamily\bfseries]
  at (\groomLabelX, \groomIImidY) {\groomIIlabel};

% ==============================
% GROOM 1
% ==============================
\referencecolumn{\rowAI}{\refimgcloseI}{\refimgmidI}{\refimgfarI}

% Experiment grid — top row (4 experiments)
\experimentcolumn{\expColA}{\compRowTopI}{\expAlabel}{\expAclose}{\expAmid}{\expAfar}
\experimentcolumn{\expColB}{\compRowTopI}{\expBlabel}{\expBclose}{\expBmid}{\expBfar}
\experimentcolumn{\expColC}{\compRowTopI}{\expClabel}{\expCclose}{\expCmid}{\expCfar}
\experimentcolumn{\expColD}{\compRowTopI}{\expDlabel}{\expDclose}{\expDmid}{\expDfar}
% Experiment grid — bottom row (3 experiments, centered)
\experimentcolumn{\botColA}{\compRowBotI}{\expElabel}{\expEclose}{\expEmid}{\expEfar}
\experimentcolumn{\botColB}{\compRowBotI}{\expFlabel}{\expFclose}{\expFmid}{\expFfar}
\experimentcolumn{\botColC}{\compRowBotI}{\expGlabel}{\expGclose}{\expGmid}{\expGfar}

% === Separator line between grooms ===
% Midpoint between groom 1 bottom (-4.8) and groom 2 top experiment labels
\pgfmathsetmacro{\groomIbottom}{\rowCI - \halfRefH}
\pgfmathsetmacro{\groomIItopLabel}{\rowAII + \halfRefH + 0.1}
\pgfmathsetmacro{\sepY}{(\groomIbottom + \groomIItopLabel) / 2}
\draw[black!40, line width=0.8pt] (\groomLabelX - 0.3, \sepY) -- (\rightEdge, \sepY);

% ==============================
% GROOM 2
% ==============================
\referencecolumn{\rowAII}{\refimgcloseII}{\refimgmidII}{\refimgfarII}

% Experiment grid — top row (4 experiments)
\experimentcolumn{\expColA}{\compRowTopII}{\expIIAlabel}{\expIIAclose}{\expIIAmid}{\expIIAfar}
\experimentcolumn{\expColB}{\compRowTopII}{\expIIBlabel}{\expIIBclose}{\expIIBmid}{\expIIBfar}
\experimentcolumn{\expColC}{\compRowTopII}{\expIIClabel}{\expIICclose}{\expIICmid}{\expIICfar}
\experimentcolumn{\expColD}{\compRowTopII}{\expIIDlabel}{\expIIDclose}{\expIIDmid}{\expIIDfar}
% Experiment grid — bottom row (3 experiments, centered)
\experimentcolumn{\botColA}{\compRowBotII}{\expIIElabel}{\expIIEclose}{\expIIEmid}{\expIIEfar}
\experimentcolumn{\botColB}{\compRowBotII}{\expIIFlabel}{\expIIFclose}{\expIIFmid}{\expIIFfar}
\experimentcolumn{\botColC}{\compRowBotII}{\expIIGlabel}{\expIIGclose}{\expIIGmid}{\expIIGfar}

\end{tikzpicture}
\caption{Rendering quality comparison at close, mid, and far distance. SWR\,=\,software rasterizer, F\,=\,reconstruction filter, L\,=\,LOD, \rev{1, 2, 4, 8 = number of MSAA samples}{}. Zoom insets show center crop from the current view. $\lambda$=3.0 for Groom 1 and $\lambda$=7.0 for Groom 2.}
\label{fig:benchmark}
\end{figure*}

% \todo{
% \begin{itemize}
%     \item big scene with many grooms and many lights (¿teaser?)

%     \item spherical harmonics lighting, baked lighting (visual + performance)

% \end{itemize}
% }

% \subsection{Shading}
% In order to find the fastest path, we compared shading on a per fragment, vertex and pixel basis. For per fragment, every single strand is shaded for every pixel. Per-vertex shading calculates the lighting only at the corners (vertices) of a shape (similar to Gouraud Shading) and then blends those colors across the surface like a gradient. Per-fragment shading instead calculates the light for every individual pixel in the fragment shader (similar to Phong Shading). The last method only calculates shading once per pixel for the closest strand (similar to Deferred Shading).
% \subsection{Level of Detail}

\section{Discussion}

\paragraph*{Compatibility}
Our solution relies only on widely supported features, making it applicable across a broad range of devices. In particular, it requires compute shader support with indirect dispatch functionality (available since Vulkan 1.0), as well as \rev{64-bit}{} integer atomic minimum operations. These atomic operations are supported natively from Vulkan 1.2 onward or can be enabled via extensions on earlier versions. \rev{For devices that do not support atomic operations on images, a buffer-based fallback can be used instead. Note that throughout this paper we solely use the term image for better distinction from non-framebuffer related resources.}{}

\paragraph*{Interactive Updates}
Inherited from hair meshes, we can express a large variety of hairstyles with minimal storage \revlegal{requirements}{cost}. Furthermore, both the hair mesh topology and the styling function can be efficiently modified on the fly, enabling interactive workflows, including animations and efficient physics simulations.

\paragraph*{Multiview Rendering}
Our approach is well suited for multiview rendering scenarios, including virtual reality. Strand assembly is performed only once, while projection is executed per view. Additionally, the deferred shading and filtering \revlegal{computational}{} \rev{cost remains small relative to the rasterization \revlegal{computational}{} cost}{pipeline remains efficient} when rendering to multiple targets, enabling stereo and multiview output with minimal \revlegal{computational}{} overhead.

\subsection{Limitations and Future Work}
While the proposed software rasterizer delivers high performance for far-field rendering, it becomes less efficient for near-field rendering, particularly when strands project wider than a single pixel. A hybrid approach combining software rasterization for distant hair with hardware rasterization for close-up strands could address this limitation. Additionally, although the strand assembly step distributes work well across threads, the DDA rasterization stage exhibits workload imbalance when threads draw lines of varying lengths. Increasing the number of control points alleviates this, but a more balanced work distribution scheme would further improve utilization. \rev{Similar, when the number of control points is significantly lower than the GPU subgroup size, processing multiple strands per workgroup could result in better SIMD lane utilization.}{} Finally, there is room to enhance filtering quality and performance, notably through the integration of temporal filtering to reduce flickering during camera motion.

\begin{acks}
We would like to thank Olivier Maury for his continued support throughout this project. Lukas Lipp and Michael Wimmer have been supported by the Austrian Science Fund (FWF, Grant F77).
\end{acks}

\bibliographystyle{ACM-Reference-Format}
\bibliography{references}
\clearpage
\appendix

\section{Appendix}

\subsection{Comparison: Hardware vs.\ Software Interpolation}
\begin{figure*}[!t]
    \includegraphics[width=\textwidth]{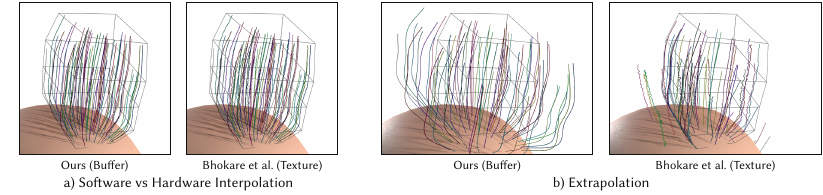}
  \caption{The hardware-supported strand evaluation approach proposed by \citet{bhokareRealTimeHairRendering2024a} suffers from precision errors, which manifest as “jagged” or stepped artifacts (a). This issue arises because hardware interpolation is limited to 256 discrete steps between pixels (i.e., 8-bit subpixel precision). The resulting lack of resolution prevents hair strands from forming smooth, continuous curves, leading to a staggered appearance. Furthermore, hardware samplers can only interpolate within a predefined range and cannot extrapolate beyond it (b). Consequently, hair strands abruptly degenerate near bundle boundaries rather than continuing linearly. Our software-based solution does not exhibit these limitations.}
  \label{fg:hairbug}
\end{figure*}

In our implementation, we evaluated the benefits of using hardware-supported strand evaluation as proposed by \citet{bhokareRealTimeHairRendering2024a}. Their approach approximates a cubic spline using two quadratic splines, which are then evaluated via the de Casteljau algorithm. The expected performance gain arises from offloading a significant portion of the interpolation work to the GPU's texture hardware.

We compared our buffer-based software interpolation approach against the 3D texture-based method of \citet{bhokareRealTimeHairRendering2024a}. From a performance perspective, the hardware interpolation path is consistently slightly slower than the software-based approach (see Table~\ref{tab:interp_perf}).

From a visual standpoint, the hardware method exhibits precision issues and does not support extrapolation, leading to noticeable artifacts (see Figure~\ref{fg:hairbug}).

\begin{table}[h]
\caption{Performance comparison of hardware (texture-based) interpolation and software (buffer-based) interpolation across compute and mesh shader pipeline.}
\label{tab:interp_perf}
\centering
\begin{tabular}{lccc}
\hline
\textbf{Method} & \textbf{MSAA} & \textbf{Interpolation} & \textbf{Time (ms)} \\
\hline
Software Rasterizer & 1 & Hardware (Texture) & 2.8 \\
Software Rasterizer & 1 & Software           & 2.5 \\
Mesh Pipeline       & 1 & Hardware (Texture) & 13.5 \\
Mesh Pipeline       & 1 & Software           & 13.0 \\
\end{tabular}

% \begin{tabular}{c c S[table-format=2.3] S[table-format=2.3]}
%     \toprule
%     \textbf{} & \textbf{MSAA} & {\textbf{Texture}} & {\textbf{Software}} \\
%     \midrule
%     \multicolumn{4}{l}{\textit{Software Rasterizer -- Compute}} \\
%     \addlinespace[2pt]
%                & 1 & 3.5 & 3.1 \\
%     \midrule
%     \multicolumn{4}{l}{\textit{Hardware Rasterizer -- Mesh Shader}} \\
%     \addlinespace[2pt]
%      & 1 & 14.0 & 13.5 \\
% \end{tabular}

\end{table}

% \newpage
\section{Pseudocode}

\begin{code}
    % \captionsetup{type=listing}
    \centering

\begin{codebox}{hlsl}
float3 calcTangent(int idx, int count, float3 prev, float3 curr, float3 next) {
    if (idx == 0) return normalize(next - curr); // Root
    else if (idx == count - 1) return normalize(curr - prev); // Tip
    else return normalize(next - prev);
}
// We use a WG size of 32 since NVIDIA has a subgroup size of 32
[shader("compute"), numthreads(32, 1, 1)]
void main(uint3 dispatchThreadID : SV_DispatchThreadID) {
    int tid = int(dispatchThreadID.x); int WG = 32; int lane = WaveGetLaneIndex();
    BundleStrandInfo b = getBundleStrandInfo();
    PCG rand(b.strandIndex);
    float2 uv = rand.next2();

    // Pass 1: Load base strand layers into shared memory
    int S = (b.numLayers + WG - 1) / WG; // Layers per lane
    int first = tid * S;
    @$\text{L}_{\text{first}}$@ = @$\text{L}_{\text{curr}}$@ = getBundleLayer(b, first, uv); // First of current lane
    @$\text{L}_{\text{last}}$@ = (S > 1) ? getBundleLayer(b, first + S - 1, uv) : @$\text{L}_{\text{curr}}$@; // Last of current lane
    @$\text{L}_{\text{prev}}$@ = subgroupShuffleUp(@$\text{L}_{\text{last}}$@); // Tail of previous lane
    for (int i = 0; i < S && (first + i) < b.numLayers; ++i) {
        int idx = first + i;
        @$\text{L}_{\text{next}}$@ = (i == S - 1) ? subgroupShuffleDown(@$\text{L}_{\text{first}}$@) : ((i == S - 2) ? @$\text{L}_{\text{last}}$@ : getBundleLayer(b, idx + 1, uv)); // Next, last or shuffle
        smLayer[idx].tangent = calcTangent(idx, b.numLayers, @$\text{L}_{\text{prev}}$@.pos, @$\text{L}_{\text{curr}}$@.pos, @$\text{L}_{\text{next}}$@.pos);
        smLayer[idx].pos = @$\text{L}_{\text{curr}}$@.pos; // Write all other attributes...
        @$\text{L}_{\text{prev}}$@ = @$\text{L}_{\text{curr}}$@; @$\text{L}_{\text{curr}}$@ = @$\text{L}_{\text{next}}$@;
    }
    GroupMemoryBarrierWithGroupSync();

    // Pass 2: Evaluate styled strand and draw segments
    int @$N_{\text{cp}}$@ = (b.numLayers - 1) * b.segmentsPerInterval + 1;
    S = (@$N_{\text{cp}}$@ - 1 + WG - 1) / WG; // CPs per lane
    first = tid * S;
    
    @$\text{CP}_{\text{first}}$@ = @$\text{CP}_{\text{curr}}$@ = evalStyledCP(first); // First of current lane
    @$\text{CP}_{\text{last}}$@ = (S > 1) ? evalStyledCP(first + S - 1) : @$\text{CP}_{\text{curr}}$@; // Last of current lane
    @$\text{CP}_{\text{prev}}$@ = subgroupShuffleUp(@$\text{CP}_{\text{last}}$@); // Tail of previous lane
    @$\text{CP}_{\text{next}}$@ = (S == 1) ? subgroupShuffleDown(@$\text{CP}_{\text{first}}$@) : ((S == 2) ? @$\text{CP}_{\text{last}}$@ : evalStyledCP(first + 1)); // Next, last or shuffle
    @$\text{T}_{\text{first}}$@ = @$\text{T}_{\text{curr}}$@ = calcTangent(first, @$N_{\text{cp}}$@, @$\text{CP}_{\text{prev}}$@, @$\text{CP}_{\text{curr}}$@, @$\text{CP}_{\text{next}}$@); @$\text{T}_{\text{prev}}$@ = {};
    for (int i = 1; i <= S; ++i) {
        int idx = first + i;
        @$\text{CP}_{\text{prev}}$@ = @$\text{CP}_{\text{curr}}$@; @$\text{CP}_{\text{curr}}$@ = @$\text{CP}_{\text{next}}$@; @$\text{T}_{\text{prev}}$@ = @$\text{T}_{\text{curr}}$@; 
        @$\text{CP}_{\text{next}}$@ = (i >= S - 1) ? subgroupShuffleDown(@$\text{CP}_{\text{first}}$@) : ((i == S - 2) ? @$\text{CP}_{\text{last}}$@ : evalStyledCP(idx + 1)); // Next, last or shuffle
        @$\text{T}_{\text{curr}}$@ = (i >= S) ? subgroupShuffleDown(@$\text{T}_{\text{first}}$@) : calcTangent(idx, @$N_{\text{cp}}$@, @$\text{CP}_{\text{prev}}$@, @$\text{CP}_{\text{curr}}$@, @$\text{CP}_{\text{next}}$@);
        // Handle both center and conservative layer writes
        if(idx < @$N_{\text{cp}}$@) drawSegment(@$\text{CP}_{\text{prev}}$@, @$\text{CP}_{\text{curr}}$@, @$\text{T}_{\text{prev}}$@, @$\text{T}_{\text{curr}}$@);
    }
}
\end{codebox}
    \caption{Simplified Software Hair Rasterizer. Dispatched per strand.}
    \label{cod:swr_code}
\end{code}

\begin{code}
    % \captionsetup{type=listing}
    \centering
    
\begin{codebox}{hlsl}
RWTexture2D<uint64_t> C; // Center-Color
Texture2D<uint64_t> G; // Center-G-Buffer
Texture2D<uint64_t> A; // Conservative-G-Buffer
bool isEmpty(uint64_t data);

[shader("compute"), numthreads(8, 4, 1)] // in case of subgroup size of 32
void main(uint3 dispatchThreadID : SV_DispatchThreadID)
{
    int2 P = int2(dispatchThreadID.xy);
    if (P >= C) return;

    uint64_t @$G_P$@ = G[P], @$A_P$@ = A[P];
    if (isEmpty(@$G_P$@) && isEmpty(@$A_P$@)) return; // Early out
    float4 @$C_P$@ = C[P]; // Center Color

    bool isCenter = !isEmpty(@$G_P$@), isCons = !isEmpty(@$A_P$@);
    float3 @$\mathbf{t}_P$@; float @$z_P$@;
    if (isCenter) @$\mathbf{t}_P$@, @$z_P$@ = unpackTangentAndDepth(@$G_P$@) // Center Sample
    else if (isCons) @$\mathbf{t}_P$@, @$z_P$@ = unpackTangentAndDepth(@$A_P$@); // Conservative Sample
    float2 @$T_P$@ = getScreenTangent(@$t_P$@);

    float4 @$S$@ = {}; float @$W$@ = 0.0, @$H_\text{center}$@ = 0.0, @$H_\text{cons}$@ = 0.0, @$w_\text{strip}$@ = 0.5; int @$r_\text{filter}$@ = 7;
    for (int dy = -@$r_\text{filter}$@; dy <= @$r_\text{filter}$@; ++dy) {
        for (int dx = -@$r_\text{filter}$@; dx <= @$r_\text{filter}$@; ++dx) {
            int2 Q = P + int2(dx, dy);
            if (Q < 0 || Q >= C) continue; // Bounds check

            float2 @$\mathbf{d}$@ = float2(dx, dy);
            float @$d_\parallel$@ = @$\mathbf{d}\!\cdot\!T_P$@;
            float @$d_\perp$@ = @$\|\mathbf{d} - d_\parallel\mathbf{T}_P\|$@;

            if (@$d_\perp$@ <= @$w_\text{strip}$@) { // Within Strip
                if (isCenter) @$H_\text{center}$@++;
                else if (isCons) @$H_\text{cons}$@++;
            }

            float4 @$C_Q$@ = isEmpty(C[Q]) ? float4(0.0) : unpackColor(C[Q]); // Neighbor color
            float @$w_s$@ = @$\exp\bigl({-d_\parallel^2/\sigma_\parallel^2 - d_\perp^2/\sigma_\perp^2}\bigr)$@; // Spatial
            float @$w_c$@ = @$\exp\bigl({-\|C_P\!-\!C_Q\|^2/\sigma_c^2}\bigr)$@; // Color
            float @$w$@ = @$w_s$@ * @$w_c$@; @$S$@ += @$\mathbf{C}_Q$@ * @$w$@; @$W$@ += @$w$@;
        }
    }

    float4 @$\hat{\mathbf{C}}_P$@ = (@$W$@ > 0.0) ? (@$S$@ / @$W$@) : @$C_P$@;
    float @$\alpha$@ = ((@$H_\text{center}$@ + @$H_\text{cons}$@) > 0.0) ? @$H_\text{center}$@ / (@$H_\text{center}$@ + @$H_\text{cons}$@) : 1;
    C[P] = packColor({@$\hat{\mathbf{C}}_P, \alpha$@}); // Write back to the Center-Color texture
}
\end{codebox}
    \caption{Simplified Orientation-Aware Anisotropic Hair Reconnection Filter. Dispatched per pixel.}
    \label{cod:filter_code}
\end{code}

\end{document}